\documentclass[journal, twocolumn]{IEEEtran}
\ifCLASSINFOpdf

\else

\fi
\usepackage{mathrsfs}

\usepackage{cuted}
\usepackage{amsmath}
\usepackage{ntheorem}

\usepackage{makeidx}  
\usepackage{algorithm}
\usepackage{algorithmicx}
\usepackage{graphicx}
\usepackage{subfigure}
\usepackage{epstopdf}
\usepackage{bm}
\usepackage{cite}
\usepackage{stfloats}

\usepackage{amssymb}
\usepackage{graphicx}
\newtheorem{theorem}{Theorem}
\usepackage{url}

\usepackage{tabularx,booktabs}
\usepackage{caption}

\newcolumntype{C}{>{\centering\arraybackslash}X} 
\usepackage{lipsum}

\usepackage{makecell} 

\usepackage{graphicx}
\usepackage{color}

\newtheorem{lem}{Lemma}

\usepackage{diagbox}
\usepackage{multirow}

\usepackage{algpseudocode}

\begin{document}

\title{Transmission Design for XL-RIS-Aided Massive MIMO System with Visibility Regions}

\author{Luchu Li, Kangda Zhi, Cunhua Pan}

\maketitle

\begin{abstract}
	This paper proposes a two-timescale transmission scheme for extremely large-scale (XL)-reconfigurable intelligent surfaces (RIS)-aided massive multi-input multi-output (MIMO) systems considering visibility regions (VRs). The beamforming of base stations (BS) is designed based on rapidly changing instantaneous channel state information (CSI), while the phase shifts of RIS are configured based on slowly changing statistical CSI. Specifically, we first formulate a system model with spatially correlated Rician fading channels and introduce the concept of VRs. Then, we derive a closed-form approximate expression for the achievable rate applicable to any number of BS antennas and RIS elements, and analyze the impact of VRs on system performance and complexity. Next, we solve the problem of maximizing the minimum user rate by optimizing the phase shifts of RIS through an algorithm based on accelerated gradient ascent. Finally, we present numerical results to demonstrate the performance of the gradient algorithm from different aspects and reveal the low system complexity of deploying XL-RIS in massive MIMO systems with the help of VRs.
\end{abstract}


\IEEEpeerreviewmaketitle

\section{Introduction}
In recent years, Reconfigurable Intelligent Surface (RIS), also known as Intelligent Reflecting Surface (IRS), has been seen as a key technology for next-generation communication systems and extensively studied by academia and industry \cite{5}, \cite{6}. It is an array composed of a large number of reconfigurable subwavelength units, which can achieve signal enhancement, attenuation, focusing, scattering, and other functions. 
Specifically, each RIS element can independently induce changes in the amplitude and phase of the incident signal, allowing the reflected signal to be constructively added to signals from other paths. Therefore, it can customize the wireless environment by manually controlling electromagnetic waves, thereby improving spectrum efficiency. Meanwhile, the RIS can achieve significant performance gains especially when the direct links between BS and users are obstructed. Therefore, compared to existing multi-antenna systems, RIS has the potential to achieve better performance in terms of cost and energy consumption. 

Recently, many researchers have studied the application of RIS in wireless systems from different perspectives. The free-space path loss models for RIS-assisted wireless communications were developed for different scenarios by studying the physics and electromagnetic nature of RISs \cite{a1}. In \cite{a2}, the authors provided a tutorial overview of IRS-aided wireless communications and elaborated its reflection and channel models, hardware architecture, and practical constraints, as well as various appealing applications in wireless networks. The authors in \cite{a3} provided a comprehensive overview of recent advances in RIS/IRS-aided wireless systems from the signal processing perspective. 

RISs can also bring gains to various communication scenarios. Specifically, a novel cascaded channel estimation strategy with low pilot overhead was proposed by exploiting the sparsity and the correlation of multiuser cascaded channels in millimeter-wave (mmWave) MISO systems \cite{a4}. In \cite{a5}, the authors proposed a novel two-stage uplink channel estimation strategy with reduced pilot overhead and error propagation for RIS-aided multi-user (MU) mmWave multiple-antenna systems. The authors of \cite{a6} proposed a three-dimensional (3D) physics-based double RIS cooperatively assisted MIMO stochastic channel model for unmanned aerial vehicle (UAV)-to-ground communication scenarios. The authors of \cite{a7} derived the approximate expression of the sum achievable security data rate (ASDR) with statistical channel state information (CSI) and demonstrated the effectiveness of using RIS in improving security performance.
Furthermore, researchers have conducted numerous studies and improvements on RIS-aided systems to meet the different needs of actual environments. A more efficient channel estimator was developed by leveraging the unitary approximate message passing (UAMP), which facilitates the applications of existing algorithms to a general RIS-aided MIMO system with a larger number of RIS elements \cite{a8}. Considering radio frequency damage and phase noise, the authors of \cite{21} conducted a theoretical study on the fundamental trade-off between spectrum and energy efficiency of RIS communication networks. In \cite{a9}, a model-free cross-entropy (CE) algorithm was proposed to optimize the binary RIS configuration for improving the signal-to-noise ratio (SNR) at the receiver. The authors in \cite{a10} demonstrated that holographic RIS can become an attractive way of achieving high spatial resolutions in satellite networks. In \cite{a11}, the authors presented a double-RIS cooperatively-assisted MIMO communication system, where two distributed RISs are respectively deployed near a nearby user group and a multi-antenna BS.

Although the advantages of RIS have been demonstrated in the aforementioned studies, most of them considered estimating instantaneous CSI in each channel coherent interval. However, in practice, transmission design schemes based on instantaneous CSI face two challenges \cite{27}. Firstly, the RIS typically consists of a large number of reflective elements to ensure the expected gains, while the pilot overhead of channel estimation schemes could be proportional to the number of RIS elements \cite{28}, which leads to excessive pilot overhead. Secondly, the estimating of instantaneous CSI requires frequent beamforming calculations and information feedback, resulting in high computational complexity, feedback overhead, and energy consumption.

Therefore, to address the above issues, researchers are currently considering exploiting the two-timescale scheme to make the application of RIS more practical. Specifically, in a two-timescale scheme, BS beamforming is designed based on instantaneous aggregated CSI, while optimization of RIS parameters is decided only based on long-term statistical CSI, such as position and angle information \cite{27}. As a result, it is only necessary to update the phase shift of RIS components when the statistical CSI changes \cite{12}. 
Specifically, \cite{30} and \cite{12} respectively studied RIS-aided single-user communication systems from the perspectives of interference and non-interference. In \cite{36}, the author considered a multi-RIS multi-user MIMO system and optimized the reconfigurable elements of RIS, as well as beamforming vectors at BS and user devices, without the need for instantaneous CSI or second-order channel statistics, to maximize system and rate. \cite{37} proposed an effective passive reflection beamforming design for RIS, which utilizes statistical CSI to analyze the achievable rate of the network while considering the impact of CSI estimation errors. In \cite{b1}, the authors proposed a two-timescale channel estimation framework and a dual-link pilot transmission scheme, where the BS transmits downlink pilots and receives uplink pilots reflected by the RIS. To maximize the average achievable rate of RIS-assisted MIMO systems, \cite{b2} presented a new two-timescale beamforming approach, where the IRS is configured relatively infrequently based on statistical CSI, and the base station precoder and power allocation are updated frequently based on quickly outdated instantaneous CSI. 
In the presence of spatial correlation and electromagnetic interference (EMI), the authors in \cite{27} investigated the two-timescale transmission scheme for RIS-aided MIMO systems, 
For cell-free systems, a low-complexity algorithm via the two-timescale transmission protocol is proposed in \cite{b3}, where the joint beamforming at BSs and RISs is facilitated via an alternating optimization framework. By using the active RIS architectures, a multi-user two-timescale channel estimation protocol is proposed to minimize the pilot overhead in \cite{b4}.

As massive MIMO technology is being commercially deployed in 5G communication systems, one research direction is to utilize extra-large (XL) passive reflection arrays for RIS. Owing to the low-cost feature, the size of an RIS can be made large to extend the service region and acquire the ability for localization enhancement \cite{40}. Therefore, the XL-RIS can achieve unprecedented gains in spatial multiplexing, SNR, energy efficiency, and coverage probability. Considering that the total path loss in the RIS-aided link is the product of those in the separate channels, the RIS is generally deployed near the users to avoid a large path loss. As a result, the users are in the near-field region of the XL-RIS, and the practical spherical electromagnetic (EM) wavefront can no longer be approximated as a planar wavefront [c1]. Moreover, the channel between the XL-RIS and a user shows spatial non-stationarity due to the near-field effect and the obstacles in the environment, according to the research product on extra-large MIMO (XL-MIMO) systems [c2]. As unequal path loss is experienced by different RIS elements, and some RIS elements suffer from blockage caused by obstacles, the channel power will be concentrated on a portion of the RIS elements, which is indicated by the visibility region (VR) \cite{40}. Therefore, as the XL-RIS is always employed near the users to avoid a large path loss, the channel between the RIS and a user is determined by the user's position and the VR. As a result, it is necessary to reset typical spatially correlated models to capture spatially non-stationary fading between XL arrays \cite{43}, \cite{46}. 
 
Taking the existence of VR into consideration, \cite{47} proposed a polar simultaneous orthogonal matching pursuit (OMP) algorithm for near-field channel estimation in XL-MIMO systems. \cite{51} proposed a simple non-stationary channel model with VR and analyzed the performance of conjugate beamforming and zero-forcing precoding in the downlink multi-user XL-MIMO systems. Considering the non-stationarity between RIS and users, \cite{40} proposed a recognition module that can estimate the VR of each user with line-of-sight (LoS) components between RIS and users. For XL-RIS-aided systems, \cite{48} proposed a low-cost near-field beam training scheme. In \cite{50}, the authors studied the channel estimation problem of XL-RIS-aided millimeter wave communication systems considering near-field effects and spatial non-stationary effects. The authors in \cite{52} investigated the achievable rate of XL-RIS-aided downlink massive MIMO in the presence of wholly/partial VRs, spatial correlation, and imperfectly estimated CSI.

However, in most of these works, the focus is on channel estimating and spatial non-stationary effects caused by XL-MIMO. Moreover, \cite{30}–\cite{b4} did not consider the employment of XL-RIS and ignored the VRs of users. As far as we know, works on the two-timescale transmission design of spatially non-stationary XL-RIS-aided massive MIMO systems are seldom conducted.

In this article, we analyze the uplink (UL) two-timescale transmission scheme of XL-RIS-assisted massive MIMO systems with VRs, considering the spatial non-stationary effects caused by XL-RIS, and use the Rician channel model to evaluate the impact of LoS and non-LoS (NLOS) channel components.
\begin{itemize}
	\item By employing the low-complexity MRC detector at the BS, we derive a closed-form expression of the uplink achievable rate under the correlated Rician channel model. This expression applies to any finite number of BS antennas and RIS elements and depends only on the statistical CSI. Then, we analyze the impact of VRs and evaluate its contribution on reducing channel matrix dimensions.
	
	\item Based on the derived expression, we propose an algorithm dependent on accelerated gradient ascent to solve the problem of maximizing the minimum user rate with statistical CSI.
	
	\item To unveil the property and benefit of integrating XL-RIS into massive MIMO networks, we provide simulation results about various important parameters. The results reveal that the existence of VR leads to a decrease in performance while the exploitation of its properties is beneficial to reduce the computational complexity of the system. Besides, it is shown that the proposed gradient algorithm can effectively solve the RIS phase optimization problem and has higher efficiency compared to GA.
\end{itemize}

The rest of this article is organized as follows. Section \ref{two} introduces the system model. Section \ref{three} derives a closed-form analytical expression for achievable rates and analyzes the impact of VRs. Section \ref{four} presents the optimization problem of maximizing the minimum user rate and introduces a complete solution method based on accelerated gradient ascent. Section \ref{fifth} analyzes a large number of numerical results, and Section \ref{sixth} draws conclusions.

\section{System Model}\label{two}

\begin{figure}[t]
	\setlength{\abovecaptionskip}{0pt}
	\setlength{\belowcaptionskip}{0pt}
	\centering
	\includegraphics[width= 0.45\textwidth]{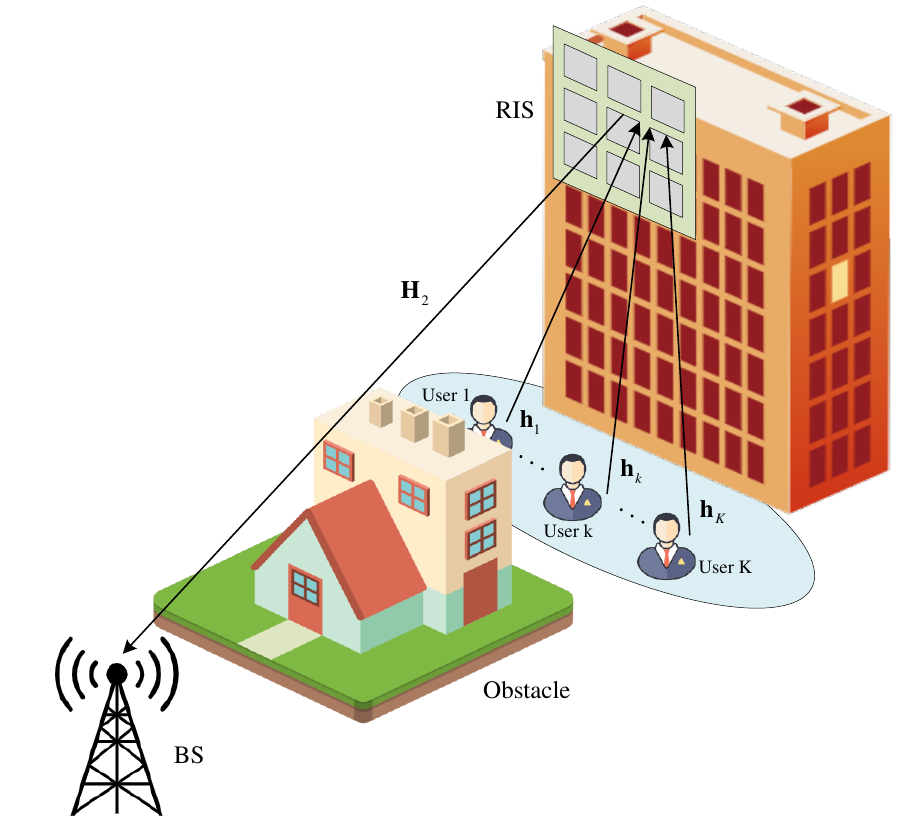}
	\DeclareGraphicsExtensions.
	\caption{An XL-RIS-aided massive MIMO system.}
	\label{picture1}
\end{figure}

As shown in Fig. \ref{picture1}, we consider a typical XL-RIS-aided massive MIMO communication system with BS, XL-RIS, and ${K}$ single antenna users. The BS and the XL-RIS are equipped with ${M}$ antennas and $N_{1}\times N_{2}=N$ passive reflective elements, using uniform linear array (ULA) and uniform square plane array (USPA) models ($N_{1}$ rows and  $N_{2}$ columns) respectively. The RIS is connected to the BS through a dedicated transmission link. Since the ground communication links can be obstructed by buildings and other objects, we assume that the direct link from the BS to users will be blocked \cite{58}.

Considering the fact that the RIS can be installed on the wall of tall buildings, it can assist in creating channels dominated by LoS propagation along with a few scatters \cite{d1}. As scattering is much weaker far from the ground, we adopt the Rician fading model for the channels between the BS and the RIS and the channel between users and the RIS \cite{d2}. We denote $\mathbf{H}_{2}\in\mathbb{C}^{M\times N}$ and $\mathbf{H}_{1}=[\mathbf{h}_{1},\mathbf{h}_{2},...,\mathbf{h}_{K}]\in\mathbb{C}^{N\times K}$ as the channel between the BS and the RIS, and the channel between the RIS and users, respectively. Specifically, these channels can be modeled as:

\begin{align}
	&\mathbf{h}_{k}=\sqrt{\alpha_{k}}\left(\sqrt{\frac{\varepsilon_{k}}{\varepsilon_{k}+1}}\mathbf{D}_{k}^{1/2}\mathbf{\bar{h}}_{k}+\sqrt{\frac{1}{\varepsilon_{k}+1}}\mathbf{R}_{VR,k}^{1/2}\mathbf{\tilde{h}}_{k}\right), \label{1}\\
	&\mathbf{H}_{2}=\sqrt{\beta}\Bigg(\sqrt{\frac\delta{\delta+1}}\bar{\mathbf{H}}_{2}+\sqrt{\frac1{\delta+1}}\tilde{\mathbf{H}}_{2}\mathbf{R}_{ris}^{1/2}\Bigg), \label{2}
\end{align}

\noindent where $\alpha_k$ and $\beta$ represent the large-scale path loss factors of the user k-RIS link and BS-RIS link, respectively. $\varepsilon_{k}$ and $\delta$ are the Rician factors of the user ${k}$-RIS link and the RIS-BS link, respectively. $\mathbf{D}_k$ denotes the visibility indicator matrix. $\mathbf{R}_{ris}$ and $\mathbf{R}_{VR,k}$ are the spatial correlation matrices of the RIS and the VRs, respectively. In addition, $\mathbf{\bar{h}}_k$ and $\mathbf{\bar{H}}_2$ denote the LoS components, while $\tilde{\mathbf{h}}_k$ and $\tilde{\mathbf{H}}_2$ denote the NLoS components.

For the NLoS components, $\tilde{\mathbf{h}}_k$ and $\tilde{\mathbf{H}}_2$ are independent identically distributed (i.i.d.) complex Gaussian random variables with zero mean and unit variance. For the LoS components, $\mathbf{\bar{h}}_k$ and $\mathbf{\bar{H}}_2$ are modeled as follows:
\begin{align}
	&\mathbf{\bar{h}}_{k}=\mathbf{a}_N(\varphi_{kr}^a,\varphi_{kr}^e),\label{3}\\
	&\mathbf{\bar{H}}_2=\mathbf{a}_M(\phi_r^a,\phi_r^e)\mathbf{a}_N^H(\varphi_t^a,\varphi_t^e),\label{4}
\end{align}

\noindent where  $\varphi_{kr}^{a}$ and $\varphi_{kr}^{e}$ are the azimuth and elevation angles of arrival (AoA) at the RIS from user ${k}$, respectively. $\varphi_{t}^{a}$ and $\varphi_{t}^{e}$ are the azimuth and elevation angles of departure (AoD) towards the BS from the RIS, respectively. $\varphi_{r}^{a}$ and $\varphi_{r}^{e}$ are the azimuth and elevation angles AoA at the BS from the RIS, respectively. Besides, we assume that these angles can be calculated by using the locations obtained from the global position system (GPS) \cite{d1}. In addition, $\mathbf{a}_X(\vartheta^a,\vartheta^e)\in\mathbb{C}^{X\times1}$ represents the array response vector, where the ${x}$-th term is:
\begin{align}\label{uspa}
	\begin{aligned}
		\left[ {{\bf a}_M}\left( {\vartheta _{}^a,\vartheta _{}^e} \right) \right]_{ x} &= \exp\left\{j2\pi \frac{d_{bs}}{\lambda }
		(x-1)\sin \vartheta _{}^e\sin \vartheta _{}^a\right\},\\
		\left[ {{\bf a}_N}\left( {\vartheta _{}^a,\vartheta _{}^e} \right) \right]_{ x} &= \exp\! \left\{\! j2\pi \frac{d_{ris}}{\lambda }
		\left(    \lfloor \left({ x} - 1 \right)/\sqrt{N}\rfloor \sin \vartheta _{}^e\sin \vartheta _{}^a\right.\right.\\
		&\quad\left.	\left.	+ \left(\left({x}-1\right)\bmod \sqrt{N}\right)  \cos \vartheta _{}^e \right) \right\},
	\end{aligned}
\end{align}

\noindent where $d_{bs}$, $d_{ris}$, and $\lambda$ represent the BS antenna spacing, RIS element spacing, and wavelength, respectively. In order to simplify the symbols, in the remainder of this article, $\mathbf{a}_M(\phi_r^a,\phi_r^e)$ and $\mathbf{a}_N(\phi_t^a,\phi_t^e)$ are simply denoted by $\mathbf{a}_{M}$ and $\mathbf{a}_{N}$, respectively.

\begin{figure}[H]
	\setlength{\abovecaptionskip}{0pt}
	\setlength{\belowcaptionskip}{0pt}
	\centering
	\includegraphics[width= 0.45\textwidth]{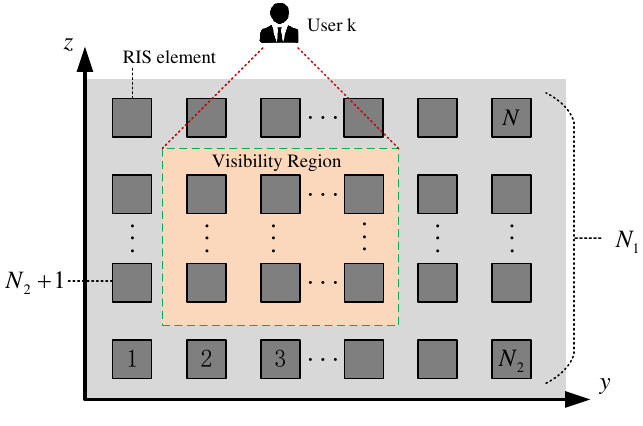}
	\DeclareGraphicsExtensions.
	\caption{The 2D geometry of an RIS with VR consisting of $N_2$ elements per row and $N_1$ elements per column.}
	\label{picture2}
\end{figure}

As shown in Fig. \ref{picture2}, the elements of the RIS are indexed row-by-row by $n\in[1, N]$. We denote $\mathbf{u}_p$ as the location of the $p$-th element with respect to the origin in Fig. \ref{picture2}, which can be represented by the following model \cite{67}:
\begin{align}
	\mathbf{u}_{p}&=[0,y(p)d_{H},z(p)d_{V}]^{\mathsf{T}},\label{7}\\
	y(p)&=\mathrm{mod}(p-1,N_{2}),\label{8}\\
	z(p)&=\lfloor p-1/N_{2} \rfloor,\label{9}
\end{align}
\noindent where $d_{H}$ and $d_{V}$ respectively denote the width and height of a single RIS element, which can be assumed to be $d_{H}=d_{V}=d_{ris}$. $y(p)$ and $z(p)$ are the horizontal and vertical indices of element $n$, respectively, on the two-dimensional grid. Therefore, the elements inside the spatial correlation matrix of the RIS can be defined as:

\begin{align}
	[\mathbf{R}_{ris}]_{p,q}=\text{sinc}\left(\frac{2\left|\left|\mathbf{u}_p-\mathbf{u}_q\right|\right|}{\lambda}\right)\ \text{for}\ p,q=1,\ldots,N,\label{6}
\end{align}
\noindent where $\lambda$ is the wavelength, and $\left|\left|\mathbf{u}_p-\mathbf{u}_q\right|\right|$ represents the distance between the $p$-th and $q$-th elements of the RIS. the width and height of a single RIS element, which can be assumed to be $d_{H}=d_{V}=d_{ris}$. As illustrated in Fig. \ref{picture2}, we use VRs to capture spatially non-stationary fading characteristics. The visibility indicator matrix $\mathbf{D}_k\in\mathbb{C}^{N\times N}$ is defined as: \begin{align}\label{D_k}
	\mathbf{D}_k=\mathrm{diag}(\mathbf{d}_k),
\end{align}
\begin{equation}\label{d_k}
	[\mathbf{d}_k]_n=\left\{
	\begin{aligned}
		1 & , & \mathrm{if}\quad n \in \mathbf{\nu}_k \\
		0 & , & \mathrm{otherwise}.
	\end{aligned}
	\right.
\end{equation} 

$\mathbf{d}_k\in\mathbb{C}^{N}$ is the visibility indicator vector. $\mathbf{\nu}_k$ is a set containing the indices of RIS elements, which are visible to the $k$-th user. Then, the RIS covariance matrix for the $k$-th user with partial visibility is defined as \cite{51}: 
\begin{align}
	\mathbf{R}_{VR,k}=\mathbf{D}_{k}^{1/2}\mathbf{R}_{ris}\mathbf{D}_{k}^{1/2}.\label{11}
\end{align}

With the help of XL-RIS, the received signal at the BS can be expressed as:
\begin{align}
	\mathbf{y}=\sqrt{p}\mathbf{Q}\mathbf{x}+\mathbf{n},\label{13}
\end{align}

\noindent where $p$ is the average transmit power of each user. $\mathbf{x}=[x_{1},x_{2},\ldots,x_{K}]^{T}$ represents the information sequence from ${K}$ users, satisfying $\mathbb{E}\{|x_k|^2\}=1$. $\mathbf{n}\sim\mathcal{CN}(\mathbf{0},\sigma^2\mathbf{I}_M)$ is the additional white Gaussian noise (AWGN).  $\mathbf{Q}\triangleq\mathbf{H}_{2}\mathbf{\Phi}\mathbf{H}_{1}\in\mathbb{C}^{M\times K}$ is the total aggregation channel from ${K}$ users to the BS. 

We adopt the low-complexity maximal-ratio-combining (MRC) technique, due to its simple construction and near-optimal performance for massive MIMO \cite{d3}.
The BS performs MRC by multiplying the received signal $\mathbf{y}$ with $\mathbf{Q}^{H}$, as follows
\begin{align}
	\mathbf{r}_{VR}=\mathbf{Q}^{H}\mathbf{y}=\sqrt{p}\mathbf{Q}^{H}\mathbf{Q}\mathbf{x}+\mathbf{Q}^{H}\mathbf{n}.\label{13}
\end{align}

Before calculating the achievable rate, the signal of user k can be written as:
\begin{align}
	r_{VR,k}=\sqrt{p} \ \mathbf{q}_k^H\mathbf{q}_kx_k+\sqrt{p}\sum_{i=1,i\neq k}^K\mathbf{q}_k^H\mathbf{q}_ix_i+\mathbf{q}_k^H\mathbf{n},\label{14}
\end{align}
\noindent where $\mathbf{q}_k\triangleq\mathbf{H}_2\mathbf{\Phi}\mathbf{h}_k\in\mathbb{C}^{M\times 1}$ is the ${k}$-th column of $\mathbf{Q}$ denoting the cascaded channel from user ${k}$ to the BS. Specifically, $\mathbf{q}_k$ can be represented as:
\begin{align}
	\mathbf{q}_k=&\mathbf{H}_2\mathbf{\Phi}\mathbf{h}_k\notag\\
	=&\sqrt{c_k\delta\varepsilon_k}\left(\mathbf{\bar{H}}_2\mathbf{\Phi}\mathbf{D}_{k}^{\frac{1}{2}}\mathbf{\bar{h}}_k\right)+\sqrt{c_k\delta}\left(\mathbf{\bar{H}}_2\mathbf{\Phi}\mathbf{R}_{VR,k}^{\frac{1}{2}}\mathbf{\tilde{h}}_k\right)\notag\\
	&+\sqrt{c_{k}\varepsilon_{k}}\left(\tilde{\mathbf{H}}_{2}\mathbf{R}_{ris}^{\frac{1}{2}}\mathbf{\Phi D}_{k}^{\frac{1}{2}}\bar{\mathbf{h}}_{k}\right)\notag\\
	&+\sqrt{c_{k}}\left(\tilde{\mathbf{H}}_{2}\mathbf{R}_{ris}^{\frac{1}{2}}\mathbf{\Phi R}_{VR,k}^{\frac{1}{2}}\tilde{\mathbf{h}}_{k}\right),\label{12}
\end{align}

\noindent where $c_{k}\triangleq\frac{\beta\alpha_{k}}{(\delta+1)(\varepsilon_{k}+1)}$, $\boldsymbol{\Phi}=\mathrm{diag}\big\{e^{j\theta_{1}},e^{j\theta_{2}},...,e^{j\theta_{N}}\big\}$ denotes the phase shift matrix of the RIS, $\theta_n\in[0,2\pi)$ denotes the phase shift of the ${n}$-th RIS element.

Due to the ergodic channel, we can express the uplink achievable rate of user $k$ as $R_{VR,k}=\mathbb{E}\left\{\log_2\left(1+\mathrm{SINR}_k\right)\right\}$, and the SINR of user $k$ is given by \cite{a}:
\begin{align}
	\mathrm{SINR}_k=\frac{p\parallel\mathbf{q}_k\parallel^4}{\sum_{i=1,i\neq k}^Kp|\mathbf{q}_k^H\mathbf{q}_i|^2+\sigma^2\parallel\mathbf{q}_k\parallel^2}.\label{15}
\end{align}

\section{Achievable Rate Analysis}\label{three}

Considering the VRs, the approximate closed-form expression of the achievable rate in the XL-RIS-aided massive MIMO system is derived and analyzed in this section. We will also present an asymptotic expression with reduced dimensions utilizing the features of VRs.

\begin{theorem}\label{theorem1}
	In the XL-RIS-aided massive MIMO systems, the uplink ergodic achievable rate of user $k$ can be approximated as
	\begin{align}
		&R_{VR,k}\notag\\
		&\approx\mathbb{E}\left\{\log_2\left(1+\frac{pE_{VR,k}^\mathrm{signal}(\mathbf\Phi)}{p\sum_{i=1,i\neq k}^KI_{VR,ki}(\mathbf\Phi)+\sigma^2E_{VR,k}^\mathrm{noise}(\mathbf\Phi)}\right)\right\},\label{rate}
	\end{align}
	\noindent where the first, second, and third items are the desired signal received by the BS, the multi-user interference, and the thermal noise, respectively. The values of $E_{VR,k}^{\mathrm{noise}}(\boldsymbol\Phi)$, $E_{VR,k}^\mathrm{signal}(\boldsymbol{\Phi})$ and $I_{VR,ki}(\boldsymbol{\Phi})$ are given in (\ref{noise})-(\ref{f_old}) at the bottom of the next page.
	\begin{figure*}[b]
		\hrulefill
		\begin{align}
			&E_{VR,k}^{\mathrm{noise}}(\boldsymbol\Phi)=c_{k}\big(M\delta\varepsilon_{k}|f_{k}(\boldsymbol\Phi)|^{2}+\delta f_{k,1,1}(\boldsymbol\Phi)+\varepsilon_{k}f_{kk,2}(\boldsymbol\Phi)M+f_{k,3,1}(\boldsymbol\Phi)M\big),\label{noise}\\
			&E_{VR,k}^\mathrm{signal}(\boldsymbol{\Phi})=M^2(c_k\delta\varepsilon_k)^2|f_k(\boldsymbol{\Phi})|^4+(c_k\delta)^2\left(f_{kk,1,2}(\boldsymbol{\Phi})+|f_{k,1,1}(\boldsymbol{\Phi})|^2\right)+(c_k\varepsilon_k)^2M(M+1)|f_{kk,2}(\boldsymbol\Phi)|^2\notag\\
			&+c_{k}^{2}M(M+1)\left(f_{kk,3,2}(\boldsymbol\Phi)+\left|f_{k,3,1}(\boldsymbol\Phi)\right|^{2}\right)+4M(c_{k}\delta)^{2}\varepsilon_{k}|f_{k}(\boldsymbol\Phi)|^{2}f_{k,1,1}(\boldsymbol\Phi)+2M(M+1)(c_{k}\varepsilon_{k})^{2}\delta|f_{k}(\boldsymbol\Phi)|^{2}f_{kk,2}(\boldsymbol\Phi)\notag\\
			&+2M(M+1)c_k^2\delta\varepsilon_k|f_k(\boldsymbol\Phi)|^2f_{k,3,1}(\boldsymbol\Phi)+2(M+1)c_k^2\delta\varepsilon_kf_{k,1,1}(\boldsymbol\Phi)f_{kk,2}(\boldsymbol\Phi)+2(M+1)c_k^2\delta\left(f_{k,1,1}(\boldsymbol\Phi)f_{k,3,1}(\boldsymbol\Phi)+f_{kk,5}(\Phi)\right)\notag\\
			&+2M(M+1)c_k^2\varepsilon_kf_{kk,2}(\Phi)f_{k,3,1}(\boldsymbol\Phi)+2M(M+1)c_k^2\varepsilon_kf_{kk,4}(\boldsymbol\Phi)+4(M+1)c_k^2\delta\varepsilon_k\mathrm{Re}\{f_{kk,6}(\boldsymbol\Phi)\},\label{signal}
		\end{align}
		\begin{align}
			&I_{VR,ki}(\boldsymbol{\Phi})=c_{k}c_{i}\delta^{2}\varepsilon_{k}\varepsilon_{i}M^{2}|f_{k}(\boldsymbol{\Phi})|^{2}|f_{i}(\boldsymbol{\Phi})|^{2}+c_{k}c_{i}\delta^{2}f_{ki,1,2}(\boldsymbol{\Phi})+c_{k}c_{i}\varepsilon_{k}\varepsilon_{i}\left(Mf_{kk,2}(\boldsymbol{\Phi})f_{ii,2}(\boldsymbol{\Phi})+M^{2}\big|f_{ki,2}(\boldsymbol{\Phi})\big|^{2}\right)\notag\\
			&+c_kc_iM\left(f_{k,3,1}(\boldsymbol\Phi)f_{i,3,1}(\boldsymbol\Phi)+Mf_{ki,3,2}(\boldsymbol\Phi)\right)+c_kc_i\delta\varepsilon_k\varepsilon_iM\left(|f_k(\boldsymbol{\Phi})|^2f_{ii,2}(\boldsymbol{\Phi})+|f_i(\boldsymbol{\Phi})|^2f_{kk,2}(\boldsymbol{\Phi})\right)\notag\\
			&+c_kc_i\delta^2M\left(\varepsilon_k|f_k(\boldsymbol{\Phi})|^2f_{i,1,1}(\boldsymbol{\Phi})+\varepsilon_i|f_i(\boldsymbol{\Phi})|^2f_{k,1,1}(\boldsymbol{\Phi})\right)+c_kc_i\delta\varepsilon_k\left(M|f_k(\boldsymbol{\Phi})|^2f_{i,3,1}(\boldsymbol{\Phi})+f_{i,1,1}(\boldsymbol{\Phi})f_{kk,2}(\boldsymbol{\Phi})\right)\notag\\
			&+c_kc_i\delta\varepsilon_i\left(\boldsymbol{M}|f_i(\boldsymbol{\Phi})|^2f_{k,3,1}(\boldsymbol{\Phi})+f_{k,1,1}(\boldsymbol{\Phi})f_{ii,2}(\boldsymbol{\Phi})\right)+c_kc_i\delta\left(f_{k,1,1}(\boldsymbol{\Phi})f_{i,3,1}(\boldsymbol{\Phi})+f_{i,1,1}(\boldsymbol{\Phi})f_{k,3,1}(\boldsymbol{\Phi})\right)\notag\\
			&+c_kc_i\varepsilon_kM\left(Mf_{ki,4}(\boldsymbol\Phi)+f_{kk,2}(\boldsymbol\Phi)f_{i,3,1}(\boldsymbol\Phi)\right)+c_kc_i\varepsilon_iM\left(Mf_{ik,4}(\boldsymbol\Phi)+f_{ii,2}(\boldsymbol\Phi)f_{k,3,1}(\boldsymbol\Phi)\right)+\varepsilon_{i}\mathrm{Re}\big\{f_{ik,6}(\boldsymbol\Phi)\big\}\big)\notag\\
			&+c_kc_i\delta\varepsilon_k\varepsilon_iM^2\left(f_{ki,7}(\boldsymbol{\Phi})f_{ik,2}(\boldsymbol{\Phi})+f_{ki,2}(\boldsymbol{\Phi})f_{ik,7}(\boldsymbol{\Phi})\right)+2c_kc_i\delta M\mathrm{Re}\{f_{ki,5}(\boldsymbol\Phi)\}+2c_{k}c_{i}\delta M\big(\varepsilon_{k}\mathrm{Re}\big\{f_{ki,6}(\boldsymbol\Phi)\big\},\label{I}
		\end{align}
		where
		\begin{align}
		&f_k(\mathbf{\Phi})\triangleq\mathbf{a}_N^H\mathbf{\Phi}\mathbf{D}_k^{1/2}\mathbf{\bar{h}}_k,\ f_{k,1,1}(\boldsymbol{\Phi})\triangleq\mathrm{Tr}(\mathbf{\bar{H}}_2\boldsymbol{\Phi}\mathbf{R}_{VR,k}\boldsymbol{\Phi}^H\mathbf{\bar{H}}_2^H),\notag\\
		&f_{ki,1,2}(\boldsymbol{\Phi})\triangleq\mathrm{Tr}\{(\bar{\mathbf{H}}_2\boldsymbol{\Phi}\mathbf{R}_{VR,k}\boldsymbol{\Phi}^H\bar{\mathbf{H}}_2^H)(\bar{\mathbf{H}}_2\boldsymbol{\Phi}\mathbf{R}_{VR,i}\boldsymbol{\Phi}^H\bar{\mathbf{H}}_2^H)\},\ f_{ki,2}(\mathbf{\Phi})\triangleq\mathbf{\bar{h}}_k^H\mathbf{D}_k^{1/2}\mathbf{\Phi}^H\mathbf{R}_{ris}\mathbf{\Phi}\mathbf{D}_i^{1/2}\mathbf{\bar{h}}_i,\notag\\
		&f_{k,3,1}(\boldsymbol{\Phi})\triangleq\mathrm{Tr}\big(\mathbf{R}_{ris}\boldsymbol{\Phi}\mathbf{R}_{VR,k}\boldsymbol{\Phi}^{H}\big),\ f_{ki,3,2}(\boldsymbol{\Phi})\triangleq\mathrm{Tr}\{\left(\mathbf{R}_{ris}\boldsymbol{\Phi}\mathbf{R}_{VR,k}\boldsymbol{\Phi}^{H}\right)(\mathbf{R}_{ris}\boldsymbol{\Phi}\mathbf{R}_{VR,i}\boldsymbol{\Phi}^{H})\},\notag\\
		&f_{ki,4}(\boldsymbol{\Phi})\triangleq\mathbf{\bar{h}}_{k}^{H}\mathbf{D}_{k}^{1/2}\boldsymbol{\Phi}^{H}\mathbf{R}_{ris}\boldsymbol{\Phi}\mathbf{R}_{VR,i}\boldsymbol{\Phi}^{H}\mathbf{R}_{ris}\boldsymbol{\Phi}\mathbf{D}_{k}^{1/2}\mathbf{\bar{h}}_{k},\ 	f_{ki,5}(\boldsymbol{\Phi})\triangleq\mathrm{Tr}\big(\bar{\mathbf{H}}_{2}\boldsymbol{\Phi}\mathbf{R}_{VR,k}\boldsymbol{\Phi}^{H}\mathbf{R}_{ris}\boldsymbol{\Phi}\mathbf{R}_{VR,i}\boldsymbol{\Phi}^{H}\bar{\mathbf{H}}_{2}^{H}\big),\notag\\
		&f_{ki,6}(\mathbf{\Phi})\triangleq\mathbf{\bar{h}}_{k}^{H}\mathbf{D}_{k}^{1/2}\mathbf{\Phi}^{H}\mathbf{\bar{H}}_{2}^{H}\mathbf{\bar{H}}_{2}\mathbf{\Phi}\mathbf{R}_{VR,i}\mathbf{\Phi}^{H}\mathbf{R}_{ris}\mathbf{\Phi}\mathbf{D}_{k}^{1/2}\mathbf{\bar{h}}_{k},\ f_{ki,7}(\boldsymbol\Phi)\triangleq\bar{\mathbf{h}}_{k}^{H}\mathbf{D}_{k}^{1/2}\mathbf{\Phi}^{H}\mathbf{a}_{N}\mathbf{a}_{N}^{H}\mathbf{\Phi}\mathbf{D}_{i}^{1/2}\bar{\mathbf{h}}_{i}.\label{f_old}
		\end{align}
	\end{figure*}

\end{theorem}

\itshape {Proof:}  \upshape Please refer to Appendix \ref{appendixA}.

\subsection{Visibility Region Analysis}
  
Considering the VRs, the derivation process is difficult due to the addition of the visibility indicator matrix $\mathbf{D}_k$ and the spatial correlation matrices $\mathbf{R}_{ris}$ and $\mathbf{R}_{VR,k}$. However, the expression (\ref{rate}) in Theorem \ref{theorem1} can be calculated without using numerical calculations
of inverse matrices and integrals, resulting in lower computational complexity. Besides, Theorem \ref{theorem1} does not rely on the instantaneous CSI, i.e., $\tilde{\mathbf{h}}_k$, and $\tilde{\mathbf{H}}_2$, but it only depends on the statistical CSI, i.e., the AoA and AoD in $\mathbf{\bar{h}}_k$ and $\mathbf{\bar{H}}_2$, the locations of the RIS, the BS, and users, which change slowly and therefore could remain invariant in a long term. As a result, by using the theoretical expression of the rate in (\ref{rate}) as an objective function for a two-timescale transmission scheme, we can optimize the phase shifts of the XL-RIS only determined by statistical CSI.

On the other hand, by analyzing the derived achievable user rate expression, it can be found that some zero terms exist in the diagonal elements of the visibility indicator matrix $\mathbf{D}_k\in\mathbb{C}^{N\times N}$, which indicates that the RIS elements corresponding to these terms are not visible to the user $k$. Due to this special property, the dimension of matrix $\mathbf{D}_k$ can be further compressed into $\mathbf{D}_{k_{*}}=\mathbf{I}_{{N_k}_{*}\times {N_k}_{*}}$, assuming that the VR of the $k$-th user contains ${N_k}_{*}$ RIS elements (${N_k}_{*}\leq{N}$). Based on expressions (1)-(12), we can reduce the dimension of other matrices, thereby decreasing the computational complexity of the entire system. We use $\mathbf{a}_{N,k_{*}}\in\mathbb{C}^{1\times {N_k}_{*}}$, $\bar{\mathbf{h}}_{k_{*}}\in\mathbb{C}^{{N_k}_{*}\times 1}$, $\bar{\mathbf{H}}_{2, k_{*}}\in\mathbb{C}^{M\times {N_k}_{*}}$, $\mathbf{R}_{VR, k_{*}}\in\mathbb{C}^{{N_k}_{*}\times {N_k}_{*}}$, $\mathbf{R}_{ris, k_{*}, i_{*}}\in\mathbb{C}^{{N_k}_{*}\times {N_i}_{*}}$, and $\mathbf{\Phi}_{k_{*}}\in\mathbb{C}^{{N_k}_{*}\times {N_k}_{*}}$ to represent the dimension-reduced matrices of $\mathbf{a}_{N}$, $\bar{\mathbf{h}}_{k}$, $\bar{\mathbf{H}}_{2}$, $\mathbf{R}_{VR}$, $\mathbf{R}_{ris}$, and $\mathbf{\Phi}$, respectively. The detailed results are given by:

\begin{align}\label{reduced_matrix}
	&[{\mathbf{a}}_{N,k_{*}}]_q=[\mathbf{a}_{N}]_{[\mathbf{\nu}_k]_q},\notag\\
	&[\bar{\mathbf{h}}_{k_{*}}]_q=[\bar{\mathbf{h}}_{k}]_{[\mathbf{\nu}_k]_q},\notag\\
	&[\bar{\mathbf{H}}_{2, k_{*}}]_{(:,q)}=[\bar{\mathbf{H}}_2]_{(:,[\mathbf{\nu}_k]_q)},\notag\\
	&[\mathbf{R}_{VR, k_{*}}]_{(p,q)}=[\mathbf{R}_{VR}]_{([\mathbf{\nu}_k]_p,[\mathbf{\nu}_k]_q)},\notag\\
	&[\mathbf{R}_{ris, i_{*}, k_{*}}]_{(p,q)}=[\mathbf{R}_{ris}]_{([\mathbf{\nu}_i]_p,[\mathbf{\nu}_k]_q)},\notag\\
	&[\mathbf{\Phi}_{k_{*}}]_{(:,q)}=[\mathbf{\Phi}]_{(:,[\mathbf{\nu}_k]_q)},
\end{align}

\noindent where $q\in[1,{N_k}_{*}]$ and $p\in[1,{N_i}_{*}]$. Besides, ${[\mathbf{\nu}_k]_q}$ presents the $q$-th element of the set $\mathbf{\nu}_k$. As a result, the expressions in (\ref{f_old}) can be rewritten as:

\begin{align}
	f_{{k_{*}}}(\mathbf{\Phi})&\triangleq\mathbf{a}_{N,{k_{*}}}^{H}\mathbf{\Phi}_{{k_{*}}}\bar{\mathbf{h}}_{k_{*}},\notag\\
	f_{{k_{*}},1,1}(\boldsymbol{\Phi})&\triangleq\mathrm{Tr}\big(\mathbf{\bar{H}}_{2,{k_{*}}}\boldsymbol{\Phi}_{{k_{*}}}\mathbf{R}_{VR,{k_{*}}}\boldsymbol{\Phi}_{{k_{*}}}^{H}\mathbf{\bar{H}}_{2,{k_{*}}}^{H}\big),\notag\\
	f_{{k_{*}},{i_{*}},1,2}(\boldsymbol{\Phi})&\triangleq\mathrm{Tr}\left\{\big(\bar{\mathbf{H}}_{2,{k_{*}}}\boldsymbol{\Phi}_{{k_{*}}}\mathbf{R}_{VR,{k_{*}}}\boldsymbol{\Phi}_{{k_{*}}}^{H}\bar{\mathbf{H}}_{2,{k_{*}}}^{H}\big)\right.\notag\\
	&\quad\left.\times\big(\bar{\mathbf{H}}_{2,{i_{*}}}\boldsymbol{\Phi}_{{i_{*}}}\mathbf{R}_{VR,{i_{*}}}\boldsymbol{\Phi}_{{i_{*}}}^{H}\bar{\mathbf{H}}_{2,{i_{*}}}^{H}\big)\right\},\notag\\
	f_{{k_{*}},{i_{*}},2}(\mathbf{\Phi})&\triangleq\mathbf{\bar{h}}_{k_{*}}^{H}\mathbf{\Phi}_{{k_{*}}}^{H}\mathbf{R}_{ris,{k_{*}},{i_{*}}}\mathbf{\Phi}_{{i_{*}}}\bar{\mathbf{h}}_{i_{*}},\notag\\
	f_{{k_{*}},3,1}(\boldsymbol{\Phi})&\triangleq\mathrm{Tr}(\mathbf{R}_{ris,{k_{*}},{k_{*}}}\boldsymbol{\Phi}_{{k_{*}}}\mathbf{R}_{VR,{k_{*}}}\boldsymbol{\Phi}_{{k_{*}}}^{H}),\notag\\
	f_{{k_{*}},{i_{*}},3,2}(\boldsymbol{\Phi})&\triangleq\mathrm{Tr}\left\{(\mathbf{R}_{ris,{i_{*}},{k_{*}}}\boldsymbol{\Phi}_{k_{*}}\mathbf{R}_{VR,{k_{*}}}\boldsymbol{\Phi}_{k_{*}}^H)\right.\notag\\
	&\quad\left.\times(\mathbf{R}_{ris,{k_{*}},{i_{*}}}\boldsymbol{\Phi}_{i_{*}}\mathbf{R}_{VR,{i_{*}}}\boldsymbol{\Phi}_{i_{*}}^H)\right\},\notag\\
	f_{{k_{*}},{i_{*}},4}(\mathbf{\Phi})&\triangleq\mathbf{\bar{h}}_{{k_{*}}}^{H}\mathbf{\Phi}_{{k_{*}}}^{H}\mathbf{R}_{ris,{k_{*}},{i_{*}}}\mathbf{\Phi}_{{i_{*}}}\mathbf{R}_{VR,{i_{*}}}\mathbf{\Phi}_{{i_{*}}}^{H}\mathbf{R}_{ris,{i_{*}},{k_{*}}}\mathbf{\Phi}_{{k_{*}}}\mathbf{\bar{h}}_{k_{*}},\notag\\
	f_{{k_{*}},{i_{*}},5}(\boldsymbol{\Phi})&\triangleq\mathrm{Tr}\left(\mathbf{\bar{H}}_{2,{k_{*}}}\boldsymbol{\Phi}_{{k_{*}}}\mathbf{R}_{VR,{k_{*}}}\boldsymbol{\Phi}_{{k_{*}}}^{H}\mathbf{R}_{ris,{k_{*}},{i_{*}}}\right.\notag\\
	&\quad\left.\times\boldsymbol{\Phi}_{{i_{*}}}\mathbf{R}_{VR,{i_{*}}}\boldsymbol{\Phi}_{{i_{*}}}^{H}\mathbf{\bar{H}}_{2,{i_{*}}}^{H}\right),\notag\\
	f_{{k_{*}},{i_{*}},6}(\Phi)&\triangleq\bar{\mathbf{h}}_{k_{*}}^{H}\boldsymbol\Phi_{{k_{*}}}^{H}\bar{\mathbf{H}}_{2,{k_{*}}}^{H}\bar{\mathbf{H}}_{2,{i_{*}}}\boldsymbol\Phi_{{i_{*}}}R_{VR,{i_{*}}}\boldsymbol\Phi_{{i_{*}}}^{H}R_{ris,{i_{*}},{k_{*}}}\boldsymbol\Phi_{{k_{*}}}\bar{\mathbf{h}}_{k_{*}},\notag\\
	f_{{k_{*}},{i_{*}},7}(\Phi)&\triangleq\bar{\mathbf{h}}_{k_{*}}^{H}\boldsymbol\Phi_{{k_{*}}}^{H}\mathbf{a}_{N,{k_{*}}}\mathbf{a}_{N,{i_{*}}}^{H}\mathbf{\Phi}_{{i_{*}}}\bar{\mathbf{h}}_{i_{*}}.\label{f_new}
\end{align}

Due to the lower dimensionality of the participating matrices, the computations in (\ref{f_new}) are comparatively simpler than those in (\ref{f_old}). Specifically, the complexity reduction is attributed to VR, where a certain user $k$ can only observe a subset of RIS elements (${N_k}_{*}\leq{N}$). By substituting (\ref{f_new}) into expressions (\ref{rate})-(\ref{I}), the achievable user rate can be conveniently computed.

\section{RIS Phase Shifts Design}\label{four}

In this section, we optimize the phase shifts of the RIS to maximize the achievable rate derived from (\ref{rate}). In a multi-user scenario, it is necessary to ensure fairness among different users. Therefore, we consider the minimum achievable user rate maximization problem, where $\mathbf{\Phi}$ is the independent variable. Specifically, this problem can be formulated as:
\begin{align}\label{problem}
	&\max_\mathbf{\Phi}\quad\min_{k}\quad R_{VR,k},\notag\\
	&\mathrm{s.t.~}\quad\left|{[\mathbf{\Phi}]}_{(n,n)}\right|=1,\forall n,
\end{align}

\noindent where $R_{VR,k}$ is the achievable user rate obtained in (\ref{rate}).

From (\ref{rate})-(\ref{f_old}), it can be observed that the objective function of the optimization problem becomes more complex in the presence of spatial correlation and VRs. Besides, when $N$ is large, the dimension of matrices in calculation can be extremely high, which leads to low algorithm efficiency. To address these issues, we have investigated the beneficial effects of VRs in section \ref{three} and found that they can reduce the computational complexity of the entire system. Furthermore, we also need a new algorithm to better solve the considered minimum user rate maximization problem.

In this section, we propose a gradient-based algorithm with respect to real variable $\theta_n$ instead of complex variable $e^{j\theta_{n}}$, where $\theta_n$ is the phase of the $n$-th RIS element. Since the objective function in (\ref{problem}) contains a non-differentiable minimum function, to conduct the gradient algorithm, we first approximate the objective function in (\ref{problem}) as:
\begin{align}
	\min_k \ R_{VR,k}(\boldsymbol\theta)&\approx-\frac1\mu\mathrm{ln}\left\{\sum_{k=1}^K\exp\{-\mu R_{VR,k}(\boldsymbol\theta)\}\right\}\notag\\
	&\triangleq f_{VR}(\boldsymbol\theta),\label{23}
\end{align}

\noindent where $\boldsymbol\theta=[\theta_1,\theta_2,...,\theta_N]^T$ is the new optimization variable. $\mu$ is a constant value that controls the approximation accuracy, which can be proven that the approximation error is less than $\frac{\ln{\mathrm{K}}}{\mu}$ \cite{d4}. Therefore, the problem (\ref{problem}) can be rewritten as:
\begin{align}\label{problem_new}
	&\max_\mathbf{\Phi}\quad\ \quad f_{VR}(\boldsymbol{\theta}),\notag\\
	&\mathrm{s.t.~}\quad\theta_n\in[0,2\pi),\forall n.
\end{align}

Due to the periodicity of the objective function $f_{VR}(\boldsymbol{\theta})$ relative to $\boldsymbol{\theta}$, the constraints in (\ref{problem_new}) can be ignored, which effectively tackles the unit-modulus constraint in (\ref{problem}). After updating the variable $\boldsymbol{\theta}$, the gradient of $f_{VR}(\boldsymbol{\theta})$ needs to be calculated:
\begin{align}\label{f_VR_gradient}
   \frac{\partial f_{VR}(\boldsymbol{\theta})}{\partial\boldsymbol{\theta}}=\frac{\Sigma_{k=1}^{K}\left\{\frac{\exp\left\{-\mu R_{VR,k}(\boldsymbol{\theta})\right\}}{1+\mathrm{SINR}_{VR,k}(\boldsymbol{\theta})}\frac{\partial\operatorname{SINR}_{VR,k}(\boldsymbol{\theta})}{\partial\boldsymbol{\theta}}\right\}}{(\ln2)\left(\sum_{k=1}^{K}\exp\{-\mu R_{VR,k}(\boldsymbol{\theta})\}\right)},
\end{align}

\noindent where
\begin{align}\label{SINR_VR_gradient}
	&\frac{\partial\operatorname{SINR}_{VR,k}(\boldsymbol{\theta})}{\partial\boldsymbol{\theta}}=\frac{p\frac{\partial E_{VR,k}^\mathrm{signal}}{\partial\boldsymbol{\theta}}}{p\sum_{i=1,i\neq k}^KI_{VR,ki}+\sigma^2E_{VR,k}^\mathrm{noise}}\notag\\
	&-pE_{VR,k}^\text{signal}\frac{p\sum_{i=1,i\neq k}^K\frac{\partial I_{VR,ki}}{\partial\boldsymbol{\theta}}+\sigma^2\frac{\partial E_{VR,k}^\mathrm{noise}}{\partial\boldsymbol{\theta}}}{\left(p\sum_{i=1,i\neq k}^KI_{VR,ki}+\sigma^2E_{VR,k}^\mathrm{noise}\right)^2}.
\end{align}

According to (\ref{f_VR_gradient}) and (\ref{SINR_VR_gradient}), the gradient of $f_{VR}(\boldsymbol{\theta})$ can be obtained by calculating $E_{VR,k}^{\mathrm{noise}}(\boldsymbol\Phi)$, $E_{VR,k}^\mathrm{signal}(\boldsymbol{\Phi})$ and $\sum_{i=1,i\neq k}^{K}I_{VR,ki}(\boldsymbol{\Phi})$. By using the conclusion of \cite[Lemma 1]{27}, we define
\begin{align}\label{f_a}
	\boldsymbol{f}_a(\mathbf{A},\mathbf{B})&\triangleq\frac{\partial\operatorname{Tr}\{\mathbf{A\Phi B\Phi}^H\}}{\partial\boldsymbol{\theta}}\notag\\
	&=j\mathbf{\Phi}^T(\mathbf{A}^T\odot\mathbf{B})\boldsymbol{c}^*-j\mathbf{\Phi}^H(\mathbf{A}\odot\mathbf{B}^T)\boldsymbol{c},
\end{align}
\noindent where $\mathbf{A}$ and $\mathbf{B}$ are definite matrices. Then, the gradients of the auxiliary functions in (\ref{f_old}) can be calculated as:
\begin{align}\label{fk_VR_gradient}
	f_k^{\prime}(\boldsymbol{\theta})\triangleq&\boldsymbol{f}_a(\mathbf{a}_N\mathbf{a}_N^H,\mathbf{D}_k^{1/2}\mathbf{\bar{h}}_k\mathbf{\bar{h}}_k^H\mathbf{D}_k^{1/2}),\notag\\
	f_{k,1,1}^{\prime}(\boldsymbol{\theta})\triangleq&\boldsymbol{f}_{a}(\mathbf{\bar{H}}_{2}^{H}\mathbf{\bar{H}}_{2},\mathbf{R}_{VR,k}),\notag\\
	f_{ki,1,2}^{\prime}(\boldsymbol{\theta})\triangleq&\boldsymbol{f}_a(\bar{\mathbf{H}}_2^H\bar{\mathbf{H}}_2\boldsymbol{\Phi}\mathbf{R}_{VR,i}\boldsymbol{\Phi}^H\bar{\mathbf{H}}_2^H\bar{\mathbf{H}}_2,\mathbf{R}_{VR,k})\notag\\
	&+\boldsymbol{f}_{a}(\bar{\mathbf{H}}_{2}^{H}\bar{\mathbf{H}}_{2}\boldsymbol{\Phi}\mathbf{R}_{VR,k}\boldsymbol{\Phi}^{H}\bar{\mathbf{H}}_{2}^{H}\bar{\mathbf{H}}_{2},\mathbf{R}_{VR,i})\notag\\
	f_{ki,2}^{\prime}(\boldsymbol{\theta})\triangleq&\boldsymbol{f}_{a}(\mathbf{R}_{ris},\mathbf{D}_{i}^{1/2}\mathbf{\bar{h}}_{i}\mathbf{\bar{h}}_{k}^{H}\mathbf{D}_{k}^{1/2}),\notag\\
	f_{k,3,1}^{\prime}(\boldsymbol{\theta})\triangleq&\boldsymbol{f}_{a}(\mathbf{R}_{ris},\mathbf{R}_{VR,k}),\notag\\
	f_{ki,3,2}^{\prime}(\boldsymbol{\theta})\triangleq&\boldsymbol{f}_{a}\big(\mathbf{R}_{ris}\boldsymbol{\Phi}\mathbf{R}_{VR,i}\boldsymbol{\Phi}^{H}\mathbf{R}_{ris},\mathbf{R}_{VR,k}\big)\notag\\
	&+\boldsymbol{f}_a(\mathbf{R}_{ris}\mathbf{\Phi}\mathbf{R}_{VR,k}\mathbf{\Phi}^H\mathbf{R}_{ris},\mathbf{R}_{VR,i})\notag\\
	f_{ki,4}^{\prime}(\boldsymbol{\theta})\triangleq&\boldsymbol{f}_{a}\big(\mathbf{R}_{ris}\boldsymbol{\Phi}\mathbf{D}_{k}^{1/2}\bar{\mathbf{h}}_{k}\bar{\mathbf{h}}_{k}^{H}\mathbf{D}_{k}^{1/2}\boldsymbol{\Phi}^{H}\mathbf{R}_{ris},\mathbf{R}_{VR,i}\big)\notag\\
	&+\boldsymbol{f}_{a}(\mathbf{R}_{ris}\mathbf{\Phi}\mathbf{R}_{VR,i}\mathbf{\Phi}^{H}\mathbf{R}_{ris},\mathbf{D}_{k}^{1/2}\mathbf{\bar{h}}_{k}\mathbf{\bar{h}}_{k}^{H}\mathbf{D}_{k}^{1/2})\notag\\
	f_{ki,5}^{\prime}(\boldsymbol{\theta})\triangleq&\boldsymbol{f}_a(\bar{\mathbf{H}}_2^H\bar{\mathbf{H}}_2\boldsymbol{\Phi}\mathbf{R}_{VR,k}\boldsymbol{\Phi}^H\mathbf{R}_{ris},\mathbf{R}_{VR,i})\notag\\
	&+\boldsymbol{f}_a(\mathbf{R}_{ris}\mathbf{\Phi}\mathbf{R}_{VR,i}\mathbf{\Phi}^H\mathbf{\bar{H}}_2^H\mathbf{\bar{H}}_2,\mathbf{R}_{VR,k})\notag\\
	f_{ki,6}^{\prime}(\boldsymbol{\theta})\triangleq&\boldsymbol{f}_a(\mathbf{R}_{ris}\boldsymbol{\Phi}\mathbf{D}_k^{1/2}\bar{\mathbf{h}}_k\bar{\mathbf{h}}_k^H\mathbf{D}_k^{1/2}\boldsymbol{\Phi}^H\bar{\mathbf{H}}_2^H\bar{\mathbf{H}}_2,\mathbf{R}_{VR,i})\notag\\
	&+\boldsymbol{f}_a(\bar{\mathbf{H}}_2^H\bar{\mathbf{H}}_2\boldsymbol{\Phi}\mathbf{R}_{VR,i}\boldsymbol{\Phi}^H\mathbf{R}_{ris},\mathbf{D}_k^{1/2}\bar{\mathbf{h}}_k\bar{\mathbf{h}}_k^H\mathbf{D}_k^{1/2})\notag\\
	f_{ik,6}^*{^{\prime}}(\boldsymbol{\theta})\triangleq&\boldsymbol{f}_a(\bar{\mathbf{H}}_2^H\bar{\mathbf{H}}_2\boldsymbol{\Phi}\mathbf{D}_k^{1/2}\bar{\mathbf{h}}_k\bar{\mathbf{h}}_k^H\mathbf{D}_k^{1/2}\boldsymbol{\Phi}^H\mathbf{R}_{ris},\mathbf{R}_{VR,i})\notag\\
	&+\boldsymbol{f}_a(\mathbf{R}_{ri\mathbf{s}}\boldsymbol{\Phi}\mathbf{R}_{VR,i}\boldsymbol{\Phi}^H\bar{\mathbf{H}}_2^H\bar{\mathbf{H}}_2,\mathbf{D}_k^{1/2}\bar{\mathbf{h}}_k\bar{\mathbf{h}}_k^H\mathbf{D}_k^{1/2})\notag\\
	f_{ki,7}^{\prime}(\boldsymbol{\theta})\triangleq&\boldsymbol{f}_{a}\big(\mathbf{a}_{N}\mathbf{a}_{N}^{H},\mathbf{D}_{i}^{1/2}\mathbf{\bar{h}}_{i}\mathbf{\bar{h}}_{k}^{H}\mathbf{D}_{k}^{1/2}\big).
\end{align}

The specific calculation details are presented in Appendix \ref{appendixB}. Then, $\frac{\partial E_{VR,k}^{\mathrm{noise}}}{\partial\boldsymbol{\theta}}$, $\frac{\partial E_{VR,k}^\mathrm{signal}}{\partial\boldsymbol{\theta}}$ and $\frac{\partial I_{VR,ki}}{\partial\boldsymbol{\theta}}$ can be derived using the above results and the chain rule. The results are given in (\ref{noise_VR_gradient})-(\ref{i_VR_gradient}) at the bottom of the next page.
\begin{figure*}[b]
	\hrulefill	
	\begin{align}
		&\frac{\partial E_{VR,k}^{\mathrm{noise}}}{\partial\boldsymbol{\theta}}=c_k\left\{M\delta\varepsilon_kf_k^{\prime}(\boldsymbol{\theta})+\delta f_{k,1,1}^{\prime}(\boldsymbol{\theta})+\varepsilon_kf_{kk,2}^{\prime}(\boldsymbol{\theta})M+f_{k,3,1}^{\prime}(\boldsymbol{\theta})M\right\},\label{noise_VR_gradient}\\
		&\frac{\partial E_{VR,k}^\mathrm{signal}}{\partial\boldsymbol{\theta}}=M^2(c_k\delta\varepsilon_k)^2|f_k(\boldsymbol\Phi)|^2f_k^{\prime}(\boldsymbol{\theta})+(c_k\delta)^2\left(f_{kk,1,2}^{\prime}(\boldsymbol{\theta})+2f_{k,1,1}(\boldsymbol\Phi)f_{k,1,1}^{\prime}(\boldsymbol{\theta})\right)+2(c_k\varepsilon_k)^2M(M+1)f_{kk,2}(\boldsymbol\Phi)f_{kk,2}^{\prime}(\boldsymbol{\theta})\notag\\
		&+c_k^2M(M+1)\left(f_{kk,3,2}^{\prime}(\boldsymbol\theta)+2f_{k,3,1}^{\prime}(\boldsymbol\theta)f_{k,3,1}(\boldsymbol\Phi)\right)+4M(c_k\delta)^2\varepsilon_k\left(f_{k,1,1}^{\prime}(\boldsymbol\theta)|f_k(\boldsymbol\Phi)|^2+f_{k,1,1}(\boldsymbol\Phi)f_k^{\prime}(\boldsymbol\theta)\right)\notag\\
		&+2M(M+1)(c_k\varepsilon_k)^2\delta f_{kk,2}^{\prime}(\boldsymbol\theta)|f_k(\boldsymbol\Phi)|^2+2M(M+1)(c_k\varepsilon_k)^2\delta f_{kk,2}(\boldsymbol\Phi)f_k^{\prime}(\boldsymbol\theta)+2M(M+1)c_k^2\delta\varepsilon_kf_{k,3,1}^{\prime}(\boldsymbol\theta)|f_k(\boldsymbol\Phi)|^2\notag\\
		&+2M(M+1)c_k^2\delta\varepsilon_kf_{k,3,1}(\boldsymbol\Phi)f_k^{\prime}(\boldsymbol\theta)+2(M+1)c_k^2\delta\varepsilon_kf_{kk,2}^{\prime}(\boldsymbol\theta)f_{k,1,1}(\boldsymbol\Phi)+2(M+1)c_k^2\delta\varepsilon_kf_{kk,2}(\boldsymbol\Phi)f_{k,1,1}^{\prime}(\boldsymbol\theta)\notag\\
		&+2(M+1)c_k^2\delta f_{k,3,1}^{\prime}(\boldsymbol\theta)f_{k,1,1}(\boldsymbol\Phi)+2(M+1)c_k^2\delta\left(
		f_{k,3,1}(\boldsymbol\Phi)f_{k,1,1}^{\prime}(\boldsymbol\theta)+f_{kk,5}^{\prime}(\boldsymbol\theta)\right)+2M(M+1)c_k^2\varepsilon_kf_{k,3,1}^{\prime}(\boldsymbol{\theta})f_{kk,2}(\boldsymbol{\Phi})\notag\\
		&+2M(M+1)c_k^2\varepsilon_k\left(
		f_{k,3,1}(\boldsymbol{\Phi})f_{kk,2}^{\prime}(\boldsymbol{\theta})+f_{kk,4}^{\prime}(\boldsymbol{\theta})\right)+2(M+1)c_k^2\delta\varepsilon_k\left(f_{kk,6}^{\prime}(\boldsymbol\theta)+f_{kk,6}^*(\boldsymbol\theta)\right),\label{signal_VR_gradient}
	\end{align}
	
	\begin{align}
		&\frac{\partial I_{VR,ki}}{\partial\boldsymbol{\theta}}=c_kc_i\delta^2\varepsilon_k\varepsilon_iM^2\left(f_k^{\prime}(\boldsymbol{\theta})|f_i(\boldsymbol{\Phi})|^2+|f_k(\boldsymbol{\Phi})|^2f_i^{\prime}(\boldsymbol{\theta})\right)+c_kc_i\delta^2f_{ki,1,2}^{\prime}(\boldsymbol{\theta})\notag\\
		&+c_kc_i\varepsilon_k\varepsilon_iM\left(f_{kk,2}(\boldsymbol\Phi)f_{ii,2}^{\prime}(\boldsymbol\theta)+f_{kk,2}^{\prime}(\boldsymbol\theta)f_{ii,2}(\boldsymbol\Phi)\right)+c_kc_i\varepsilon_k\varepsilon_iM^2\left(f_{ik,2}(\boldsymbol{\Phi})f_{ki,2}^{\prime}(\boldsymbol{\theta})+f_{ki,2}(\boldsymbol{\Phi})f_{ik,2}^{\prime}(\boldsymbol{\theta})\right)\notag\\
		&+c_kc_iM\left(f_{k,3,1}(\boldsymbol{\Phi})f_{i,3,1}^{\prime}(\boldsymbol{\theta})+f_{k,3,1}^{\prime}(\boldsymbol{\theta})f_{i,3,1}(\boldsymbol{\Phi})\right)+c_kc_i{M^2}f_{ki,3,2}^{\prime}(\boldsymbol{\theta})+c_kc_i\delta\varepsilon_k\varepsilon_iM\left(|f_k(\boldsymbol\Phi)|^2f_{ii,2}^{\prime}(\boldsymbol\theta)+f_k^{\prime}(\boldsymbol\theta)f_{ii,2}(\boldsymbol\Phi)\right)\notag\\
		&+c_kc_i\delta\varepsilon_k\varepsilon_iM\left(|f_i(\boldsymbol{\Phi})|^2f_{kk,2}^{\prime}(\boldsymbol{\theta})+f_i^{\prime}(\boldsymbol{\theta})f_{kk,2}(\boldsymbol{\Phi})\right)+c_kc_i\delta^2\varepsilon_kM\left(|f_k(\boldsymbol{\Phi})|^2f_{i,1,1}^{\prime}(\boldsymbol{\theta})+f_k^{\prime}(\boldsymbol{\theta})f_{i,1,1}(\boldsymbol{\Phi})\right)\notag\\
		&+c_kc_i\delta^2\varepsilon_iM\left(|f_i(\boldsymbol{\Phi})|^2f_{k,1,1}^{\prime}(\boldsymbol{\theta})+f_i^{\prime}(\boldsymbol{\theta})f_{k,1,1}(\boldsymbol{\Phi})\right)+c_kc_i\delta\varepsilon_kM\left(|f_k(\boldsymbol{\Phi})|^2f_{i,3,1}^{\prime}(\boldsymbol{\theta})+f_k^{\prime}(\boldsymbol{\theta})f_{i,3,1}(\boldsymbol{\Phi})\right)\notag\\
		&+c_kc_i\delta\varepsilon_k\left(f_{i,1,1}(\boldsymbol{\Phi})f_{kk,2}^{\prime}(\boldsymbol{\theta})+f_{i,1,1}^{\prime}(\boldsymbol{\theta})f_{kk,2}(\boldsymbol{\Phi})\right)+c_kc_i\delta\varepsilon_iM\left(|f_i(\boldsymbol{\Phi})|^2f_{k,3,1}^{\prime}(\boldsymbol{\theta})+f_i^{\prime}(\boldsymbol{\theta})f_{k,3,1}(\boldsymbol{\Phi})\right)\notag\\
		&+c_kc_i\delta\varepsilon_i\left(f_{ii,2}^{\prime}(\boldsymbol{\theta})f_{k,1,1}(\boldsymbol{\Phi})+f_{ii,2}(\boldsymbol{\Phi})f_{k,1,1}^{\prime}(\boldsymbol{\theta})\right)+c_kc_i\delta\left(f_{i,3,1}^{\prime}(\boldsymbol{\theta})f_{k,1,1}(\boldsymbol{\Phi})+f_{i,3,1}(\boldsymbol{\Phi})f_{k,1,1}^{\prime}(\boldsymbol{\theta})\right)\notag\\
		&+c_kc_i\delta\left(f_{i,1,1}(\boldsymbol\Phi)f_{k,3,1}^{\prime}(\boldsymbol\theta)+f_{i,1,1}^{\prime}(\boldsymbol\theta)f_{k,3,1}(\boldsymbol\Phi)\right)+c_kc_i\varepsilon_kM\left(f_{i,3,1}^{\prime}(\boldsymbol{\theta})f_{kk,2}(\boldsymbol{\Phi})+f_{i,3,1}(\boldsymbol{\Phi})f_{kk,2}^{\prime}(\boldsymbol{\theta})\right)+c_kc_i\varepsilon_k{M^2}f_{ki,4}^{\prime}(\boldsymbol{\theta})\notag\\
		&+c_kc_i\varepsilon_iM\left(f_{k,3,1}^{\prime}(\boldsymbol{\theta})f_{ii,2}(\boldsymbol{\Phi})+f_{k,3,1}(\boldsymbol{\Phi})f_{ii,2}^{\prime}(\boldsymbol{\theta})\right)+c_kc_i\varepsilon_i{M^2}f_{ik,4}^{\prime}(\boldsymbol{\theta})+c_kc_i\delta\varepsilon_k\varepsilon_iM^2\left(f_{ik,2}^{\prime}(\boldsymbol{\theta})f_{ki,7}(\boldsymbol{\Phi})+f_{ik,2}(\boldsymbol{\Phi})f_{ki,7}^{\prime}(\boldsymbol{\theta})\right)\notag\\
		&+c_{k}c_{i}\delta\varepsilon_{k}\varepsilon_{i}M^{2}\left(f_{ki,2}^{\prime}(\boldsymbol\theta)f_{ik,7}(\boldsymbol\Phi)+f_{ki,2}(\boldsymbol\Phi)f_{ik,7}^{\prime}(\boldsymbol\theta)\right)+c_kc_i\delta M\left(f_{ki,5}^{\prime}(\boldsymbol\theta)+f_{ik,5}^{\prime}(\boldsymbol\theta)\right)+2c_kc_i\delta\varepsilon_kM\left(f_{ki,6}^{\prime}(\boldsymbol\theta)+f_{ki,6}^*(\boldsymbol\theta)\right)\notag\\
		&+2c_kc_i\delta\varepsilon_iM\left(f_{ik,6}^{\prime}(\boldsymbol\theta)+f_{ik,6}^{*}(\boldsymbol\theta)\right).\label{i_VR_gradient}
	\end{align}
\end{figure*}

\begin{algorithm}[!ht]
	\renewcommand{\algorithmicrequire}{\textbf{Input:}}
	\renewcommand{\algorithmicensure}{\textbf{Output:}}
	\caption{gradient algorithm}
	\label{algorithm1}
	\begin{algorithmic}[1]
		\State Randomly initialize $\boldsymbol\theta_0$ and set $i=0,e_0=1,\boldsymbol{x}_{-1}=\boldsymbol{\theta}_0$;
		\While {1} 
		\State Calculate the gradient vector $f_{VR}^{\prime}(\boldsymbol{\theta}_i)$;
		\State Obtain the step size $\kappa_i$ based on the backtracking line search;
		
		\State $\boldsymbol{x}_i=\boldsymbol\theta_i+\kappa_if_{VR}^{\prime}(\boldsymbol\theta_i)$;
		\State $e_{i+1}=(1+\sqrt{4e_i^2+1})/2$;
		\State $\boldsymbol\theta_{i+1}=\boldsymbol{x}_i+(e_i-1)(\boldsymbol{x}_i-\boldsymbol{x}_{i-1})/e_{i+1}$
		
		\If {$f_{VR}(\boldsymbol{\theta}_{i+1})-f_{VR}(\boldsymbol{\theta}_i)<10^{-4}$}
		\State $\boldsymbol{\theta}^*=\boldsymbol{\theta}_{i+1}$, break;
		\EndIf
		
		\State $i=1+1$;
		
		\EndWhile	
	\end{algorithmic}
\end{algorithm}

The complete process for solving problem (\ref{problem_new}) is provided in Algorithm \ref{algorithm1}. In order to improve performance, the Nesterov accelerated gradient method as shown in steps 6 and 7 is used in the algorithm.

In Algorithm 1, $\boldsymbol{\theta}^*$ is the optimal solution, from which the RIS phase shift matrix $\boldsymbol{\Phi}$ can be calculated. After obtaining $\boldsymbol{\Phi}$, it can be applied to the original problem (\ref{problem}) to obtain the maximum result of the minimum achievable rate.

\section{Simulation and Results}\label{fifth}

\begin{table*}[t]
	\renewcommand\arraystretch{0.9}
	\centering 
	\captionsetup{font={small}}
	\caption{Simulation parameters.}	
	\begin{tabular}{|c|c|c|c|c|c|c}
		\hline
		BS antennas & $M=64$ & RIS elements &	$N=200$\\	
		\hline
		RIS element spacing & $d_{ris}=\lambda/2$  & Antenna spacing& $d_{bs}=\lambda/2$\\
		\hline
		Rician factors &   {$\delta=1$, $\varepsilon_{k}=10,\forall k$}&Transmit power & $p=30$ dBm\\
		\hline
	\end{tabular}\label{table1}
\end{table*}

In this section, we evaluate the performance of XL-RIS-aided massive MIMO systems and validate the correctness of our analysis. We also illustrate the impact of key parameters in this section. Firstly, a typical XL-RIS-aided scenario is considered where the XL-RIS is deployed near some cell-edge users. We assume that $k=4$ users are evenly located on a semicircle centered at the XL-RIS with a radius of $d_{UI}=15$m. The distance between the BS and the XL-RIS is $d_{IB}=800$m. All AoA and AoD of BS, XL-RIS, and users are randomly generated from $[0,2\pi)$, and these parameters will be fixed as known after initial generation. Besides, we set the distance-dependent large-scale path-loss factors equal to $\alpha_{k}=10^{-3}d_{UI}^{\alpha_{-}k_{-}UR}, \forall k $ and $\beta=10^{-3}d_{IB}^{\beta_{-}RB}$, and the path-loss exponents are $\alpha_{-}k_{-}UR=2, \forall k $ and $\beta_{-}RB=2.5$.  Considering the impact of VRs, we assume that each user can only see the XL-RIS elements within a certain region of the entire USPA model. To simulate the complexity of the actual environment, we randomly set the size and location of the VRs for different users. The other simulation parameters are presented in Table \ref{table1}.

\begin{figure}[t]
	\centering
	\includegraphics[width= 0.45\textwidth]{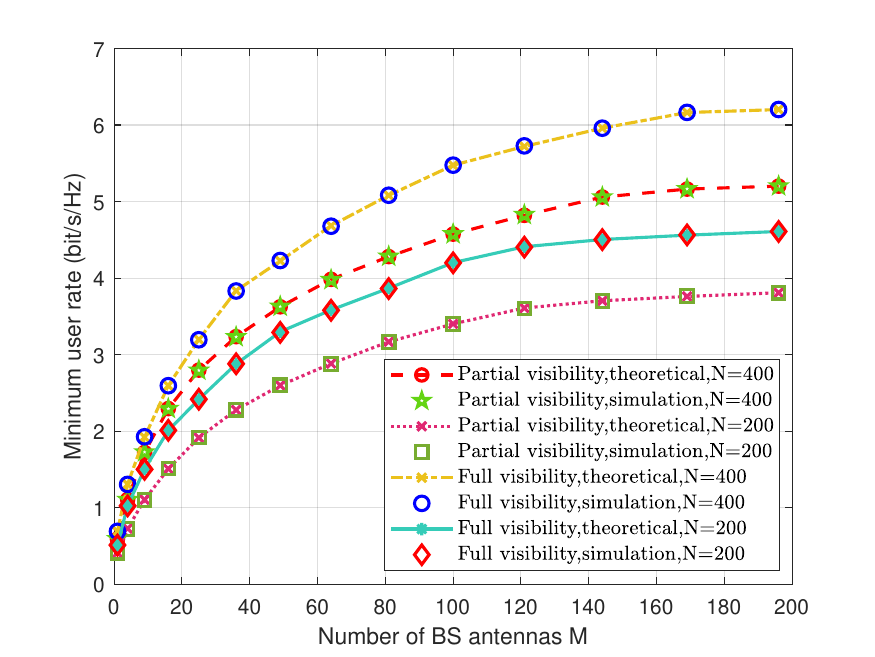}
	\caption{Minimum user rate versus BS antennas number $M$}
	\label{figure1-1}
\end{figure}

\begin{figure}[t]
	\centering
	\includegraphics[width= 0.45\textwidth]{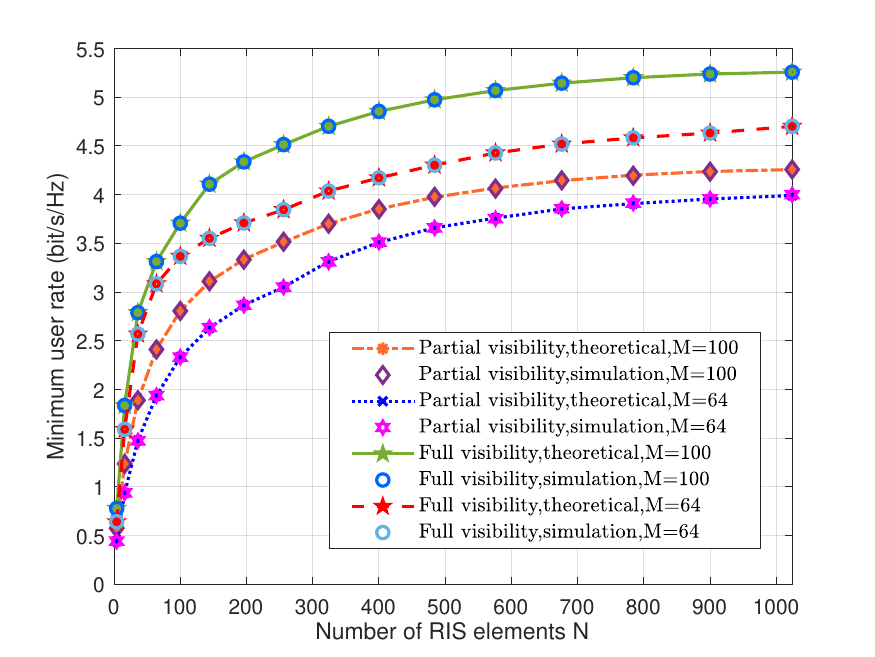}
	\caption{Minimum user rate versus RIS elements number $N$}
	\label{figure1-2}
\end{figure}

Fig. \ref{figure1-1} and Fig. \ref{figure1-2} respectively show the relationship between $M$ or $N$ and the minimum user rate when considering the visibility region of RIS. The comparison between theoretical and simulation results demonstrates that the obtained approximate achievable rate expression in (\ref{rate}) is in good agreement with the Monte Carlo simulation results, confirming the accuracy of the mathematical derivations. In addition, as shown in Fig. \ref{figure1-1}, it can be confirmed that as the number of antennas M increases, the minimum user rate gradually increases. Although the rate eventually converges to a theoretical limit, which is due to the inter-user interference, it can be significantly improved by increasing the number of $N$. On the other hand, deploying RIS results in promising throughput with a moderate number of antennas, which reduces power consumption and hardware costs. For instance, 60 antennas with 400 RIS elements could perform better than 120 antennas with 200 RIS elements, thanks to the RIS’s passive beamforming gain.

In Fig. \ref{figure1-1} and Fig. \ref{figure1-2}, “partial visibility” indicates that each user can only see partial non-overlapping RIS elements, while “full visibility” indicates that each user can see all RIS elements. It can be observed that the minimum user rate is relatively lower when considering VR. This is due to a reduction in the number of available RIS elements per user, resulting in a corresponding decline in channel transmission performance. In addition, although the VRs lead to performance degradation, it can be compensated by increasing $N$. For example, by analyzing Fig. \ref{figure1-2} we can find that a partial visibility system with 400 RIS elements can outperform a full visibility system with about 100 RIS elements, assuming the two systems have the same number of antennas. Therefore, with the help of increasing $N$, we can achieve the same rate as full visibility systems while still maintaining network capacity requirements.

\begin{figure}[t]
	\centering
	\includegraphics[width= 0.45\textwidth]{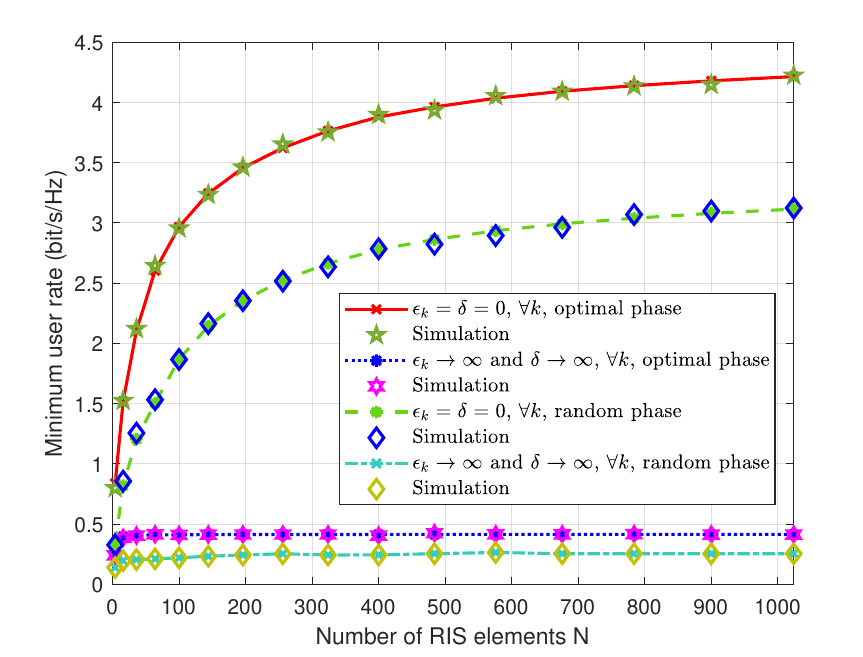}
	\caption{Max-min achievable rate versus $N$ when $\delta=\varepsilon_k=0$ or $\delta\rightarrow\infty $ and $\varepsilon_k\rightarrow\infty $}
	\label{figure2-1}
\end{figure}

\begin{figure}[t]
	\centering
	\includegraphics[width= 0.45\textwidth]{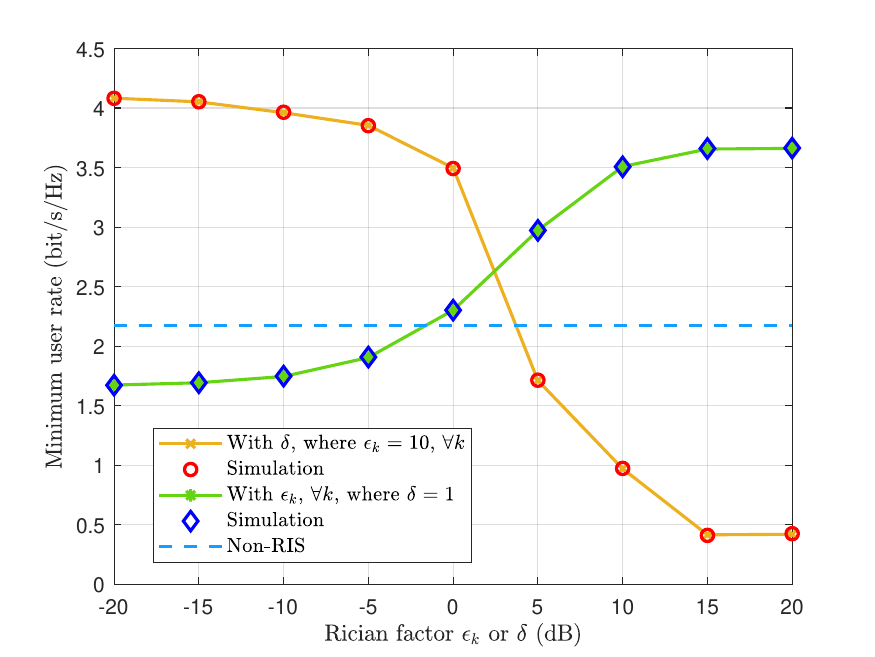}
	\caption{Max-min achievable rate versus $\delta$ or $\varepsilon_k$}
	\label{figure2-2}
\end{figure}

Fig. \ref{figure2-1} shows that when $\delta$ and $\varepsilon_k$ approach zero or infinity, the minimum user rate gradually converges to the theoretical limit, respectively. Besides, by optimizing the phase shifts of the RIS based on statistical CSI, the transmit power of each user can be further reduced compared to using random phase shifts. Fig. \ref{figure2-2} shows the influence of the Rician factors, indicating that the minimum user rate is a decreasing function of $\delta$, but it is also an increasing function of $\varepsilon_k$. 

By analyzing Fig. \ref{figure2-1} and Fig. \ref{figure2-2}, it can be concluded that as $\delta$ gradually increases, the impact of the LoS component $\mathbf{\bar{H}}_2$ in the RIS-BS channel is more significant. It increases the correlation of channels between different users and intensifies multi-user interference, thereby reducing spatial multiplexing gain. When both $\delta$ and $\varepsilon_k$ are large, it is not conducive to achieving efficient communication. On the other hand, positioning the RIS at a higher elevation compared to the ground level can increase the value of $\varepsilon_k$. This, in turn, enhances the proportion of the LoS component $\mathbf{\bar{h}}_k$ in the user $k$-RIS channel, thereby achieving higher system throughput with fairness in mind.

\begin{figure}[t]
	\centering
	\includegraphics[width= 0.45\textwidth]{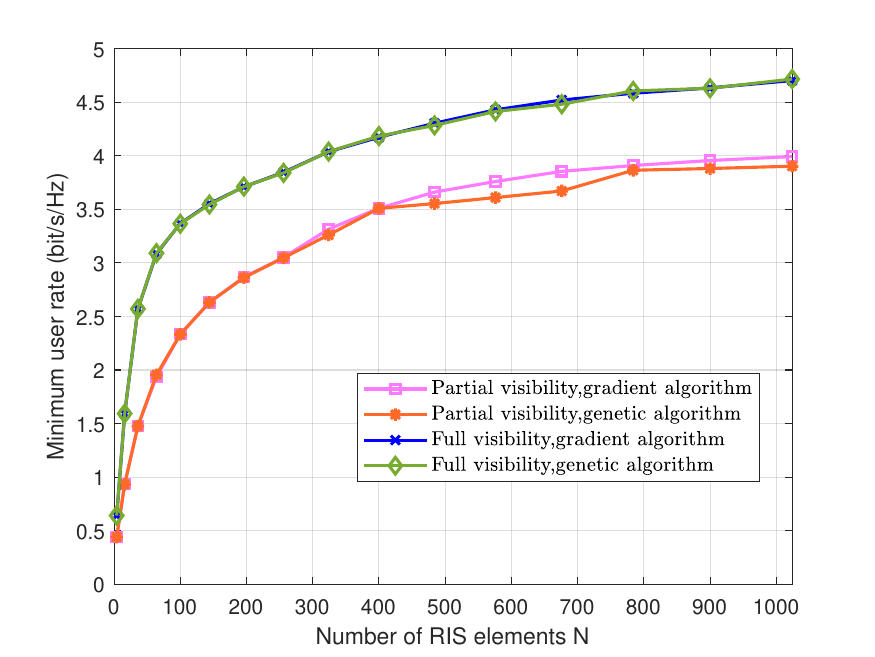}
	\caption{Comparison of results between different optimization algorithms}
	\label{figure3-1}
\end{figure}

\begin{figure}[t]
	\centering
	\includegraphics[width= 0.45\textwidth]{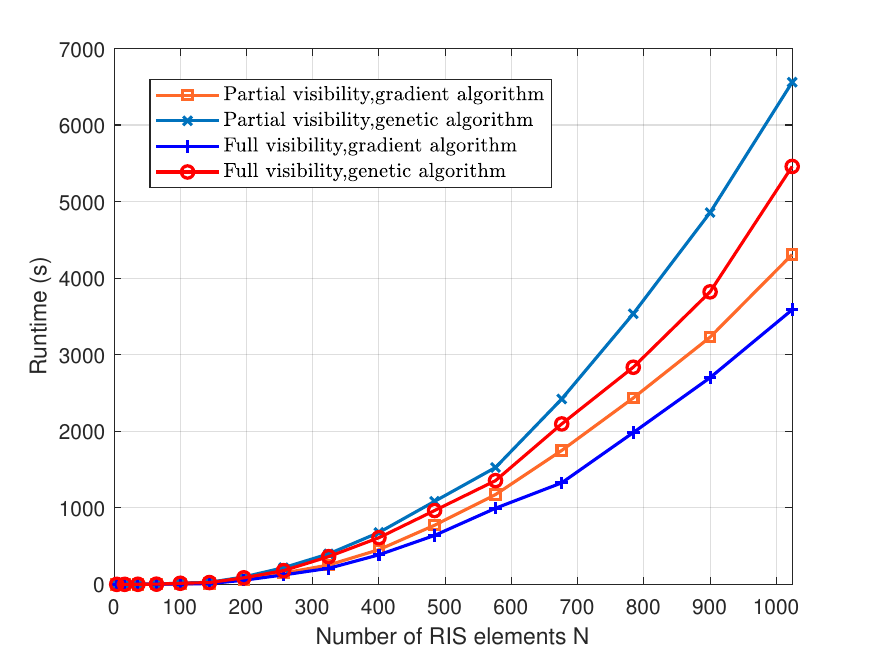}
	\caption{Comparison of runtime between different optimization algorithms}
	\label{figure3-2}
\end{figure}

In Fig. \ref{figure3-1} and Fig. \ref{figure3-2}, we compared the results and runtime of the gradient algorithm and the genetic algorithm, respectively. The analysis shows that the two algorithms have similar computational results, which proves the accuracy of the gradient algorithm. However, the runtime of the genetic algorithm is always longer than that of the gradient algorithm, especially when considering the VRs. This is because in the presence of spatial correlation and VRs, the objective function of the optimization problem becomes more complex and the local search ability of the genetic algorithm is poor, resulting in long time consumption and low search efficiency in the later stage of evolution. In this case, the genetic algorithm can no longer provide perfect operation results, while the gradient algorithm can still play a better role. Moreover, the efficiency of the gradient algorithm doesn’t decrease as dramatically as the genetic algorithm when $N$ is large. The reason is that the gradient algorithm considers angles as optimization variables, thus avoiding performance losses caused by projection operations. As a result, it can be seen that the proposed method outperforms the genetic algorithm.

\begin{figure}[t]
	\centering
	\includegraphics[width= 0.45\textwidth]{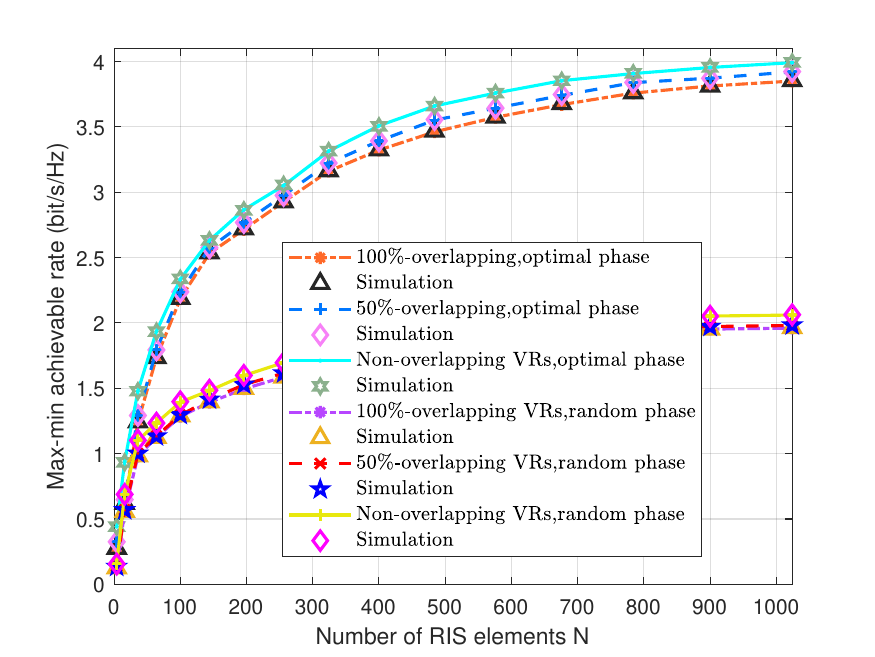}
	\caption{Max-min achievable rate versus $N$ for overlapping and non-overlapping VRs}
	\label{figure4}
\end{figure}

In Fig. \ref{figure4}, the influence of overlapping VRs was analyzed. Firstly, it is assumed that non-overlapping means that each user’s VR does not overlap with other users’, while overlapping means that each user’s VR will overlap with other users' to varying degrees. It can be seen that the minimum user rate of a 100\%-overlapping system is reduced by 0.1408 bits/s/Hz compared with a non-overlapping system, assuming $N=1024$. Therefore, the impact of overlapping VRs on the minimum user rate is relatively small and can be almost ignored when the rate is large. It indicates that by using the proposed phase shifts optimization algorithm, the interference at overlapping VRs can be reduced. As a result, it can significantly minimize the performance gap and ensure fairness in the public user rate of the entire system.

\begin{figure}[t]
	\centering
	\includegraphics[width= 0.45\textwidth]{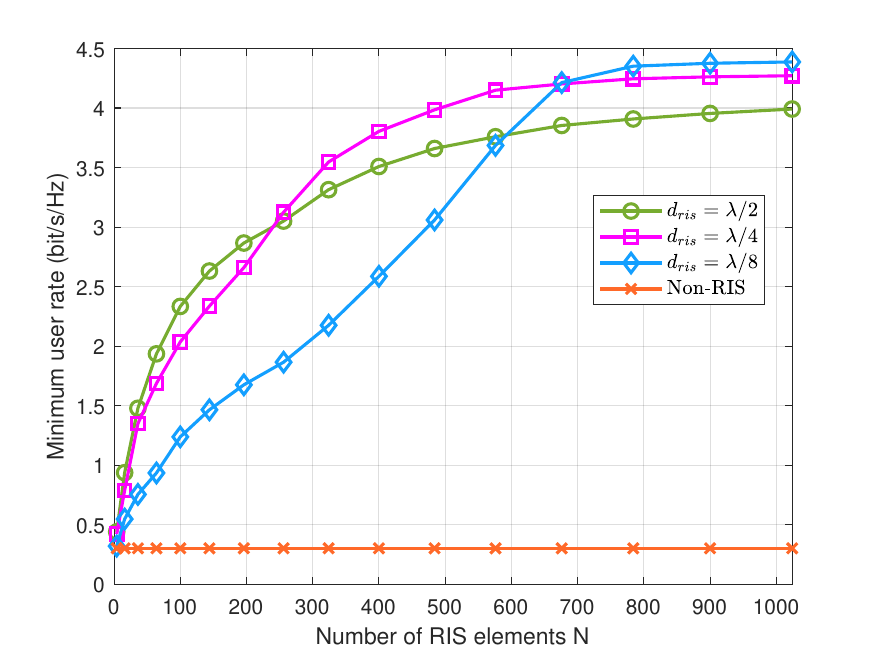}
	\caption{Max-min achievable rate versus $N$ for different values of the RIS element spacing $d_ris$}
	\label{figure5}
\end{figure}

In this paper, the objective function of the optimization problem is $f_{VR}(\boldsymbol{\theta})$, where spatial correlation is considered in the RIS phase shifts design. From Fig. \ref{figure5}, it can be observed that as the spacing between RIS components decreases ($d_{ris}=\lambda/2,\lambda/4,\lambda/8$), the influence of spatial correlation on the minimum user rate cannot be ignored. Specifically, for cases where the value of $N$ is small, the rate decreases as the distance decreases, which is attributed to the decrease in channel rank. For cases where the value of $N$ is relatively large, the channel rank still decreases. However, by increasing the number of RIS elements and utilizing channel correlation to enhance the customized wireless channel capability provided by RIS, the beamforming gain provided by optimized RIS can exceed the negative impact of spatial correlation, thereby achieving higher user rates.

\begin{figure}[t]
	\centering
	\includegraphics[width= 0.45\textwidth]{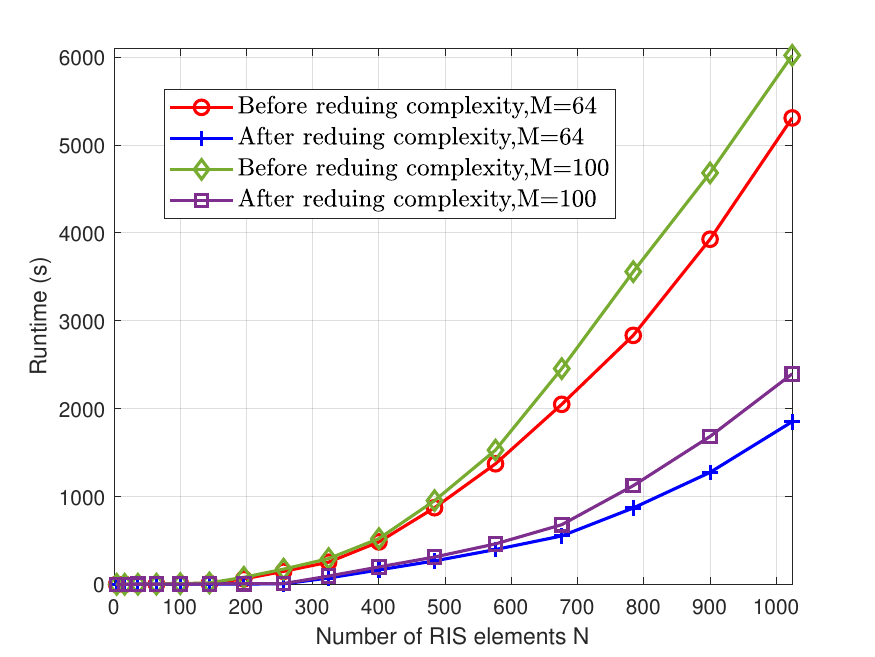}
	\caption{Comparison of runtime before and after reducing complexity}
	\label{figure6-1}
\end{figure}

In Fig. \ref{figure6-1}, it can be observed that when considering the influence of RIS visibility, as $N$ increases, the time for solving the Max-Min problem with the gradient algorithm also gradually increases. However, as the solution process for low complexity rate expressions is much simpler, the runtime can be significantly reduced. Especially when $N$ is large in XL-RIS scenarios, such as $N=1024$, reducing complexity can increase computational speed by nearly three times. 

\begin{figure}[t]
	\centering
	\includegraphics[width= 0.45\textwidth]{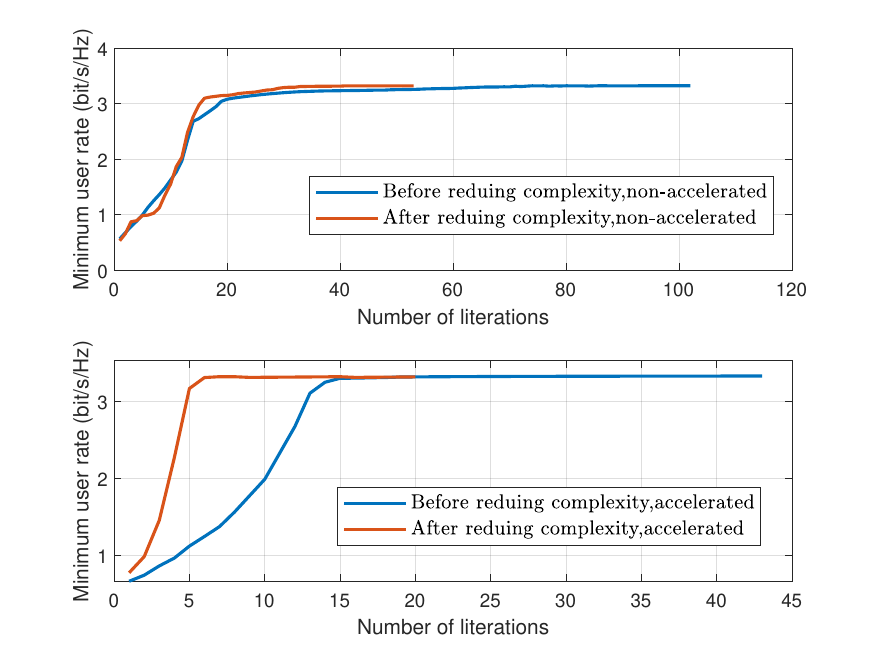}
	\caption{Comparison of system solution iteration times before and after reducing complexity}
	\label{figure6-2}
\end{figure}

On the other hand, Fig. \ref{figure6-2} investigated the convergence behavior of the proposed low-complexity gradient algorithm compared to the ordinary method. By applying the proposed complexity reduction method, it can be observed that when $M=64$ and $N=400$, the algorithm converges within 20 iterations. Compared with the 43 iterations that use the usual gradient algorithm, there is a slight improvement in convergence speed. Besides, although there is not much difference in the number of iterations before and after reducing complexity, the efficiency is significantly increased and the total time consumption is decreased due to the greatly reduced complexity of single iteration calculations. Moreover, the convergence speed can be further effectively improved by applying the proposed acceleration method. Hence, the above results confirm that redundant components can be removed through the proposed method in (\ref{reduced_matrix}). As a result, the achievable rate expression with dimension-reduced matrices can effectively improve system efficiency and save resources.

\section{Conclusion}\label{sixth}
In this article, we mainly studied the two-timescale design of XL-RIS-aided massive MIMO communication systems, taking into account the influence of VRs. Firstly, we considered a spatially-correlated channel model in the presence of VRs. A closed-form expression of the achievable user rate was derived and the optimization problem was clarified. At the same time, we analyzed the impact of the visibility region of RIS on system complexity and simplified the user rate expression. Secondly, we proposed a gradient algorithm different from the genetic algorithm and validated the potential of applying it to the phase shifts optimization problem. Finally, we conducted simulations and analysis about XL-RIS-aided massive MIMO systems, and the impact of VRs on system performance was verified from different perspectives.

\begin{appendices}
\section{}\label{appendixA}
In the derivation of user achievable rate, since the full visibility model used in section 1.3 is a special case of the partial visibility model, we mainly discuss the calculation process of equation (16) in this section. Before the formal derivation process, we first prove a mathematical conclusion required for the subsequent part.

\begin{lem}\label{lemma_A1}
	For deterministic matrices $\mathbf{A},\mathbf{B}\in\mathbb{C}^{M\times M}$ and $\mathbf{W}\in\mathbb{C}^{N\times N}$, while $\mathbf{A}$ and $\mathbf{B}$ are both unitary matrices, there is:
	\begin{align}\label{lemmaA}
		\begin{aligned}
			&\mathbb{E}\{\widetilde{\mathbf{H}}_2^H\mathbf{A}\widetilde{\mathbf{H}}_2\mathbf{W}\widetilde{\mathbf{H}}_2^H\mathbf{B}\widetilde{\mathbf{H}}_2\}\\
			&=\mathrm{Tr}\{\mathbf{W}\}\mathrm{Tr}\{\mathbf{A}\mathbf{B}\}\boldsymbol{I}_N+\mathrm{Tr}\{\mathbf{A}\}\mathrm{Tr}\{\mathbf{B}\}\mathbf{W}.
		\end{aligned}
	\end{align}
\end{lem}

\itshape {Proof:}  \upshape 
We define $\widetilde{\mathbf{H}}_{2}=[\mathbf{J}_{1},\ldots,\mathbf{J}_{N}]$, where $\mathbf{J}_n\in\mathbb{C}^{M\times1}\text{,}1\leq n\leq N$, each component is independent and satisfies $\mathbf{J}_{n}\sim\mathcal{CN}(\mathbf{0},\mathbf{I}_M) $. Therefore, the $(i,j)$-th element in $\widetilde{\mathbf{H}}_2^H\mathbf{A}\widetilde{\mathbf{H}}_2\mathbf{W}\widetilde{\mathbf{H}}_2^H\mathbf{B}\widetilde{\mathbf{H}}_2$ can be written as:
\begin{align}
	\begin{aligned}
		&\left[\widetilde{\mathbf{H}}_2^H\mathbf{A}\widetilde{\mathbf{H}}_2\mathbf{W}\widetilde{\mathbf{H}}_2^H\mathbf{B}\widetilde{\mathbf{H}}_2\right]_{i,j}\\
		&=\sum_{h=1}^N\sum_{m=1}^N\mathbf{J}_i^H\mathbf{A}\mathbf{J}_mw_{mh}\mathbf{J}_h^H\mathbf{B}\mathbf{J}_j,
	\end{aligned}
\end{align}

\noindent where $w_{mh}$ is the $(w,h)$-th element of $\mathbf{W}$. Since $\mathbb{E}\{\mathbf{J}_{k}^{H}\mathbf{A}\mathbf{J}_{k}\mathbf{J}_{k}^{H}\mathbf{B}\mathbf{J}_{k}\}=\mathrm{Tr}\{\mathbf{A}\mathbf{B}\}+\mathrm{Tr}\{\mathbf{A}\}\mathrm{Tr}\{\mathbf{B}\}$, the expectation of diagonal elements in $\widetilde{\mathbf{H}}_2^H\mathbf{A}\widetilde{\mathbf{H}}_2\mathbf{W}\widetilde{\mathbf{H}}_2^H\mathbf{B}\widetilde{\mathbf{H}}_2$ can be calculated as follows:
\begin{align}
	\begin{aligned}
		&\left[\widetilde{\mathbf{H}}_2^H\mathbf{A}\widetilde{\mathbf{H}}_2\mathbf{W}\widetilde{\mathbf{H}}_2^H\mathbf{B}\widetilde{\mathbf{H}}_2\right]_{i,i}\\
		=&\mathbb{E}\{\mathbf{J}_i^H\mathbf{A}\mathbf{J}_iw_{ii}\mathbf{J}_i^H\mathbf{B}\mathbf{J}_i\}\\
		&+\mathbb{E}\left\{\sum_{m=1,m\neq i}^N\mathbf{J}_i^H\mathbf{A}\mathbf{J}_mw_{mm}\mathbf{J}_m^H\mathbf{B}\mathbf{J}_i\right\}\\
		=&w_{ii}\mathbb{E}\{\mathbf{J}_i^H\mathbf{A}\mathbf{J}_i\mathbf{J}_i^H\mathbf{B}\mathbf{J}_i\}\\
		&+\mathbb{E}\left\{\sum_{m=1,m\neq i}^Nw_{mm}\mathbf{J}_i^H\mathbf{A}\mathbb{E}\{\mathbf{J}_m\mathbf{J}_m^H\}\mathbf{B}\mathbf{J}_i\right\}\\
		=&w_{ii}(\mathrm{Tr}\{\mathbf{AB}\}+\mathrm{Tr}\{\mathbf{A}\}\mathrm{Tr}\{\mathbf{B}\})+\sum_{m=1,m\neq i}^{N}w_{mm}\mathrm{Tr}\{\mathbf{AB}\}\\
		=&w_{ii}\operatorname{Tr}\{\mathbf{A}\}\mathrm{Tr}\{\mathbf{B}\}+\operatorname{Tr}\{\mathbf{AB}\}\mathrm{Tr}\{\mathbf{W}\}.
	\end{aligned}
\end{align}

The expectation of non-diagonal elements in $\widetilde{\mathbf{H}}_2^H\mathbf{A}\widetilde{\mathbf{H}}_2\mathbf{W}\widetilde{\mathbf{H}}_2^H\mathbf{B}\widetilde{\mathbf{H}}_2$ can be derived through the following process:
\begin{align}
	\begin{aligned}
		&\begin{bmatrix}\widetilde{\mathbf{H}}_2^H\mathbf{A}\widetilde{\mathbf{H}}_2\mathbf{W}\widetilde{\mathbf{H}}_2^H\mathbf{B}\widetilde{\mathbf{H}}_2\end{bmatrix}_{i,j}\\
		&=\mathbb{E}\{\mathbf{J}_i^H\mathbf{A}\mathbf{J}_iw_{ij}\mathbf{J}_j^H\mathbf{B}\mathbf{J}_j\}\\
		&=w_{ij}\mathbb{E}\{\mathbf{J}_i^H\mathbf{A}\mathbf{J}_i\}\mathbb{E}\{\mathbf{J}_j^H\mathbf{B}\mathbf{J}_j\}\\
		&=w_{ij}\text{Tr}\{\mathbf{A}\}\text{Tr}\{\mathbf{B}\}.
	\end{aligned}
\end{align}

Therefore, by combining the above two equations, the conclusion in Lemma \ref{lemma_A1} can be obtained.

\subsection{The derivation of noise term $E_{VR,k}^{\mathrm{noise}}(\boldsymbol{\Phi})$}

As $\mathbf{q}_k$ can be written in the form of equation (\ref{12}), we divide it into four components:
\begin{align}\label{q_k_div}
	\begin{aligned}
		\mathbf{q}_{k}^{1} &=\sqrt{c_k\delta\varepsilon_k}\left(\bar{\mathbf{H}}_2\mathbf{\Phi}\mathbf{D}_k^{\frac12}\bar{\mathbf{h}}_k\right), \\
		\mathbf{q}_{k}^{2} &=\sqrt{c_k\delta}\left(\bar{\mathbf{H}}_2\boldsymbol{\Phi}\mathbf{R}_{VR,k}^{\frac12}\tilde{\mathbf{h}}_k\right), \\
		\mathbf{q}_{k}^{3} &=\sqrt{c_{k}\varepsilon_{k}}\left(\widetilde{\mathbf{H}}_{2}\mathbf{R}_{ris}^{\frac12}\mathbf{\Phi}\mathbf{D}_{k}^{\frac12}\mathbf{\bar{h}}_{k}\right), \\
		\mathbf{q}_{k}^{4} &=\sqrt{c_{k}}\left(\widetilde{\mathbf{H}}_{2}\mathbf{R}_{ris}^{\frac{1}{2}}\mathbf{\Phi R}_{VR,k}^{\frac{1}{2}}\tilde{\mathbf{h}}_{k}\right),
	\end{aligned}
\end{align}

\noindent where $\mathbf{q}_{k}=\mathbf{q}_{k}^{1}+\mathbf{q}_{k}^{2}+\mathbf{q}_{k}^{3}+\mathbf{q}_{k}^{4}$. The expansion of  $E_{VR,k}^{\mathrm{noise}}(\boldsymbol{\Phi})$ can be written as:
\begin{align}\label{A.2}
	\begin{aligned}
		E_{VR,k}^{\mathrm{noise}}(\boldsymbol{\Phi})&=\mathbb{E}\parallel\mathbf{q}_k\parallel^2=\mathbb{E}\{\mathbf{q}_k^H\mathbf{q}_k\}\\
		&=\sum_{\omega=1}^4\sum_{\psi=1}^4\mathbb{E}\{\mathbf{q}_k^{\omega H}\mathbf{q}_k^{\psi}\}.
	\end{aligned}
\end{align}

Due to the assumption that the components of $\tilde{\mathbf{h}}_{k}$ and $\widetilde{\mathbf{H}}_{2}$ are i.i.d. complex Gaussian random variables with zero mean and unit variance, $E_{VR,k}^{\mathrm{noise}}(\boldsymbol{\Phi})$ can be simplified as follows:
\begin{align}\label{A.3}
	\begin{aligned}
		E_{VR,k}^\mathrm{noise}(\boldsymbol{\Phi})=\sum_{\omega=1}^4\mathbb{E}\{\mathbf{q}_k^{\omega H}\mathbf{q}_k^\omega\},
	\end{aligned}
\end{align}

where
\begin{align}\label{qk1-qk1}
	\begin{aligned}
		&\mathbb{E}\left\{\mathbf{q}_k^1{}^H\mathbf{q}_k^1\right\}\\
		&=c_k\delta\varepsilon_k\mathbb{E}\{\bar{\mathbf{h}}_k^H\mathbf{D}_k^{1/2}\mathbf{\Phi}^H\bar{\mathbf{H}}_2^H\bar{\mathbf{H}}_2\mathbf{\Phi}\mathbf{D}_k^{1/2}\bar{\mathbf{h}}_k\}\\
		&=c_k\delta\varepsilon_kM(\bar{\mathbf{h}}_k^H\mathbf{D}_k^{1/2}\mathbf{\Phi}^H\mathbf{a}_N\mathbf{a}_N^H\mathbf{\Phi}\mathbf{D}_k^{1/2}\mathbf{\bar{h}}_k)\\
		&=c_k\delta\varepsilon_kM|f_k(\mathbf{\Phi})|^2,
	\end{aligned}
\end{align}
\begin{align}\label{qk2-qk2}
	\begin{aligned}
		&\mathbb{E}\left\{\mathbf{q}_{k}^{2}{}^{H}\mathbf{q}_{k}^{2}\right\}\\
		&=c_{k}\delta\mathbb{E}\{\tilde{\mathbf{h}}_{k}^{H}\mathbf{R}_{VR,k}^{1/2}\mathbf{\Phi}^{H}\bar{\mathbf{H}}_{2}^{H}\bar{\mathbf{H}}_{2}\mathbf{\Phi}\mathbf{R}_{VR,k}^{1/2}\tilde{\mathbf{h}}_{k}\}\\
		&=c_k\delta M\mathrm{Tr}(\bar{\mathbf{H}}_2\mathbf{\Phi R}_{VR,k}\mathbf{\Phi}^H\bar{\mathbf{H}}_2^H)\\
		&=c_k\delta Mf_{k,1,1}(\mathbf{\Phi}),
	\end{aligned}
\end{align}
\begin{align}\label{qk3-qk3}
	\begin{aligned}
		&\mathbb{E}\left\{\mathbf{q}_k^3{}^H\mathbf{q}_{k}^{3}\right\}\\
		&=c_{k}\varepsilon_{k}\mathbb{E}\{\bar{\mathbf{h}}_{k}^{H}\mathbf{D}_{k}^{1/2}\boldsymbol{\Phi}^{H}\mathbf{R}_{ris}^{1/2}\widetilde{\mathbf{H}}_{2}^{H}\widetilde{\mathbf{H}}_{2}\mathbf{R}_{ris}^{1/2}\boldsymbol{\Phi}\mathbf{D}_{k}^{1/2}\bar{\mathbf{h}}_{k}\}\\
		&=c_{k}\varepsilon_{k}M\bar{\mathbf{h}}_{k}^{H}\mathbf{D}_{k}^{1/2}\mathbf{\Phi}^{H}\mathbf{R}_{ris}\mathbf{\Phi}\mathbf{D}_{k}^{1/2}\mathbf{\bar{h}}_{k}\\
		&=c_{k}\varepsilon_{k}Mf_{kk,2}(\mathbf{\Phi}),
	\end{aligned}
\end{align}
\begin{align}\label{qk4-qk4}
	\begin{aligned}
		&\mathbb{E}\left\{\mathbf{q}_{k}^{4H}\mathbf{q}_{k}^{4}\right\}\\
		&=c_{k}\operatorname{E}\{\tilde{\mathbf{h}}_{k}^{H}\mathbf{R}_{VR,k}^{1/2}\boldsymbol{\Phi}^{H}\mathbf{R}_{ris}^{1/2}\widetilde{\mathbf{H}}_{2}^{H}\widetilde{\mathbf{H}}_{2}\mathbf{R}_{ris}^{1/2}\boldsymbol{\Phi}\mathbf{R}_{VR,k}^{1/2}\tilde{\mathbf{h}}_{k}\}\\
		&=c_kM\mathrm{Tr}(\mathbf{R}_{ris}\boldsymbol{\Phi}\mathbf{R}_{VR,k}\boldsymbol{\Phi}^H)\\
		&=c_{k}Mf_{k,3,1}(\mathbf{\Phi}).
	\end{aligned}
\end{align}

The values of $f_k(\mathbf{\Phi})$, $f_{k,1,1}(\mathbf{\Phi})$, $f_{kk,2}(\mathbf{\Phi})$, $f_{k,3,1}(\mathbf{\Phi})$ are given in (\ref{f_old}). In summary, the result of the noise term $E_{VR,k}^{\mathrm{noise}}(\boldsymbol{\Phi})$ is:
\begin{align}\label{noise_result}
	\begin{aligned}
		E_{VR,k}^{\mathrm{noise}}(\boldsymbol{\Phi})=&c_k\left\{\delta\varepsilon_kM|f_k(\boldsymbol{\Phi})|^2+\delta Mf_{k,1,1}(\boldsymbol{\Phi}) \right.\\
		&\left. +\varepsilon_kMf_{kk,2}(\boldsymbol{\Phi})+Mf_{k,3,1}(\boldsymbol{\Phi})\right\}.
	\end{aligned}
\end{align}

\subsection{The derivation of signal term $E_{VR,k}^{\mathrm{signal}}(\boldsymbol{\Phi})$}

Due to $E_{VR,k}^{\mathrm{signal}}(\boldsymbol{\Phi})=\mathbb{E}\parallel\mathbf{q}_{k}\parallel^{4}=\mathbb{E}\{\mathbf{q}_{k}^{H}\mathbf{q}_{k}^{H}\mathbf{q}_{k}^{H}\mathbf{q}_{k}\}$, we use (\ref{q_k_div}) to expand it:
\begin{align}\label{A.9}
	\begin{aligned}
		E_{VR,k}^{\mathrm{signal}}(\mathbf{\Phi}) =&\sum_{\omega_1,\psi_1}^4\sum_{\omega_2,\psi_2}^4\operatorname{E}\left\{\left(\mathbf{q}_k^{\omega_1H}\mathbf{q}_k^{\psi_1}\right)\left(\mathbf{q}_k^{\omega_2H}\mathbf{q}_k^{\psi_2}\right)^H\right\}  \\
		=&\sum_{\omega=1}^4\sum_{\psi=1}^4\mathbb{E}\left\{(\mathbf{q}_k^{\omega H}\mathbf{q}_k^{\psi})(\mathbf{q}_k^{\omega H}\mathbf{q}_k^{\psi})^H\right\} \\
		&+2\sum_{\omega_1=1}^4\sum_{\psi_1=\omega_1+1}^4\mathbb{E}\left\{\left(\mathbf{q}_k^{\omega_1H}\mathbf{q}_k^{\omega_1}\right)\left(\mathbf{q}_k^{\psi_1H}\mathbf{q}_k^{\psi_1}\right)^H\right\}\\
		&+2\mathrm{Re}\left\{\mathbb{E}\left\{\left(\mathbf{q}_k^1{}^H\mathbf{q}_k^2\right)\left(\mathbf{q}_k^3{}^H\mathbf{q}_k^4\right)^H\right\}\right\}\\
		&+2\mathrm{Re}\left\{\mathbb{E}\left\{\left(\mathbf{q}_k^1{}^H\mathbf{q}_k^3\right)\left(\mathbf{q}_k^2{}^H\mathbf{q}_k^4\right)^H\right\}\right\}\\
		&+2\mathrm{Re}\left\{\mathbb{E}\left\{\left(\mathbf{q}_k^2{}^H\mathbf{q}_k^1\right)\left(\mathbf{q}_k^4{}^H\mathbf{q}_k^3\right)^H\right\}\right\}\\
		&+2\mathrm{Re}\left\{\mathbb{E}\left\{\left(\mathbf{q}_k^2{}^H\mathbf{q}_k^4\right)\left(\mathbf{q}_k^1{}^H\mathbf{q}_k^3\right)^H\right\}\right\}.
	\end{aligned}
\end{align}

Then we classify different values of $\omega_1,\psi_1,\omega,\psi$, and first calculate the result of $\sum_{\omega=1}^4\sum_{\psi=1}^4\mathbb{E}\left\{(\mathbf{q}_k^{\omega H}\mathbf{q}_k^{\psi})(\mathbf{q}_k^{\omega H}\mathbf{q}_k^{\psi})^H\right\}$. The formula can be further simplified into the following form:
\begin{align}\label{A.10}
	\begin{aligned}
		&\sum_{\omega=1}^4\sum_{\psi=1}^4\mathbb{E}\left\{(\mathbf{q}_k^{\omega H}\mathbf{q}_k^{\psi})(\mathbf{q}_k^{\omega H}\mathbf{q}_k^{\psi})^H\right\}\\
		&=\sum_{\omega=1}^4\mathbb{E}\left\{(\mathbf{q}_k^{\omega H}\mathbf{q}_k^\omega)(\mathbf{q}_k^{\omega H}\mathbf{q}_k^\omega)^H\right\}\\
		&+2\sum_{\omega=1}^4\sum_{\psi=\omega+1}^4\mathbb{E}\left\{(\mathbf{q}_k^{\omega H}\mathbf{q}_k^{\psi})(\mathbf{q}_k^{\omega H}\mathbf{q}_k^{\psi})^H\right\}.
	\end{aligned}
\end{align}

The specific calculation of the first part $\sum_{\omega=1}^4\mathbb{E}\left\{(\mathbf{q}_k^{\omega H}\mathbf{q}_k^\omega)(\mathbf{q}_k^{\omega H}\mathbf{q}_k^\omega)^H\right\}$ is as follows. When $\omega=1$, we have
\begin{align}\label{A.11}
	\begin{aligned}
		&\mathbb{E}\left\{\left(\mathbf{q}_k^1{}^H\mathbf{q}_k^1\right)\left(\mathbf{q}_k^1{}^H\mathbf{q}_k^1\right)^H\right\} \\
		&=(c_k\delta\varepsilon_k)^2\mathbb{E}\left\{|\bar{\mathbf{h}}_k^H\mathbf{D}_k^{1/2}\Phi^H\bar{\mathbf{H}}_2^H\bar{\mathbf{H}}_2\Phi\mathbf{D}_k^{1/2}\bar{\mathbf{h}}_k|^2\right\} \\
		&=(c_k\delta\varepsilon_k)^2\mathbb{E}\left\{\left|\left(\bar{\mathbf{h}}_k^H\mathbf{D}_k^{1/2}\mathbf{\Phi}^H\mathbf{a}_N\right)\mathbf{a}_M^H\mathbf{a}_M(\mathbf{a}_N^H\mathbf{\Phi}\mathbf{D}_k^{1/2}\mathbf{\bar{h}}_k)\right|^2\right\} \\
		&=M^{2}(c_{k}\delta\varepsilon_{k})^{2}|f_{k}(\mathbf{\Phi})|^{4}.
	\end{aligned}
\end{align}

The value of $f_k(\mathbf{\Phi})$ is given in (\ref{f_old}). When $\omega=2$, we have
\begin{align}\label{A.12}
	\begin{aligned}
		&\mathbb{E}\left\{\left(\mathbf{q}_k^2{}^H\mathbf{q}_k^2\right)\left(\mathbf{q}_k^2{}^H\mathbf{q}_k^2\right)^H\right\}\\
		=&(c_k\delta)^2\mathbb{E}\left\{|\tilde{\mathbf{h}}_k^H\mathbf{R}_{VR,k}^{1/2}\mathbf{\Phi}^H\bar{\mathbf{H}}_2^H\bar{\mathbf{H}}_2\mathbf{\Phi}\mathbf{R}_{VR,k}^{1/2}\tilde{\mathbf{h}}_k|^2\right\}\\
		=&(c_k\delta)^2\left\{\mathrm{Tr}\left(\left(\mathbf{R}_\mathrm{VR,k}^{1/2}\boldsymbol{\Phi}^\mathrm{H}\bar{\mathbf{H}}_2^\mathrm{H}\bar{\mathbf{H}}_2\boldsymbol{\Phi}\mathbf{R}_\mathrm{VR,k}^{1/2}\right)^2\right) \right.\\
		&\left.+\left|\mathrm{Tr}(\mathbf{R}_\mathrm{VR,k}^{1/2}\boldsymbol{\Phi}^\mathrm{H}\bar{\mathbf{H}}_2^\mathrm{H}\bar{\mathbf{H}}_2\boldsymbol{\Phi}\mathbf{R}_\mathrm{VR,k}^{1/2})\right|^2\right\}\\
		=&(c_k\delta)^2\left(f_{kk,1,2}(\mathbf\Phi)+\left|f_{k,1,1}(\mathbf\Phi)\right|^2\right).
	\end{aligned}
\end{align}

The values of $f_{kk,1,2}(\mathbf\Phi)$ and $f_{k,1,1}(\mathbf\Phi)$ are given in (\ref{f_old}). When $\omega=3$, we have
\begin{align}\label{A.13}
	\begin{aligned}
		&\mathbb{E}\left\{\left(\mathbf{q}_k^3{}^H\mathbf{q}_k^3\right)\left(\mathbf{q}_k^3{}^H\mathbf{q}_k^3\right)^H\right\}\\
		=&(c_{k}\varepsilon_{k})^{2}\mathbb{E}\left\{|\bar{\mathbf{h}}_{k}^{H}\mathbf{D}_{k}^{1/2}\mathbf{\Phi}^{H}\mathbf{R}_{ris}^{1/2}\tilde{\mathbf{H}}_{2}^{H}\tilde{\mathbf{H}}_{2}\mathbf{R}_{ris}^{1/2}\mathbf{\Phi}\mathbf{D}_{k}^{1/2}\bar{\mathbf{h}}_{k}|^{2}\right\}\\
		=&(c_k\varepsilon_k)^2\bar{\mathbf{h}}_k^H\mathbf{D}_k^{1/2}\boldsymbol{\Phi}^H\mathbf{R}_{ris}^{1/2}\\
		&\times\begin{pmatrix}\boldsymbol{M}\mathrm{Tr}(\mathbf{R}_{ris}^{1/2}\boldsymbol{\Phi}\mathbf{D}_k^{1/2}\bar{\mathbf{h}}_k\bar{\mathbf{h}}_k^H\mathbf{D}_k^{1/2}\boldsymbol{\Phi}^H\mathbf{R}_{ris}^{1/2})\mathbf{I}_N\\+M^2\mathbf{R}_{ris}^{1/2}\boldsymbol{\Phi}\mathbf{D}_k^{1/2}\bar{\mathbf{h}}_k\bar{\mathbf{h}}_k^H\mathbf{D}_k^{1/2}\boldsymbol{\Phi}^H\mathbf{R}_{ris}^{1/2}\end{pmatrix}\\
		&\times\mathbf{R}_{ris}^{1/2}\boldsymbol{\Phi}\mathbf{D}_k^{1/2}\mathbf{\bar{h}}_k\\
		=&(c_{k}\varepsilon_{k})^{2}M(M+1)|\bar{\mathbf{h}}_{k}^{H}\mathbf{D}_{k}^{1/2}\mathbf{\Phi}^{H}\mathbf{R}_{ris}\mathbf{\Phi}\mathbf{D}_{k}^{1/2}\mathbf{\bar{h}}_{k}|^{2} \\
		=&(c_{k}\varepsilon_{k})^{2}M(M+1)|f_{kk,2}(\mathbf{\Phi})|^{2}.
	\end{aligned}
\end{align}

This derivation uses the conclusion (\ref{lemmaA}) of Lemma \ref{lemma_A1}, where $\mathbf{A}=\mathbf{B}=\mathbf{I}_{N}$ and $\mathbf{W}=\mathbf{R}_{ris}^{1/2}\mathbf{\Phi}\mathbf{D}_{k}^{1/2}\mathbf{\bar{h}}_{k}\mathbf{\bar{h}}_{k}^{H}\mathbf{D}_{k}^{1/2}\mathbf{\Phi}^{H}\mathbf{R}_{ris}^{1/2}$. The value of $f_{kk,2}(\mathbf{\Phi})$ is given in (\ref{f_old}). When $\omega=4$, we have
\begin{align}\label{A.14}
	\begin{aligned}
		&\mathbb{E}\left\{\left(\mathbf{q}_k^4{}^H\mathbf{q}_k^4\right)\left(\mathbf{q}_k^4{}^H\mathbf{q}_k^4\right)^H\right\}\\
		=&c_k^2\mathbb{E}\left\{|\tilde{\mathbf{h}}_k^H\mathbf{R}_{VR,k}^{1/2}\boldsymbol{\Phi}^H\mathbf{R}_{ris}^{1/2}\tilde{\mathbf{H}}_2^H\tilde{\mathbf{H}}_2\mathbf{R}_{ris}^{1/2}\boldsymbol{\Phi}\mathbf{R}_{VR,k}^{1/2}\tilde{\mathbf{h}}_k|^2\right\}\\
		=&c_{k}^{2}\mathbb{E}\left\{\tilde{\mathbf{h}}_{k}^{H}\mathbf{R}_{VR,k}^{1/2}\boldsymbol{\Phi}^{H}\mathbf{R}_{ris}^{1/2} \right.\\
		&\times\begin{pmatrix}M\mathrm{Tr}\big(\mathbf{R}_{ris}^{1/2}\boldsymbol{\Phi}\mathbf{R}_{VR,k}^{1/2}\tilde{\mathbf{h}}_{k}\tilde{\mathbf{h}}_{k}^{H}\mathbf{R}_{VR,k}^{1/2}\boldsymbol{\Phi}^{H}\mathbf{R}_{ris}^{1/2}\big)\mathbf{I}_{N}\\+M^{2}\mathbf{R}_{ris}^{1/2}\boldsymbol{\Phi}\mathbf{R}_{VR,k}^{1/2}\mathbf{h}_{k}^{1}\tilde{\mathbf{h}}_{k}^{H}\mathbf{R}_{\nu R,k}^{1/2}\boldsymbol{\Phi}^{H}\mathbf{R}_{ris}^{1/2}\end{pmatrix}\\
		&\left. \times \mathbf{R}_{ris}^{1/2}\boldsymbol{\Phi}\mathbf{R}_{VR,k}^{1/2}\tilde{\mathbf{h}}_{k}\right\}\\
		=&c_k^2M(M+1)\mathbb{E}\left\{\tilde{\mathbf{h}}_k^H\mathbf{R}_{VR,k}^{1/2}\boldsymbol{\Phi}^H\mathbf{R}_{ris}^{1/2}\boldsymbol{\Phi}\mathbf{R}_{VR,k}^{1/2}\mathbf{\tilde{h}}_k\right.\\
		&\left.\times\mathbf{\tilde{h}}_k^H\mathbf{R}_{VR,k}^{1/2}\boldsymbol{\Phi}^H\mathbf{R}_{ris}^{1/2}\boldsymbol{\Phi}\mathbf{R}_{VR,k}^{1/2}\mathbf{\tilde{h}}_k\right\}\\
		=&c_{k}^{2}M(M+1)\left\{\mathrm{Tr}\left(\left(\boldsymbol{\Phi}^{H}\mathbf{R}_{ris}\boldsymbol{\Phi}\mathbf{R}_{VR,k}\right)^{2}\right)\right.\\
		&\left.+\left|\mathrm{Tr}(\boldsymbol{\Phi}^{H}\mathbf{R}_{ris}\boldsymbol{\Phi}\mathbf{R}_{VR,k})\right|^{2}\right\}\\
		=&c_k^2M(M+1)\left(f_{kk,3,2}(\boldsymbol{\Phi})+\left|f_{k,3,1}(\boldsymbol{\Phi})\right|^2\right).
	\end{aligned}
\end{align}

The values of $f_{kk,3,2}(\mathbf{\Phi})$ and $f_{k,3,1}(\boldsymbol{\Phi})$ are given in (\ref{f_old}). Then we calculate the result of second part $2\sum_{\omega_1=1}^4\sum_{\psi_1=\omega_1+1}^4\mathbb{E}\left\{\left(\mathbf{q}_k^{\omega_1H}\mathbf{q}_k^{\omega_1}\right)\left(\mathbf{q}_k^{\psi_1H}\mathbf{q}_k^{\psi_1}\right)^H\right\}$. Firstly, we consider the terms with $\omega=1$. When $\psi=2$, we have
\begin{align}\label{A.15}
	\begin{aligned}
		&2\mathbb{E}\left\{\left(\mathbf{q}_k^1{}^H\mathbf{q}_k^2\right)\left(\mathbf{q}_k^1{}^H\mathbf{q}_k^2\right)^H\right\} \\
		&=2(c_k\delta)^2\varepsilon_k\mathbb{E}\left\{|\bar{\mathbf{h}}_k^H\mathbf{D}_k^{1/2}\mathbf{\Phi}^H\bar{\mathbf{H}}_2^H\bar{\mathbf{H}}_2\mathbf{\Phi}\mathbf{R}_{VR,k}^{1/2}\tilde{\mathbf{h}}_k|^2\right\} \\
		&=2(c_k\delta)^2\varepsilon_k\bar{\mathbf{h}}_k^H\mathbf{D}_k^{1/2}\mathbf{\Phi}^H\bar{\mathbf{H}}_2^H\bar{\mathbf{H}}_2\mathbf{\Phi}\mathbf{R}_{VR,k}^{1/2}\mathbb{E}\{\tilde{\mathbf{h}}_k\tilde{\mathbf{h}}_k^H\}\\
		&\times\mathbf{R}_{VR,k}^{1/2}\mathbf{\Phi}^H\bar{\mathbf{H}}_2^H\bar{\mathbf{H}}_2\mathbf{\Phi}\mathbf{D}_k^{1/2}\bar{\mathbf{h}}_k \\
		&=2(c_k\delta)^2\varepsilon_k\bar{\mathbf{h}}_k^H\mathbf{D}_k^{1/2}\mathbf{\Phi}^H\bar{\mathbf{H}}_2^H\bar{\mathbf{H}}_2\mathbf{\Phi}\mathbf{R}_{VR,k}\mathbf{\Phi}^H\bar{\mathbf{H}}_2^H\bar{\mathbf{H}}_2\mathbf{\Phi}\mathbf{D}_k^{1/2}\mathbf{\bar{h}}_k \\
		&=2M(c_{k}\delta)^{2}\varepsilon_{k}|f_{k}(\mathbf{\Phi})|^{2}f_{k,1,1}(\mathbf{\Phi}).
	\end{aligned}
\end{align}

When $\psi=3$, we have
\begin{align}\label{A.16}
	\begin{aligned}
		&2\mathbb{E}\left\{\left(\mathbf{q}_k^1{}^H\mathbf{q}_k^3\right)\left(\mathbf{q}_k^1{}^H\mathbf{q}_k^3\right)^H\right\} \\
		&=2(c_k\varepsilon_k)^2\delta\mathbb{E}\left\{|\bar{\mathbf{h}}_k^H\mathbf{D}_k^{1/2}\mathbf{\Phi}^H\bar{\mathbf{H}}_2^H\tilde{\mathbf{H}}_2\mathbf{R}_{ris}^{1/2}\mathbf{\Phi}\mathbf{D}_k^{1/2}\bar{\mathbf{h}}_k|^2\right\} \\
		&=2(c_k\varepsilon_k)^2\delta\bar{\mathbf{h}}_k^H\mathbf{D}_k^{1/2}\boldsymbol{\Phi}^H\bar{\mathbf{H}}_2^H\\
		&\times\mathbb{E}\{\tilde{\mathbf{H}}_2\mathbf{R}_{ris}^{1/2}\boldsymbol{\Phi}\mathbf{D}_k^{1/2}\bar{\mathbf{h}}_k^H\mathbf{\bar{h}}_k^H\mathbf{D}_k^{1/2}\boldsymbol{\Phi}^H\mathbf{R}_{ris}^{1/2}\tilde{\mathbf{H}}_2^H\}\bar{\mathbf{H}}_2\boldsymbol{\Phi}\mathbf{D}_k^{1/2}\bar{\mathbf{h}}_k \\
		&=2(c_k\varepsilon_k)^2\delta\bar{\mathbf{h}}_k^H\mathbf{D}_k^{1/2}\mathbf{\Phi}^H\bar{\mathbf{H}}_2^H\bar{\mathbf{H}}_2\mathbf{\Phi}\mathbf{D}_k^{1/2}\bar{\mathbf{h}}_k\\
		&\times\mathrm{Tr}(\mathbf{R}_{ris}^{1/2}\mathbf{\Phi}\mathbf{D}_k^{1/2}\mathbf{\bar{h}}_k\bar{\mathbf{h}}_k^H\mathbf{D}_k^{1/2}\mathbf{\Phi}^H\mathbf{R}_{ris}^{1/2}) \\
		&=2(c_k\varepsilon_k)^2\delta(\bar{\mathbf{h}}_k^H\mathbf{D}_k^{1/2}\boldsymbol{\Phi}^H\bar{\mathbf{H}}_2^H\bar{\mathbf{H}}_2\boldsymbol{\Phi}\mathbf{D}_k^{1/2}\bar{\mathbf{h}}_k)\\
		&\times(\bar{\mathbf{h}}_k^H\mathbf{D}_k^{1/2}\boldsymbol{\Phi}^H\mathbf{R}_{ris}\boldsymbol{\Phi}\mathbf{D}_k^{1/2}\bar{\mathbf{h}}_k) \\
		&=2M(c_{k}\varepsilon_{k})^{2}\delta|f_{k}(\boldsymbol{\Phi})|^{2}f_{kk,2}(\boldsymbol{\Phi}).
	\end{aligned}
\end{align}

When $\psi=4$, we have
\begin{align}\label{A.17}
	\begin{aligned}
		&2\mathbb{E}\left\{\left(\mathbf{q}_k^1{}^H\mathbf{q}_k^4\right)\left(\mathbf{q}_k^1{}^H\mathbf{q}_k^4\right)^H\right\} \\
		&=2c_k^2\delta\varepsilon_k\mathbb{E}\left\{|\bar{\mathbf{h}}_k^H\mathbf{D}_k^{1/2}\boldsymbol{\Phi}^H\bar{\mathbf{H}}_2^H\tilde{\mathbf{H}}_2\mathbf{R}_{ris}^{1/2}\boldsymbol{\Phi}\mathbf{R}_{VR,k}^{1/2}\tilde{\mathbf{h}}_k|^2\right\} \\
		&=2c_k^2\delta\varepsilon_k\bar{\mathbf{h}}_k^H\mathbf{D}_k^{1/2}\boldsymbol{\Phi}^H\bar{\mathbf{H}}_2^H\\
		&\times\mathbb{E}\{\tilde{\mathbf{H}}_2\mathbf{R}_{ris}^{1/2}\boldsymbol{\Phi}\mathbf{R}_{VR,k}^{1/2}\tilde{\mathbf{h}}_k^H\mathbf{R}_{VR,k}^{1/2}\boldsymbol{\Phi}^H\mathbf{R}_{ris}^{1/2}\tilde{\mathbf{H}}_2^H\}\bar{\mathbf{H}}_2\boldsymbol{\Phi}\mathbf{D}_k^{1/2}\bar{\mathbf{h}}_k \\
		&=2c_k^2\delta\varepsilon_k\bar{\mathbf{h}}_k^H\mathbf{D}_k^{1/2}\mathbf{\Phi}^H\bar{\mathbf{H}}_2^H\bar{\mathbf{H}}_2\mathbf{\Phi}\mathbf{D}_k^{1/2}\mathbf{\bar{h}}_k\\
		&\times\mathbb{E}\{\mathrm{Tr}(\mathbf{R}_{ri\mathbf{s}}^{1/2}\mathbf{\Phi}\mathbf{R}_{VR,k}^{1/2}\tilde{\mathbf{h}}_k\tilde{\mathbf{h}}_k^H\mathbf{R}_{VR,k}^{1/2}\mathbf{\Phi}^H\mathbf{R}_{ri\mathbf{s}}^{1/2})\} \\
		&=2c_k^2\delta\varepsilon_k(\bar{\mathbf{h}}_k^H\mathbf{D}_k^{1/2}\mathbf{\Phi}^H\bar{\mathbf{H}}_2^H\bar{\mathbf{H}}_2\mathbf{\Phi}\mathbf{D}_k^{1/2}\bar{\mathbf{h}}_k)\\
		&\times\mathbb{E}\{\tilde{\mathbf{h}}_k^H\mathbf{R}_{VR,k}^{1/2}\boldsymbol{\Phi}^H\mathbf{R}_{ris}\boldsymbol{\Phi}\mathbf{R}_{VR,k}^{1/2}\tilde{\mathbf{h}}_k\} \\
		&=2Mc_{k}^{2}\delta\varepsilon_{k}|f_{k}(\mathbf\Phi)|^{2}f_{k,3,1}(\mathbf\Phi).
	\end{aligned}
\end{align}

Secondly, we consider the terms with $\omega=2$. When $\psi=3$, we have
\begin{align}\label{A.18}
	\begin{aligned}
		&2\mathbb{E}\left\{\left(\mathbf{q}_k^2{}^H\mathbf{q}_k^3\right)\left(\mathbf{q}_k^2{}^H\mathbf{q}_k^3\right)^H\right\} \\
		&=2c_{k}^{2}\delta\varepsilon_{k}\mathbb{E}\left\{\left|\tilde{\mathbf{h}}_{k}^{H}\mathbf{R}_{VR,k}^{1/2}\mathbf{\Phi}^{H}\bar{\mathbf{H}}_{2}^{H}\tilde{\mathbf{H}}_{2}\mathbf{R}_{ris}^{1/2}\mathbf{\Phi}\mathbf{D}_{k}^{1/2}\bar{\mathbf{h}}_{k}\right|^{2}\right\}\\
		&=2c_{k}^{2}\delta\varepsilon_{k}\mathbb{E}\left\{\tilde{\mathbf{h}}_{k}^{H}\mathbf{R}_{VR,k}^{1/2}\boldsymbol{\Phi}^{H}\bar{\mathbf{H}}_{2}^{H}\right.\\
		&\left. \times\mathbb{E}\{\tilde{\mathbf{H}}_{2}\mathbf{R}_{ris}^{1/2}\boldsymbol{\Phi}\mathbf{D}_{k}^{1/2}\bar{\mathbf{h}}_{k}^{H}\mathbf{D}_{k}^{1/2}\boldsymbol{\Phi}^{H}\mathbf{R}_{ris}^{1/2}\tilde{\mathbf{H}}_{2}^{H}\}\bar{\mathbf{H}}_{2}\boldsymbol{\Phi}\mathbf{R}_{VR,k}^{1/2}\tilde{\mathbf{h}}_{k}\right\} \\
		&=2c_k^2\delta\varepsilon_k\mathbb{E}\{\tilde{\mathbf{h}}_k^H\mathbf{R}_{VR,k}^{1/2}\mathbf{\Phi}^H\bar{\mathbf{H}}_2^H\bar{\mathbf{H}}_2\mathbf{\Phi}\mathbf{R}_{VR,k}^{1/2}\tilde{\mathbf{h}}_k\}\\
		&\times\text{Tr}(\mathbf{R}_{ris}^{1/2}\mathbf{\Phi}\mathbf{D}_k^{1/2}\mathbf{\bar{h}}_k^H\mathbf{\bar{h}}_k^{1/2}\mathbf{\Phi}^H\mathbf{R}_{ris}^{1/2}) \\
		&=2c_k^2\delta\varepsilon_k\mathrm{Tr}(\bar{\mathbf{H}}_2\boldsymbol{\Phi}\mathbf{R}_{VR,k}\boldsymbol{\Phi}^H\bar{\mathbf{H}}_2^H)\\
		&\times(\bar{\mathbf{h}}_k^H\mathbf{D}_k^{1/2}\boldsymbol{\Phi}^H\mathbf{R}_{ris}\boldsymbol{\Phi}\mathbf{D}_k^{1/2}\bar{\mathbf{h}}_k) \\
		&=2c_{k}^{2}\delta\varepsilon_{k}f_{k,1,1}(\mathbf\Phi)f_{kk,2}(\mathbf\Phi).
	\end{aligned}
\end{align}

When $\psi=4$, we have
\begin{align}\label{A.19}
	\begin{aligned}
		&2\mathbb{E}\left\{\left(\mathbf{q}_k^2{}^H\mathbf{q}_k^4\right)\left(\mathbf{q}_k^2{}^H\mathbf{q}_k^4\right)^H\right\} \\
		&=2c_k^2\delta\mathbb{E}\left\{|\tilde{\mathbf{h}}_k^H\mathbf{R}_{VR,k}^{1/2}\boldsymbol{\Phi}^H\bar{\mathbf{H}}_2^H\tilde{\mathbf{H}}_2\mathbf{R}_{ris}^{1/2}\boldsymbol{\Phi}\mathbf{R}_{VR,k}^{1/2}\tilde{\mathbf{h}}_k|^2\right\} \\
		&=2c_k^2\delta\mathbb{E}\left\{\tilde{\mathbf{h}}_k^H\mathbf{R}_{VR,k}^{1/2}\boldsymbol{\Phi}^H\bar{\mathbf{H}}_2^H\right.\\
		&\left. \times\mathbb{E}\{\tilde{\mathbf{H}}_2\mathbf{R}_{ris}^{1/2}\boldsymbol{\Phi}\mathbf{R}_{VR,k}^{1/2}\tilde{\mathbf{h}}_k\tilde{\mathbf{h}}_k^H\mathbf{R}_{VR,k}^{1/2}\boldsymbol{\Phi}^H\mathbf{R}_{ris}^{1/2}\tilde{\mathbf{H}}_2^H\}\mathbf{\bar{\mathbf{H}}}_2\boldsymbol{\Phi}\mathbf{R}_{VR,k}^{1/2}\tilde{\mathbf{h}}_k\right\} \\
		&=2c_k^2\delta\mathbb{E}\left\{\tilde{\mathbf{h}}_k^H\mathbf{R}_{VR,k}^{1/2}\mathbf{\Phi}^H\overline{\mathbf{H}}_2^H\overline{\mathbf{H}}_2\mathbf{\Phi}\mathbf{R}_{VR,k}^{1/2}\tilde{\mathbf{h}}_k\right.\\
		&\left.\times\mathrm{Tr}(\mathbf{R}_{ris}^{1/2}\mathbf{\Phi}\mathbf{R}_{VR,k}^{1/2}\tilde{\mathbf{h}}_k\tilde{\mathbf{h}}_k^H\mathbf{R}_{VR,k}^{1/2}\mathbf{\Phi}^H\mathbf{R}_{ris}^{1/2})\right\} \\
		&=2c_k^2\delta\mathbb{E}\left\{\tilde{\mathbf{h}}_k^H\mathbf{R}_{VR,k}^{1/2}\mathbf{\Phi}^H\bar{\mathbf{H}}_2^H\bar{\mathbf{H}}_2\mathbf{\Phi}\mathbf{R}_{VR,k}^{1/2}\tilde{\mathbf{h}}_k\right.\\
		&\left.\times\tilde{\mathbf{h}}_k^H\mathbf{R}_{VR,k}^{1/2}\mathbf{\Phi}^H\mathbf{R}_{ris}\mathbf{\Phi}\mathbf{R}_{VR,k}^{1/2}\tilde{\mathbf{h}}_k\right\} \\
		&=2c_k^2\delta\left\{\mathrm{Tr}\big(\mathbf{R}_{VR,k}^{1/2}\boldsymbol{\Phi}^H\bar{\mathbf{H}}_2^H\bar{\mathbf{H}}_2\boldsymbol{\Phi}\mathbf{R}_{VR,k}^{1/2}\big)\right.\\
		&\times\mathrm{Tr}\big(\mathbf{R}_{VR,k}^{1/2}\boldsymbol{\Phi}^H\mathbf{R}_{ris}\boldsymbol{\Phi}\mathbf{R}_{VR,k}^{1/2}\big)\\
		&\left.+\mathrm{Tr}\big(\mathbf{R}_{VR,k}^{1/2}\boldsymbol{\Phi}^H\bar{\mathbf{H}}_2^H\bar{\mathbf{H}}_2\boldsymbol{\Phi}\mathbf{R}_{VR,k}\boldsymbol{\Phi}^H\mathbf{R}_{ris}\boldsymbol{\Phi}\mathbf{R}_{VR,k}^{1/2}\big)\right\} \\
		&=2c_k^2\delta\left(f_{k,1,1}(\mathbf\Phi)f_{k,3,1}(\mathbf\Phi)+f_{kk,5}(\mathbf\Phi)\right).
	\end{aligned}
\end{align}

The value of $f_{kk,5}(\mathbf{\Phi})$ is given in (\ref{f_old}). Thirdly, we consider the terms with $\omega=3$. When $\psi=4$, we have
\begin{align}\label{A.20}
	\begin{aligned}
		&2\mathbb{E}\left\{\left(\mathbf{q}_k^3{}^H\mathbf{q}_k^4\right)\left(\mathbf{q}_k^3{}^H\mathbf{q}_k^4\right)^H\right\} \\
		&=2c_{k}^{2}\varepsilon_{k}\mathbb{E}\left\{\left|\bar{\mathbf{h}}_{k}^{H}\mathbf{D}_{k}^{1/2}\boldsymbol{\Phi}^{H}\mathbf{R}_{ris}^{1/2}\tilde{\mathbf{H}}_{2}^{H}\tilde{\mathbf{H}}_{ris}^{1/2}\boldsymbol{\Phi}\mathbf{R}_{VR,k}^{1/2}\tilde{\mathbf{h}}_{k}\right|^{2}\right\} \\
		&=2c_{k}^{2}\varepsilon_{k}\mathbb{E}\left\{\mathbf{\bar{h}}_{k}^{H}\mathbf{D}_{k}^{1/2}\mathbf{\Phi}^{H}\mathbf{R}_{ris}^{1/2}\mathbf{\tilde{H}}_{2}^{H}\mathbf{\tilde{H}}_{ris}^{1/2}\mathbf{\Phi}\right.\\
		&\left.\times\mathbf{R}_{VR,k}\mathbf{\Phi}^{H}\mathbf{R}_{ris}^{1/2}\mathbf{\tilde{H}}_{2}^{H}\mathbf{\tilde{H}}_{ris}\mathbf{R}_{ris}^{1/2}\mathbf{\Phi}\mathbf{D}_{k}^{1/2}\mathbf{\bar{h}}_{k}\right\} \\
		&=2c_{k}^{2}\varepsilon_{k}\bar{\mathbf{h}}_{k}^{H}\mathbf{D}_{k}^{1/2}\mathbf{\Phi}^{H}\mathbf{R}_{ris}^{1/2}\\
		&\times\left\{\begin{array}{c}{{M\mathrm{Tr}\big(\mathbf{R}_{ris}^{1/2}\mathbf{\Phi}\mathbf{R}_{VR,k}\mathbf{\Phi}^{H}\mathbf{R}_{ris}^{1/2}\big)\mathbf{I}_{N}}}\\{{+M^{2}\mathbf{R}_{ris}^{1/2}\mathbf{\Phi}\mathbf{R}_{VR,k}\mathbf{\Phi}^{H}\mathbf{R}_{ris}^{1/2}}}\\\end{array}\right\}\mathbf{R}_{ris}^{1/2}\mathbf{\Phi}\mathbf{D}_{k}^{1/2}\mathbf{\bar{h}}_{k} \\
		&=2c_{k}^{2}\varepsilon_{k}\begin{Bmatrix}M\mathrm{Tr}\big(\mathbf{R}_{ris}\mathbf{\Phi}\mathbf{R}_{VR,k}\mathbf{\Phi}^{H}\big)\bar{\mathbf{h}}_{k}^{H}\mathbf{D}_{k}^{1/2}\mathbf{\Phi}^{H}\mathbf{R}_{ris}\mathbf{\Phi}\mathbf{D}_{k}^{1/2}\bar{\mathbf{h}}_{k}\\+M^{2}\bar{\mathbf{h}}_{k}^{H}\mathbf{D}_{k}^{1/2}\mathbf{\Phi}^{H}\mathbf{R}_{ris}\mathbf{\Phi}\mathbf{R}_{VR,k}\mathbf{\Phi}^{H}\mathbf{R}_{ris}\mathbf{\Phi}\mathbf{D}_{k}^{1/2}\bar{\mathbf{h}}_{k}\end{Bmatrix} \\
		&=2c_{k}^{2}\varepsilon_{k}M\left(f_{kk,2}(\mathbf{\Phi})f_{k,3,1}(\mathbf{\Phi})+Mf_{kk,4}(\mathbf{\Phi})\right).
	\end{aligned}
\end{align}

The value of $f_{kk,4}(\mathbf{\Phi})$ is given in (\ref{f_old}). Next, we calculate the result of $2\sum_{\omega_1=1}^4\sum_{\psi_1=\omega_1+1}^4\mathbb{E}\left\{\left(\mathbf{q}_k^{\omega_1H}\mathbf{q}_k^{\omega_1}\right)\left(\mathbf{q}_k^{\psi_1H}\mathbf{q}_k^{\psi_1}\right)^H\right\}$ in (\ref{A.9}).

Firstly, we consider the terms with $\omega_1=1$. When $\psi_1=2$, we have
\begin{align}\label{A.22}
	\begin{aligned}
		&2\mathbb{E}\left\{\left(\mathbf{q}_k^1{}^H\mathbf{q}_k^1\right)\left(\mathbf{q}_k^2{}^H\mathbf{q}_k^2\right)^H\right\} \\
		=&2\mathbb{E}\left\{\left(\mathbf{q}_{k}^{1}{}^{H}\mathbf{q}_{k}^{1}\right)\right\}\mathbb{E}\left\{\left(\mathbf{q}_{k}^{2}{}^{H}\mathbf{q}_{k}^{2}\right)\right\} \\
		=&2M(c_{k}\delta)^{2}\varepsilon_{k}|f_{k}(\mathbf{\Phi})|^{2}f_{k,1,1}(\mathbf{\Phi}).
	\end{aligned}
\end{align}

When $\psi_1=3$, we have
\begin{align}\label{A.23}
	\begin{aligned}
		&2\mathbb{E}\left\{\left(\mathbf{q}_k^1{}^H\mathbf{q}_k^1\right)\left(\mathbf{q}_k^3{}^H\mathbf{q}_k^3\right)^H\right\} \\
		=&2\mathbb{E}\left\{\left(\mathbf{q}_{k}^{1}{}^{H}\mathbf{q}_{k}^{1}\right)\right\}\mathbb{E}\left\{\left(\mathbf{q}_{k}^{3}{}^{H}\mathbf{q}_{k}^{3}\right)\right\} \\
		=&2M^2(c_k\varepsilon_k)^2\delta|f_k(\mathbf{\Phi})|^2f_{kk,2}(\mathbf{\Phi}).
	\end{aligned}
\end{align}

When $\psi_1=4$, we have
\begin{align}\label{A.24}
	\begin{aligned}
		&2\mathbb{E}\left\{\left(\mathbf{q}_k^1{}^H\mathbf{q}_k^1\right)\left(\mathbf{q}_k^4{}^H\mathbf{q}_k^4\right)^H\right\} \\
		=&2\mathbb{E}\left\{\left(\mathbf{q}_{k}^{1}{}^{H}\mathbf{q}_{k}^{1}\right)\right\}\mathbb{E}\left\{\left(\mathbf{q}_{k}^{4}{}^{H}\mathbf{q}_{k}^{4}\right)\right\} \\
		=&2M^{2}c_{k}^{2}\delta\varepsilon_{k}|f_{k}(\mathbf{\Phi})|^{2}f_{k,3,1}(\mathbf{\Phi}).
	\end{aligned}
\end{align}

Secondly, we consider the terms with $\omega_1=2$. When $\psi_1=3$, we have
\begin{align}\label{A.25}
	\begin{aligned}
		&2\mathbb{E}\left\{\left(\mathbf{q}_k^2{}^H\mathbf{q}_k^2\right)\left(\mathbf{q}_k^3{}^H\mathbf{q}_k^3\right)^H\right\} \\
		=&2\mathbb{E}\left\{\left(\mathbf{q}_{k}^{2}{}^{H}\mathbf{q}_{k}^{2}\right)\right\}\mathbb{E}\left\{\left(\mathbf{q}_{k}^{3}{}^{H}\mathbf{q}_{k}^{3}\right)\right\} \\
		=&2Mc_{k}^{2}\delta\varepsilon_{k}f_{k,1,1}(\Phi)f_{kk,2}(\Phi).
	\end{aligned}
\end{align}

When $\psi_1=4$, we have
\begin{align}\label{A.26}
	\begin{aligned}
		&2\mathbb{E}\left\{\left(\mathbf{q}_k^2{}^H\mathbf{q}_k^2\right)\left(\mathbf{q}_k^4{}^H\mathbf{q}_k^4\right)^H\right\} \\
		=&2c_k^2\delta\mathbb{E}\left\{\tilde{\mathbf{h}}_k^H\mathbf{R}_{VR,k}^{1/2}\boldsymbol{\Phi}^H\bar{\mathbf{H}}_2^H\bar{\mathbf{H}}_2\boldsymbol{\Phi}\mathbf{R}_{VR,k}^{1/2}\tilde{\mathbf{h}}_k\right.\\
		&\left.\times\tilde{\mathbf{h}}_k^H\mathbf{R}_{VR,k}^{1/2}\boldsymbol{\Phi}^H\mathbf{R}_{ris}^{1/2}\mathbb{E}\{\tilde{\mathbf{H}}_2^H\tilde{\mathbf{H}}_2\}\mathbf{R}_{ris}^{1/2}\boldsymbol{\Phi}\mathbf{R}_{VR,k}^{1/2}\tilde{\mathbf{h}}_k\right\} \\
		=&2c_k^2\delta M\mathbb{E}\left\{\tilde{\mathbf{h}}_k^H\mathbf{R}_{VR,k}^{1/2}\boldsymbol{\Phi}^H\bar{\mathbf{H}}_2^H\bar{\mathbf{H}}_2\boldsymbol{\Phi}\mathbf{R}_{VR,k}^{1/2}\right.\\
		&\left.\times\tilde{\mathbf{h}}_k\tilde{\mathbf{h}}_k^H\mathbf{R}_{VR,k}^{1/2}\boldsymbol{\Phi}^H\mathbf{R}_{ris}\boldsymbol{\Phi}\mathbf{R}_{VR,k}^{1/2}\tilde{\mathbf{h}}_k\right\} \\
		=&2c_k^2\delta M\left\{\mathrm{Tr}\big(\mathbf{R}_{VR,k}^{1/2}\boldsymbol{\Phi}^H\bar{\mathbf{H}}_2^H\bar{\mathbf{H}}_2\boldsymbol{\Phi}\mathbf{R}_{VR,k}^{1/2}\big)\right.\\
		&\times\mathrm{Tr}\big(\mathbf{R}_{VR,k}^{1/2}\boldsymbol{\Phi}^H\mathbf{R}_{ris}\boldsymbol{\Phi}\mathbf{R}_{VR,k}^{1/2}\big)\\
		&\left.+\mathrm{Tr}\big(\mathbf{R}_{VR,k}^{1/2}\boldsymbol{\Phi}^H\bar{\mathbf{H}}_2^H\bar{\mathbf{H}}_2\boldsymbol{\Phi}\mathbf{R}_{VR,k}\boldsymbol{\Phi}^H\mathbf{R}_{ris}\boldsymbol{\Phi}\mathbf{R}_{VR,k}^{1/2}\big)\right\} \\
		=&2c_k^2\delta M\left(f_{k,1,1}(\boldsymbol{\Phi})f_{k,3,1}(\boldsymbol{\Phi})+f_{kk,5}(\boldsymbol{\Phi})\right).
	\end{aligned}
\end{align}

Thirdly, we consider the terms with $\omega_1=3$. When $\psi_1=4$, we have
\begin{align}\label{A.27}
	\begin{aligned}
		&2\mathbb{E}\left\{\left(\mathbf{q}_k^3{}^H\mathbf{q}_k^3\right)\left(\mathbf{q}_k^4{}^H\mathbf{q}_k^4\right)^H\right\} \\
		=&2c_{k}^{2}\varepsilon_{k}\mathbb{E}\{\bar{\mathbf{h}}_{k}^{\mathrm{H}}\mathbf{D}_{k}^{1/2}\boldsymbol{\Phi}^{H}\mathbf{R}_{ris}^{1/2}\tilde{\mathbf{H}}_{2}^{H}\tilde{\mathbf{H}}_{2}\mathbf{R}_{ris}^{1/2}\boldsymbol{\Phi}\mathbf{D}_{k}^{1/2}\bar{\mathbf{h}}_{k}^{H}\\
		&\times\tilde{\mathbf{h}}_{k}\mathbf{R}_{VR,k}^{1/2}\boldsymbol{\Phi}^{H}\mathbf{R}_{ris}^{1/2}\tilde{\mathbf{H}}_{2}^{H}\tilde{\mathbf{H}}_{2}\mathbf{R}_{ris}^{1/2}\boldsymbol{\Phi}\mathbf{R}_{VR,k}^{1/2}\tilde{\mathbf{h}}_{k}\}\\
		=&2c_{k}^{2}\varepsilon_{k}\mathbb{E}\left\{\bar{\mathbf{h}}_{k}^{H}\mathbf{D}_{k}^{1/2}\boldsymbol{\Phi}^{H}\mathbf{R}_{ris}^{1/2}\right.\\
		&\times\begin{Bmatrix}\boldsymbol{M}\mathrm{Tr}\big(\mathbf{R}_{ris}^{1/2}\boldsymbol{\Phi}\mathbf{D}_{k}^{1/2}\bar{\mathbf{h}}_{k}\tilde{\mathbf{h}}_{k}^{H}\mathbf{R}_{VR,k}^{1/2}\boldsymbol{\Phi}^{H}\mathbf{R}_{ris}^{1/2}\big)\mathbf{I}_{N}\\+\boldsymbol{M}^{2}\mathbf{R}_{ris}^{1/2}\boldsymbol{\Phi}\mathbf{D}_{k}^{1/2}\bar{\mathbf{h}}_{k}\tilde{\mathbf{h}}_{k}^{H}\mathbf{R}_{VR,k}^{1/2}\boldsymbol{R}_{ris}^{1/2}\boldsymbol{\Phi}^{H}\mathbf{R}_{ris}^{1/2}\end{Bmatrix}\\
		&\left.\times\mathbf{R}_{ris}^{1/2}\boldsymbol{\Phi}\mathbf{R}_{VR,k}^{1/2}\tilde{\mathbf{h}}_{k}\right\}\\
		=&2c_k^2\varepsilon_kM\bar{\mathbf{h}}_{k}^{H}\mathbf{D}_{k}^{1/2}\mathbf{\Phi}^{H}\mathbf{R}_{ris}\mathbf{\Phi}\mathbf{R}_{VR,k}^{1/2}\\
		&\times\mathbb{E}\{\tilde{\mathbf{h}}_{k}\tilde{\mathbf{h}}_{k}^{H}\}\mathbf{R}_{VR,k}^{1/2}\mathbf{\Phi}^{H}\mathbf{R}_{ris}\mathbf{\Phi}\mathbf{D}_{k}^{1/2}\mathbf{\bar{h}}_{k}\\
		&+2c_k^2\varepsilon_kM^2\bar{\mathbf{h}}_{k}^{H}\mathbf{D}_{k}^{1/2}\mathbf{\Phi}^{H}\mathbf{R}_{ris}\mathbf{\Phi}\mathbf{D}_{k}^{1/2}\mathbf{\bar{h}}_{k}\\
		&\times\mathbb{E}\{\tilde{\mathbf{h}}_{k}^{H}\mathbf{R}_{VR,k}^{1/2}\mathbf{\Phi}^{H}\mathbf{R}_{ris}\mathbf{\Phi}\mathbf{R}_{VR,k}^{1/2}\tilde{\mathbf{h}}_{k}\}\\
		=&2c_k^2\varepsilon_kM\left(Mf_{kk,2}(\mathbf\Phi)f_{k,3,1}(\Phi)+f_{kk,4}(\mathbf\Phi)\right).
	\end{aligned}
\end{align}

Finally, we focus on the remaining terms in (\ref{A.9}):
\begin{align}\label{A.29}
	\begin{aligned}
		&2\mathrm{Re}\left\{\mathbb{E}\left\{\left(\mathbf{q}_k^1{}^H\mathbf{q}_k^2\right)\left(\mathbf{q}_k^3{}^H\mathbf{q}_k^4\right)^H\right\}\right\}\\
		=&2\mathrm{Re}\left\{c_{k}^{2}\delta\varepsilon_{k}\bar{\mathbf{h}}_{k}^{H}\mathbf{D}_{k}^{1/2}\mathbf{\Phi}^{H}\bar{\mathbf{H}}_{2}^{H}\bar{\mathbf{H}}_{2}\mathbf{\Phi}\mathbf{R}_{VR,k}^{1/2}\mathbb{E}\{\tilde{\mathbf{h}}_{k}\tilde{\mathbf{h}}_{k}^{H}\}\right.\\
		&\left.\times\mathbf{R}_{VR,k}^{1/2}\mathbf{\Phi}^{H}\mathbf{R}_{ris}^{1/2}\mathbb{E}\{\tilde{\mathbf{H}}_{2}^{H}\tilde{\mathbf{H}}_{2}\}\mathbf{R}_{ris}^{1/2}\mathbf{\Phi}\mathbf{D}_{k}^{1/2}\bar{\mathbf{h}}_{k}\right\}\\
        =&2{M}c_k^2\delta\varepsilon_k\mathrm{Re}\left\{\mathbf{\bar{h}}_k^H\mathbf{D}_k^{1/2}\boldsymbol{\Phi}^H\mathbf{\bar{H}}_2^H\mathbf{\bar{H}}_2\boldsymbol{\Phi}\right.\\
        &\left.\times\mathbf{R}_{VR,k}\boldsymbol{\Phi}^H\mathbf{R}_{ris}\boldsymbol{\Phi}\mathbf{D}_k^{1/2}\mathbf{\bar{h}}_k\right\} \\
		=&2Mc_{k}^{2}\delta\varepsilon_{k}\mathrm{Re}\{f_{kk,6}(\mathbf{\Phi})\},
	\end{aligned}
\end{align}

\begin{align}\label{A.30}
	\begin{aligned}
		&2\mathrm{Re}\left\{\mathbb{E}\left\{\left(\mathbf{q}_k^1{}^H\mathbf{q}_k^3\right)\left(\mathbf{q}_k^2{}^H\mathbf{q}_k^4\right)^H\right\}\right\}\\
		=&2c_{k}^{2}\delta\varepsilon_{k}\mathrm{Re}\left\{\mathbb{E}\left\{\bar{\mathbf{h}}_{k}^{H}\mathbf{D}_{k}^{1/2}\mathbf{\Phi}^{H}\bar{\mathbf{H}}_{2}^{H}\tilde{\mathbf{H}}_{2}\mathbf{R}_{ris}^{1/2}\mathbf{\Phi}\mathbf{D}_{k}^{1/2}\bar{\mathbf{h}}_{k}\right.\right.\\
		&\left.\left. \times\tilde{\mathbf{h}}_{k}^{H}\mathbf{R}_{VR,k}^{1/2}\mathbf{\Phi}^{H}\mathbf{R}_{ris}^{1/2}\tilde{\mathbf{H}}_{2}^{H}\bar{\mathbf{H}}_{2}^{H}\mathbf{\Phi}\mathbf{R}_{VR,k}^{1/2}\tilde{\mathbf{h}}_{k} \right\}\right\} \\
		=&2c_k^2\delta\varepsilon_k\mathrm{Re}\left\{\overline{\mathbf{h}}_k^H\mathbf{D}_k^{1/2}\mathbf{\Phi}^H\bar{\mathbf{H}}_2^H\bar{\mathbf{H}}_2\mathbf{\Phi}\mathbf{R}_{VR,k}^{1/2}\right.\\
		&\left. \times\mathbb{E}\{\tilde{\mathbf{h}}_k\tilde{\mathbf{h}}_k^H\}\mathbf{R}_{VR,k}^{1/2}\mathbf{\Phi}^H\mathbf{R}_{ris}\boldsymbol{\Phi}\mathbf{D}_k^{1/2}\mathbf{\bar{\mathbf{h}}}_k\right\} \\
		=&2c_k^2\delta\varepsilon_k\mathrm{Re}\left\{\bar{\mathbf{h}}_k^H\mathbf{D}_k^{1/2}\boldsymbol{\Phi}^H\bar{\mathbf{H}}_2^H\bar{\mathbf{H}}_2\boldsymbol{\Phi}\right.\\
		&\left. \times\mathbf{R}_{VR,k}\boldsymbol{\Phi}^H\mathbf{R}_{ris}\boldsymbol{\Phi}\mathbf{D}_k^{1/2}\bar{\mathbf{h}}_k\right\} \\
		=&2c_{k}^{2}\delta\varepsilon_{k}\mathrm{Re}\{f_{kk,6}(\mathbf{\Phi})\},
	\end{aligned}
\end{align}

\begin{align}\label{A.31}
	\begin{aligned}
		&2\mathrm{Re}\left\{\mathbb{E}\left\{\left(\mathbf{q}_k^2{}^H\mathbf{q}_k^1\right)\left(\mathbf{q}_k^4{}^H\mathbf{q}_k^3\right)^H\right\}\right\}\\
		=&2c_{k}^{2}\delta\varepsilon_{k}\mathrm{Re}\left\{\mathrm{E}\left\{\tilde{\mathbf{h}}_{k}^{H}\mathbf{R}_{VR,k}^{1/2}\mathbf{\Phi}^{H}\bar{\mathbf{H}}_{2}^{H}\bar{\mathbf{H}}_{2}\mathbf{\Phi}\mathbf{D}_{k}^{1/2}\bar{\mathbf{h}}_{k}\right.\right.\\
		&\left.\left. \times\bar{\mathbf{h}}_{k}^{H}\mathbf{D}_{k}^{1/2}\mathbf{\Phi}^{H}\mathbf{R}_{ris}^{1/2}\mathbb{E}\{\tilde{\mathbf{H}}_{2}^{H}\tilde{\mathbf{H}}_{2}\}\mathbf{R}_{ris}^{1/2}\mathbf{\Phi}\mathbf{R}_{VR}^{1/2}\tilde{\mathbf{h}}_{k}\right\}\right\} \\
		=&2c_k^2\delta\varepsilon_kM\mathrm{Re}\left\{\mathrm{E}\left\{\tilde{\mathbf{h}}_k^H\mathbf{R}_{VR,k}^{1/2}\boldsymbol{\Phi}^H\bar{\mathbf{H}}_2^H\bar{\mathbf{H}}_2\boldsymbol{\Phi}\mathbf{D}_k^{1/2}\bar{\mathbf{h}}_k\right.\right.\\
		&\left.\left. \times\bar{\mathbf{h}}_k^H\mathbf{D}_k^{1/2}\boldsymbol{\Phi}^H\mathbf{R}_{ris}\boldsymbol{\Phi}\mathbf{R}_{VR,k}^{1/2}\tilde{\mathbf{h}}_k\right\}\right\} \\
		=&2c_k^2\delta\varepsilon_kM\mathrm{Re}\left\{\bar{\mathbf{h}}_k^H\mathbf{D}_k^{1/2}\boldsymbol{\Phi}^H\bar{\mathbf{H}}_2^H\bar{\mathbf{H}}_2\boldsymbol{\Phi}\right.\\
		&\left. \times\mathbf{R}_{VR,k}\boldsymbol{\Phi}^H\mathbf{R}_{ris}\boldsymbol{\Phi}\mathbf{D}_k^{1/2}\bar{\mathbf{h}}_k\right\} \\
		=&2Mc_{k}^{2}\delta\varepsilon_{k}\mathrm{Re}\{f_{kk,6}(\mathbf{\Phi})\},		
	\end{aligned}
\end{align}

\begin{align}\label{A.32}
	\begin{aligned}
		&2\mathrm{Re}\left\{\mathbb{E}\left\{\left(\mathbf{q}_k^2{}^H\mathbf{q}_k^4\right)\left(\mathbf{q}_k^1{}^H\mathbf{q}_k^3\right)^H\right\}\right\}\\
		=&2c_{k}^{2}\delta\varepsilon_{k}\mathrm{Re}\left\{\mathbb{E}\left\{\tilde{\mathbf{h}}_{k}^{H}\mathbf{R}_{VR,k}^{1/2}\boldsymbol{\Phi}^{H}\bar{\mathbf{H}}_{2}^{H}\tilde{\mathbf{H}}_{2}^{1/2}\mathbf{R}_{ris}^{1/2}\boldsymbol{\Phi}\mathbf{R}_{VR,k}^{1/2}\tilde{\mathbf{h}}_{k}\right.\right.\\
		&\left.\left. \times\bar{\mathbf{h}}_{k}^{H}\mathbf{D}_{k}^{1/2}\boldsymbol{\Phi}^{H}\mathbf{R}_{ris}^{1/2}\tilde{\mathbf{H}}_{2}^{H}\bar{\mathbf{H}}_{2}\boldsymbol{\Phi}\mathbf{D}_{k}^{1/2}\bar{\mathbf{h}}_{k}\right\}\right\} \\
		=&2c_k^2\delta\varepsilon_k\mathrm{Re}\left\{\mathbb{E}\left\{\mathrm{Tr}(\mathbf{R}_{ris}^{1/2}\boldsymbol{\Phi}\mathbf{R}_{VR,k}^{1/2}\tilde{\mathbf{h}}_k\bar{\mathbf{h}}_k^H\mathbf{D}_k^{1/2}\mathbf{\Phi}^H\mathbf{R}_{ris}^{1/2})\right.\right.\\
		&\left.\left. \times\tilde{\mathbf{h}}_k^H\mathbf{R}_{VR,k}^{1/2}\boldsymbol{\Phi}^H\bar{\mathbf{H}}_2^H\bar{\mathbf{H}}_2\boldsymbol{\Phi}\mathbf{D}_k^{1/2}\bar{\mathbf{h}}_k\right\}\right\} \\
		=&2c_{k}^{2}\delta\varepsilon_{k}\mathrm{Re}\left\{\bar{\mathbf{h}}_{k}^{H}\mathbf{D}_{k}^{1/2}\mathbf{\Phi}^{H}\mathbf{R}_{ris}\mathbf{\Phi}\mathbf{R}_{VR,k}^{1/2}\mathbb{E}\{\tilde{\mathbf{h}}_{k}\tilde{\mathbf{h}}_{k}^{H}\}\right.\\
		&\left. \times\mathbf{R}_{VR,k}^{1/2}\mathbf{\Phi}^{H}\bar{\mathbf{H}}_{2}^{H}\bar{\mathbf{H}}_{2}\mathbf{\Phi}\mathbf{D}_{k}^{1/2}\mathbf{\bar{\mathbf{h}}_{k}}\right\} \\
		=&2c_{k}^{2}\delta\varepsilon_{k}\mathrm{Re}\{f_{kk,6}(\mathbf{\Phi})\}.	
	\end{aligned}
\end{align}

The value of $f_{kk,6}(\mathbf{\Phi})$ is given in (\ref{f_old}). Therefore, by combining the above parts, the result of $E_{VR,k}^{\mathrm{signal}}(\boldsymbol{\Phi})$ can be obtained as shown in (\ref{signal}).

\subsection{The derivation of signal term $I_{VR,ki}(\boldsymbol{\Phi})$}

Due to $I_{VR,ki}(\boldsymbol{\Phi})=\mathbb{E}\{|\mathbf{q}_{k}^{H}\mathbf{q}_{i}|^{2}\}=\mathbb{E}\{\mathbf{q}_{k}^{H}\mathbf{q}_{i}\mathbf{q}_{i}^{H}\mathbf{q}_{k}\}$, we use (\ref{q_k_div}) to expand it:
\begin{align}\label{A.33}
	\begin{aligned}
		I_{VR,ki}(\Phi) =&\sum_{\omega_1,\psi_1}^4\sum_{\omega_2,\psi_2}^4\mathbb{E}\left\{\left(\mathbf{q}_k^{\omega_1H}\mathbf{q}_i^{\psi_1}\right)\left(\mathbf{q}_k^{\omega_2H}\mathbf{q}_i^{\psi_2}\right)^H\right\}  \\
		=&\sum_{\omega=1}^4\sum_{\psi=1}^4\mathbb{E}\left\{(\mathbf{q}_k^{\omega H}\mathbf{q}_i^{\psi})(\mathbf{q}_k^{\omega H}\mathbf{q}_i^{\psi})^H\right\} \\
		&+2\sum_{\omega_1=1}^4\sum_{\psi_1=\omega_1+1}^4\mathbb{E}\left\{\left(\mathbf{q}_k^{\omega_1H}\mathbf{q}_i^{\omega_1}\right)\left(\mathbf{q}_k^{\psi_1H}\mathbf{q}_i^{\psi_1}\right)^H\right\} \\
		&+2\mathrm{Re}\left\{\mathbb{E}\left\{\left(\mathbf{q}_k^1{}^H\mathbf{q}_i^2\right)\left(\mathbf{q}_k^3{}^H\mathbf{q}_i^4\right){}^H\right\}\right\}\\
		&+2\mathrm{Re}\left\{\mathbb{E}\left\{\left(\mathbf{q}_k^2{}^H\mathbf{q}_i^1\right)\left(\mathbf{q}_k^4{}^H\mathbf{q}_i^3\right){}^H\right\}\right\}.
	\end{aligned}
\end{align}

As the forms of $I_{VR,ki}(\boldsymbol{\Phi})$ and $E_{VR,k}^{\mathrm{signal}}(\boldsymbol{\Phi})$ are similar in (\ref{A.33}), we adopt the same processing method. Therefore, we divide $I_{VR,ki}(\boldsymbol{\Phi})$ into different parts and calculate them sequentially. In the following, we first calculate the result of $\sum_{\omega=1}^4\sum_{\psi=1}^4\mathbb{E}\left\{(\mathbf{q}_k^{\omega H}\mathbf{q}_i^{\psi})(\mathbf{q}_k^{\omega H}\mathbf{q}_i^{\psi})^H\right\}$. The formula can be further rewritten into the following form:
\begin{align}\label{A.34}
	\begin{aligned}
		&\sum_{\omega=1}^{4} \sum_{\psi=1}^4\mathbb{E}\left\{(\mathbf{q}_k^{\omega H}\mathbf{q}_i^\psi)(\mathbf{q}_k^{\omega H}\mathbf{q}_i^\psi)^H\right\}  \\
		=&\sum_{\omega=1}^4\mathbb{E}\left\{(\mathbf{q}_k^{\omega H}\mathbf{q}_i^\omega)(\mathbf{q}_k^{\omega H}\mathbf{q}_i^\omega)^H\right\} \\
		&+\sum_{\omega=1}^4\sum_{\psi\neq\omega}^4\mathbb{E}\left\{(\mathbf{q}_k^{\omega H}\mathbf{q}_i^{\psi})(\mathbf{q}_k^{\omega H}\mathbf{q}_i^{\psi})^{^H}\right\}.
	\end{aligned}
\end{align}

The specific calculation of the first part $\sum_{\omega=1}^4\mathbb{E}\left\{(\mathbf{q}_k^{\omega H}\mathbf{q}_i^\omega)(\mathbf{q}_k^{\omega H}\mathbf{q}_i^\omega)^H\right\}$ is as follows. When $\omega=1$, we have
\begin{align}\label{A.35}
	\begin{aligned}
		&\mathbb{E}\left\{\left(\mathbf{q}_k^1{}^H\mathbf{q}_i^1\right)\left(\mathbf{q}_k^1{}^H\mathbf{q}_i^1\right)^H\right\} \\
		=&c_kc_i\delta^2\varepsilon_k\varepsilon_i\mathbb{E}\left\{|\bar{\mathbf{h}}_k^H\mathbf{D}_k^{1/2}\mathbf{\Phi}^H\bar{\mathbf{H}}_2^H\bar{\mathbf{H}}_2\mathbf{\Phi}\mathbf{D}_i^{1/2}\bar{\mathbf{h}}_i|^2\right\} \\
		=&c_kc_i\delta^2\varepsilon_k\varepsilon_i\mathbb{E}\left\{\left|(\bar{\mathbf{h}}_k^H\mathbf{D}_k^{1/2}\mathbf{\Phi}^H\mathbf{a}_N)\mathbf{a}_M^H\mathbf{a}_M(\mathbf{a}_N^H\mathbf{\Phi}\mathbf{D}_i^{1/2}\mathbf{\bar{h}}_i)\right|^2\right\} \\
		=&c_kc_i\delta^2\varepsilon_k\varepsilon_iM^2\mathbb{E}\left\{(\bar{\mathbf{h}}_k^H\mathbf{D}_k^{1/2}\boldsymbol{\Phi}^H\mathbf{a}_N)(\mathbf{a}_N^H\mathbf{\Phi}\mathbf{D}_i^{1/2}\bar{\mathbf{h}}_i)\right.\\
		&\left.\times(\bar{\mathbf{h}}_i^H\mathbf{D}_i^{1/2}\boldsymbol{\Phi}^H\mathbf{a}_N)(\mathbf{a}_N^H\boldsymbol{\Phi}\mathbf{D}_k^{1/2}\bar{\mathbf{h}}_k)\right\} \\
		=&M^{2}c_{k}c_{i}\delta^{2}\varepsilon_{k}\varepsilon_{i}|f_{k}(\Phi)|^{2}|f_{i}(\Phi)|^{2}.
	\end{aligned}
\end{align}

When $\omega=2$, we have
\begin{align}\label{A.36}
	\begin{aligned}
		&\mathbb{E}\left\{\left(\mathbf{q}_k^2{}^H\mathbf{q}_i^2\right)\left(\mathbf{q}_k^2{}^H\mathbf{q}_i^2\right)^H\right\} \\
		=&c_{k}c_{i}\delta^{2}\mathbb{E}\left\{\left|\tilde{\mathbf{h}}_{k}^{H}\mathbf{R}_{VR,k}^{1/2}\mathbf{\Phi}^{H}\bar{\mathbf{H}}_{2}^{H}\bar{\mathbf{H}}_{2}\mathbf{\Phi}\mathbf{R}_{VR,i}^{1/2}\tilde{\mathbf{h}}_{i}\right|^{2}\right\} \\
		=&c_{k}c_{i}\delta^{2}\mathrm{E}\left\{\tilde{\mathbf{h}}_{k}^{H}\mathbf{R}_{VR,k}^{1/2}\mathbf{\Phi}^{H}\bar{\mathbf{H}}_{2}^{H}\bar{\mathbf{H}}_{2}\mathbf{\Phi}\mathbf{R}_{VR,i}^{1/2}\mathbb{E}\{\tilde{\mathbf{h}}_{i}\tilde{\mathbf{h}}_{i}^{H}\}\right.\\
		&\left.\times\mathbf{R}_{VR,i}^{1/2}\mathbf{\Phi}^{H}\bar{\mathbf{H}}_{2}^{H}\bar{\mathbf{H}}_{2}\mathbf{\Phi}\mathbf{R}_{VR,k}^{1/2}\tilde{\mathbf{h}}_{k}\right\}\\
		=&c_{k}c_{i}\delta^{2}\mathrm{Tr}\{\mathbf{R}_{VR,k}^{1/2}\boldsymbol{\Phi}^{H}\bar{\mathbf{H}}_{2}^{H}\bar{\mathbf{H}}_{2}\boldsymbol{\Phi}\mathbf{R}_{VR,i}\boldsymbol{\Phi}^{H}\bar{\mathbf{H}}_{2}^{H}\bar{\mathbf{H}}_{2}\boldsymbol{\Phi}\mathbf{R}_{VR,k}^{1/2}\} \\
		=&c_{k}c_{i}\delta^{2}f_{ki,1,2}(\mathbf\Phi).
	\end{aligned}
\end{align}

When $\omega=3$, we have
\begin{align}\label{A.37}
	\begin{aligned}
		&\mathbb{E}\left\{\left(\mathbf{q}_k^3{}^H\mathbf{q}_i^3\right)\left(\mathbf{q}_k^3{}^H\mathbf{q}_i^3\right)^H\right\} \\
		=&c_{k}c_{i}\varepsilon_{k}\varepsilon_{i}\mathbb{E}\left\{|\bar{\mathbf{h}}_{k}^{H}\mathbf{D}_{k}^{1/2}\mathbf{\Phi}^{H}\mathbf{R}_{ris}^{1/2}\tilde{\mathbf{H}}_{2}^{H}\tilde{\mathbf{H}}_{2}\mathbf{R}_{ris}^{1/2}\mathbf{\Phi}\mathbf{D}_{i}^{1/2}\bar{\mathbf{h}}_{i}|^{2}\right\} \\
		=&c_kc_i\varepsilon_k\varepsilon_i\bar{\mathbf{h}}_k^H\mathbf{D}_k^{1/2}\mathbf{\Phi}^H\mathbf{R}_{ris}^{1/2}\\
		&\times\begin{pmatrix}M\mathrm{Tr}\big(\mathbf{R}_{ris}^{1/2}\mathbf{\Phi}\mathbf{D}_i^{1/2}\bar{\mathbf{h}}_i\bar{\mathbf{h}}_i^H\mathbf{D}_i^{1/2}\mathbf{\Phi}^H\mathbf{R}_{ris}^{1/2}\big)\mathbf{I}_N\\+M^2\mathbf{R}_{ris}^{1/2}\mathbf{\Phi}\mathbf{D}_i^{1/2}\bar{\mathbf{h}}_i\bar{\mathbf{h}}_i^H\mathbf{D}_i^{1/2}\mathbf{\Phi}^H\mathbf{R}_{ris}^{1/2}\end{pmatrix}\\
		&\times\mathbf{R}_{ris}^{1/2}\mathbf{\Phi}\mathbf{D}_k^{1/2}\mathbf{\bar{h}}_k \\
		=&c_kc_i\varepsilon_k\varepsilon_iM\\
		&\times\begin{pmatrix}\bar{\mathbf{h}}_k^H\mathbf{D}_k^{1/2}\boldsymbol{\Phi}^H\mathbf{R}_{ris}\boldsymbol{\Phi}\mathbf{D}_k^{1/2}\bar{\mathbf{h}}_k\bar{\mathbf{h}}_i^H\mathbf{D}_i^{1/2}\boldsymbol{\Phi}^H\mathbf{R}_{ris}\boldsymbol{\Phi}\mathbf{D}_i^{1/2}\bar{\mathbf{h}}_i\\+M|\bar{\mathbf{h}}_k^H\mathbf{D}_k^{1/2}\boldsymbol{\Phi}^H\mathbf{R}_{ris}\boldsymbol{\Phi}\mathbf{D}_i^{1/2}\bar{\mathbf{h}}_i|^2\end{pmatrix} \\
		=&c_{k}c_{i}\varepsilon_{k}\varepsilon_{i}M\left(f_{kk,2}(\boldsymbol{\Phi})f_{ii,2}(\boldsymbol{\Phi})+M\big|f_{ki,2}(\boldsymbol{\Phi})\big|^{2}\right).
	\end{aligned}
\end{align}

This derivation uses the conclusion (\ref{lemmaA}) of Lemma \ref{lemma_A1}, where $\mathbf{A}=\mathbf{B}=\mathbf{I}_{N}$ and $\mathbf{W}=\mathbf{R}_{ris}^{1/2}\mathbf{\Phi}\mathbf{D}_{i}^{1/2}\mathbf{\bar{h}}_{i}\mathbf{\bar{h}}_{i}^{H}\mathbf{D}_{i}^{1/2}\mathbf{\Phi}^{H}\mathbf{R}_{ris}^{1/2}$. When $\omega=4$, we have
\begin{align}\label{A.38}
	\begin{aligned}
		&\mathbb{E}\left\{\left(\mathbf{q}_k^4{}^H\mathbf{q}_i^4\right)\left(\mathbf{q}_k^4{}^H\mathbf{q}_i^4\right)^H\right\} \\
		=&c_{k}c_{i}\mathbb{E}\left\{|\tilde{\mathbf{h}}_{k}^{H}\mathbf{R}_{VR,k}^{1/2}\boldsymbol{\Phi}^{H}\mathbf{R}_{ris}^{1/2}\widetilde{\mathbf{H}}_{2}^{H}\widetilde{\mathbf{H}}_{2}\mathbf{R}_{ris}^{1/2}\boldsymbol{\Phi}\mathbf{R}_{VR,i}^{1/2}\tilde{\mathbf{h}}_{i}|^{2}\right\} \\
		=&c_kc_i\mathbb{E}\left\{\tilde{\mathbf{h}}_k^H\mathbf{R}_{VR,k}^{1/2}\boldsymbol{\Phi}^H\mathbf{R}_{ris}^{1/2}\right.\\
		&\times\begin{pmatrix}M\mathrm{Tr}\big(\mathbf{R}_{ris}^{1/2}\boldsymbol{\Phi}\mathbf{R}_{VR,i}^{1/2}\mathbb{E}\{\tilde{\mathbf{h}}_i\tilde{\mathbf{h}}_i^H\}\mathbf{R}_{VR,i}^{1/2}\boldsymbol{\Phi}^H\mathbf{R}_{ris}^{1/2}\big)\mathbf{I}_N\\+M^2\mathbf{R}_{ris}^{1/2}\boldsymbol{\Phi}\mathbf{R}_{VR,i}^{1/2}\mathbb{E}\{\tilde{\mathbf{h}}_i\tilde{\mathbf{h}}_i^H\}\mathbf{R}_{VR,i}^{1/2}\boldsymbol{\Phi}^H\mathbf{R}_{ris}^{1/2}\end{pmatrix}\\
		&\left.\times\mathbf{R}_{ris}^{1/2}\boldsymbol{\Phi}\mathbf{R}_{VR,k}^{1/2}\tilde{\mathbf{h}}_k\right\} \\
		=&c_kc_iM\\
		&\times\mathbb{E}\begin{Bmatrix}\tilde{\mathbf{h}}_k^H\mathbf{R}_{VR,k}^{1/2}\boldsymbol{\Phi}^H\mathbf{R}_{ris}\boldsymbol{\Phi}\mathbf{R}_{VR,k}^{1/2}\tilde{\mathbf{h}}_k\mathrm{Tr}(\mathbf{R}_{ris}\boldsymbol{\Phi}\mathbf{R}_{VR,i}\boldsymbol{\Phi}^H)\\+M\tilde{\mathbf{h}}_k^H\mathbf{R}_{VR,k}^{1/2}\boldsymbol{\Phi}^H\mathbf{R}_{ris}\boldsymbol{\Phi}\mathbf{R}_{VR,i}\boldsymbol{\Phi}^H\mathbf{R}_{ris}\boldsymbol{\Phi}\mathbf{R}_{VR,k}^{1/2}\tilde{\mathbf{h}}_k\end{Bmatrix} \\
		=&c_kc_iM\left(f_{k,3,1}(\boldsymbol{\Phi})f_{i,3,1}(\boldsymbol{\Phi})+Mf_{ki,3,2}(\boldsymbol{\Phi})\right).
	\end{aligned}
\end{align}

Then we calculate the result of second part $\sum_{\omega=1}^4\sum_{\psi\neq\omega}^4\mathbb{E}\left\{(\mathbf{q}_k^{\omega H}\mathbf{q}_i^{\psi})(\mathbf{q}_k^{\omega H}\mathbf{q}_i^{\psi})^{^H}\right\}$. Firstly, we consider the terms with $\omega=1$. When $\psi=2$, we have
\begin{align}\label{A.39}
	\begin{aligned}
		&\mathbb{E}\left\{\left(\mathbf{q}_k^1{}^H\mathbf{q}_i^2\right)\left(\mathbf{q}_k^1{}^H\mathbf{q}_i^2\right)^H\right\} \\
		=&c_kc_i\delta^2\varepsilon_k\mathbb{E}\left\{|\bar{\mathbf{h}}_k^H\mathbf{D}_k^{1/2}\mathbf{\Phi}^H\bar{\mathbf{H}}_2^H\bar{\mathbf{H}}_2\mathbf{\Phi}\mathbf{R}_{VR,i}^{1/2}\tilde{\mathbf{h}}_i|^2\right\} \\
		=&c_kc_i\delta^2\varepsilon_k\bar{\mathbf{h}}_k^H\mathbf{D}_k^{1/2}\mathbf{\Phi}^H\bar{\mathbf{H}}_2^H\bar{\mathbf{H}}_2\mathbf{\Phi}\mathbf{R}_{VR,i}^{1/2}\mathbb{E}\{\tilde{\mathbf{h}}_i\tilde{\mathbf{h}}_i^H\}\\
		&\times\mathbf{R}_{VR,i}^{1/2}\mathbf{\Phi}^H\bar{\mathbf{H}}_2^H\bar{\mathbf{H}}_2\mathbf{\Phi}\mathbf{D}_k^{1/2}\bar{\mathbf{h}}_k \\
		=&c_kc_i\delta^2\varepsilon_k\bar{\mathbf{h}}_k^H\mathbf{D}_k^{1/2}\mathbf{\Phi}^H\bar{\mathbf{H}}_2^H\bar{\mathbf{H}}_2\mathbf{\Phi}\mathbf{R}_{VR,i}\mathbf{\Phi}^H\bar{\mathbf{H}}_2^H\bar{\mathbf{H}}_2\mathbf{\Phi}\mathbf{D}_k^{1/2}\mathbf{\bar{h}}_k \\
		=&Mc_{k}c_{i}\delta^{2}\varepsilon_{k}|f_{k}(\mathbf{\Phi})|^{2}f_{i,1,1}(\mathbf{\Phi}).
	\end{aligned}
\end{align}

When $\psi=3$, we have
\begin{align}\label{A.40}
	\begin{aligned}
		&\mathbb{E}\left\{\left(\mathbf{q}_k^1{}^H\mathbf{q}_i^3\right)\left(\mathbf{q}_k^1{}^H\mathbf{q}_i^3\right)^H\right\} \\
		=&c_{k}c_{i}\delta\varepsilon_{k}\varepsilon_{i}\mathbb{E}\left\{\left|\bar{\mathbf{h}}_{k}^{H}\mathbf{D}_{k}^{1/2}\mathbf{\Phi}^{H}\bar{\mathbf{H}}_{2}^{H}\tilde{\mathbf{H}}_{2}\mathbf{R}_{ris}^{1/2}\mathbf{\Phi}\mathbf{D}_{i}^{1/2}\bar{\mathbf{h}}_{i}\right|^{2}\right\} \\
		=&c_kc_i\delta\varepsilon_k\varepsilon_i\bar{\mathbf{h}}_k^H\mathbf{D}_k^{1/2}\boldsymbol{\Phi}^H\bar{\mathbf{H}}_2^H\\
		&\times\mathbb{E}\{\tilde{\mathbf{H}}_2\mathbf{R}_{ris}^{1/2}\boldsymbol{\Phi}\mathbf{D}_i^{1/2}\bar{\mathbf{h}}_i\bar{\mathbf{h}}_i^H\mathbf{D}_i^{1/2}\boldsymbol{\Phi}^H\mathbf{R}_{ris}^{1/2}\tilde{\mathbf{H}}_2^H\}\bar{\mathbf{H}}_2\boldsymbol{\Phi}\mathbf{D}_k^{1/2}\bar{\mathbf{h}}_k \\
		=&c_{k}c_{i}\delta\varepsilon_{k}\varepsilon_{i}\bar{\mathbf{h}}_{k}^{H}\mathbf{D}_{k}^{1/2}\boldsymbol{\Phi}^{H}\bar{\mathbf{H}}_{2}^{H}\bar{\mathbf{H}}_{2}\boldsymbol{\Phi}\mathbf{D}_{k}^{1/2}\bar{\mathbf{h}}_{k}\\
		&\times\mathrm{Tr}(\mathbf{R}_{ris}^{1/2}\boldsymbol{\Phi}\mathbf{D}_{i}^{1/2}\bar{\mathbf{h}}_{i}\bar{\mathbf{h}}_{i}^{H}\mathbf{D}_{i}^{1/2}\boldsymbol{\Phi}^{H}\mathbf{R}_{ris}^{1/2}) \\
		=&c_kc_i\delta\varepsilon_k\varepsilon_i(\bar{\mathbf{h}}_k^H\mathbf{D}_k^{1/2}\boldsymbol{\Phi}^H\bar{\mathbf{H}}_2^H\bar{\mathbf{H}}_2\boldsymbol{\Phi}\mathbf{D}_k^{1/2}\bar{\mathbf{h}}_k)\\
		&\times(\bar{\mathbf{h}}_i^H\mathbf{D}_i^{1/2}\boldsymbol{\Phi}^H\mathbf{R}_{ris}\boldsymbol{\Phi}\mathbf{D}_i^{1/2}\bar{\mathbf{h}}_i) \\
		=&Mc_{k}c_{i}\delta\varepsilon_{k}\varepsilon_{i}|f_{k}(\Phi)|^{2}f_{ii,2}(\Phi).
	\end{aligned}
\end{align}

When $\psi=4$, we have
\begin{align}\label{A.41}
	\begin{aligned}
		&\mathbb{E}\left\{\left(\mathbf{q}_k^1{}^H\mathbf{q}_i^4\right)\left(\mathbf{q}_k^1{}^H\mathbf{q}_i^4\right)^H\right\} \\
		=&c_{k}c_{i}\delta\varepsilon_{k}\mathbb{E}\left\{\left|\bar{\mathbf{h}}_{k}^{H}\mathbf{D}_{k}^{1/2}\boldsymbol{\Phi}^{H}\bar{\mathbf{H}}_{2}^{H}\tilde{\mathbf{H}}_{2}\mathbf{R}_{ris}^{1/2}\boldsymbol{\Phi}\mathbf{R}_{VR,i}^{1/2}\tilde{\mathbf{h}}_{i}\right|^{2}\right\} \\
		=&c_{k}c_{i}\delta\varepsilon_{k}\bar{\mathbf{h}}_{k}^{H}\mathbf{D}_{k}^{1/2}\mathbf{\Phi}^{H}\bar{\mathbf{H}}_{2}^{H}\mathbb{E}\left\{\tilde{\mathbf{H}}_{2}\mathbf{R}_{ris}^{1/2}\mathbf{\Phi}\mathbf{R}_{VR,i}^{1/2}\tilde{\mathbf{h}}_{i}\right.\\
		&\left.\times\tilde{\mathbf{h}}_{i}^{H}\mathbf{R}_{VR,i}^{1/2}\mathbf{\Phi}^{H}\mathbf{R}_{ris}^{1/2}\tilde{\mathbf{H}}_{2}^{H}\right\}\bar{\mathbf{H}}_{2}\mathbf{\Phi}\mathbf{D}_{k}^{1/2}\mathbf{\bar{\mathbf{h}}_{k}} \\
		=&c_{k}c_{i}\delta\varepsilon_{k}\bar{\mathbf{h}}_{k}^{H}\mathbf{D}_{k}^{1/2}\mathbf{\Phi}^{H}\bar{\mathbf{H}}_{2}^{H}\bar{\mathbf{H}}_{2}\mathbf{\Phi}\mathbf{D}_{k}^{1/2}\mathbf{\bar{h}}_{k}\\
		&\times\mathbb{E}\{\mathrm{Tr}(\mathbf{R}_{ris}^{1/2}\mathbf{\Phi}\mathbf{R}_{VR,i}^{1/2}\tilde{\mathbf{h}}_{i}\tilde{\mathbf{h}}_{i}^{H}\mathbf{R}_{VR,i}^{1/2}\mathbf{\Phi}^{H}\mathbf{R}_{ris}^{1/2}) \\
		=&c_kc_i\delta\varepsilon_k(\bar{\mathrm{h}}_k^H\mathbf{D}_k^{1/2}\boldsymbol{\Phi}^H\bar{\mathbf{H}}_2^H\bar{\mathbf{H}}_2\boldsymbol{\Phi}\mathbf{D}_k^{1/2}\bar{\mathrm{h}}_k)\\
		&\times\mathbb{E}\{\tilde{\mathbf{h}}_i^H\mathbf{R}_{VR,i}^{1/2}\boldsymbol{\Phi}^H\mathbf{R}_{ris}\boldsymbol{\Phi}\mathbf{R}_{VR,i}^{1/2}\tilde{\mathbf{h}}_i\} \\
		=&Mc_{k}c_{i}\delta\varepsilon_{k}|f_{k}(\mathbf\Phi)|^{2}f_{i,3,1}(\mathbf\Phi).
	\end{aligned}
\end{align}

Secondly, we consider the terms with $\omega=2$. When $\psi=1$, we have
\begin{align}\label{A.42}
	\begin{aligned}
		&\mathbb{E}\left\{\left(\mathbf{q}_k^2{}^H\mathbf{q}_i^1\right)\left(\mathbf{q}_k^2{}^H\mathbf{q}_i^1\right)^H\right\} \\
		=&\mathbb{E}\left\{\left(\mathbf{q}_i^1{}^H\mathbf{q}_k^2\right)\left(\mathbf{q}_i^1{}^H\mathbf{q}_k^2\right)^H\right\} \\
		=&Mc_{k}c_{i}\delta^{2}\varepsilon_{i}|f_{i}(\mathbf\Phi)|^{2}f_{k,1,1}(\mathbf\Phi).
	\end{aligned}
\end{align}

When $\psi=3$, we have
\begin{align}\label{A.43}
	\begin{aligned}
		&\mathbb{E}\left\{\left(\mathbf{q}_k^2{}^H\mathbf{q}_i^3\right)\left(\mathbf{q}_k^2{}^H\mathbf{q}_i^3\right)^H\right\} \\
		=&c_{k}c_{i}\delta\varepsilon_{i}\mathbb{E}\left\{|\tilde{\mathbf{h}}_{k}^{H}\mathbf{R}_{VR,k}^{1/2}\boldsymbol{\Phi}^{H}\bar{\mathbf{H}}_{2}^{H}\widetilde{\mathbf{H}}_{2}\mathbf{R}_{ris}^{1/2}\boldsymbol{\Phi}\mathbf{D}_{i}^{1/2}\bar{\mathbf{h}}_{i}|^{2}\right\} \\
		=&c_{k}c_{i}\delta\varepsilon_{i}\mathbb{E}\left\{\tilde{\mathbf{h}}_{k}^{H}\mathbf{R}_{VR,k}^{1/2}\boldsymbol{\Phi}^{H}\bar{\mathbf{H}}_{2}^{H}\right.\\
		&\left.\times\mathbb{E}\{\tilde{\mathbf{H}}_{2}\mathbf{R}_{ris}^{1/2}\boldsymbol{\Phi}\mathbf{D}_{i}^{1/2}\bar{\mathbf{h}}_{i}\bar{\mathbf{h}}_{i}^{H}\mathbf{D}_{i}^{1/2}\boldsymbol{\Phi}^{H}\mathbf{R}_{ris}^{1/2}\tilde{\mathbf{H}}_{2}^{H}\}\mathbf{\bar{\mathbf{H}}}_{2}\boldsymbol{\Phi}\mathbf{R}_{VR,k}^{1/2}\mathbf{\tilde{\mathbf{h}}}_k\right\} \\
		=&c_kc_i\delta\varepsilon_i\mathbb{E}\{\tilde{\mathbf{h}}_k^H\mathbf{R}_{VR,k}^{1/2}\mathbf{\Phi}^H\bar{\mathbf{H}}_2^H\bar{\mathbf{H}}_2\mathbf{\Phi}\mathbf{R}_{VR,k}^{1/2}\tilde{\mathbf{h}}_k\}\\
		&\times\mathrm{Tr}(\mathbf{R}_{ris}^{1/2}\mathbf{\Phi}\mathbf{D}_i^{1/2}\mathbf{\bar{h}}_i\bar{\mathbf{h}}_i^H\mathbf{D}_i^{1/2}\mathbf{\Phi}^H\mathbf{R}_{ris}^{1/2}) \\
		=&c_kc_i\delta\varepsilon_i\operatorname{Tr}(\bar{\mathbf{H}}_2\boldsymbol{\Phi}\mathbf{R}_{VR,k}\boldsymbol{\Phi}^H\bar{\mathbf{H}}_2^H)(\bar{\mathbf{h}}_i^H\mathbf{D}_i^{1/2}\boldsymbol{\Phi}^H\mathbf{R}_{ris}\boldsymbol{\Phi}\mathbf{D}_i^{1/2}\bar{\mathbf{h}}_i) \\
		=&c_{k}c_{i}\delta\varepsilon_{i}f_{k,1,1}(\mathbf\Phi)f_{ii,2}(\mathbf\Phi).
	\end{aligned}
\end{align}

When $\psi=4$, we have
\begin{align}\label{A.44}
	\begin{aligned}
		&\mathbb{E}\left\{\left(\mathbf{q}_k^2{}^H\mathbf{q}_i^4\right)\left(\mathbf{q}_k^2{}^H\mathbf{q}_i^4\right)^H\right\} \\
		=&c_{k}c_{i}\delta\mathbb{E}\left\{\left|\tilde{\mathbf{h}}_{k}^{H}\mathbf{R}_{VR,k}^{1/2}\boldsymbol{\Phi}^{H}\bar{\mathbf{H}}_{2}^{H}\tilde{\mathbf{H}}_{2}\mathbf{R}_{ris}^{1/2}\boldsymbol{\Phi}\mathbf{R}_{VR,i}^{1/2}\tilde{\mathbf{h}}_{i}\right|^{2}\right\} \\
		=&c_{k}c_{i}\delta\mathbb{E}\left\{\tilde{\mathbf{h}}_{k}^{H}\mathbf{R}_{VR,k}^{1/2}\boldsymbol{\Phi}^{H}\bar{\mathbf{H}}_{2}^{H}\mathbb{E}\left\{\tilde{\mathbf{H}}_{2}\mathbf{R}_{ris}^{1/2}\boldsymbol{\Phi}\mathbf{R}_{VR,i}^{1/2}\right.\right.\\
		&\left.\left.\times\mathbb{E}\{\tilde{\mathbf{h}}_{i}\tilde{\mathbf{h}}_{i}^{H}\}\mathbf{R}_{VR,i}^{1/2}\boldsymbol{\Phi}^{H}\mathbf{R}_{ris}^{1/2}\tilde{\mathbf{H}}_{2}^{H}\right\}\mathbf{\bar{\mathbf{H}}}_{2}\boldsymbol{\Phi}\mathbf{R}_{VR,k}^{1/2}\tilde{\mathbf{h}}_{k}\right\} \\
		=&c_{k}c_{i}\delta\mathbb{E}\left\{\tilde{\mathbf{h}}_{k}^{H}\mathbf{R}_{VR,k}^{1/2}\mathbf{\Phi}^{H}\bar{\mathbf{H}}_{2}^{H}\bar{\mathbf{H}}_{2}\mathbf{\Phi}\mathbf{R}_{VR,k}^{1/2}\tilde{\mathbf{h}}_{k}\right.\\
		&\left.\times\operatorname{Tr}(\mathbf{R}_{ris}^{1/2}\mathbf{\Phi}\mathbf{R}_{VR,k}\boldsymbol{\Phi}^{H}\mathbf{R}_{ris}^{1/2})\right\} \\
		=&c_kc_i\delta\text{Tr}(\mathbf{R}_{VR,k}^{1/2}\mathbf{\Phi}^H\bar{\mathbf{H}}_2^H\bar{\mathbf{H}}_2\mathbf{\Phi}\mathbf{R}_{VR,k}^{1/2})\\
		&\times\text{Tr}(\mathbf{R}_{ris}^{1/2}\mathbf{\Phi}\mathbf{R}_{VR,k}\mathbf{\Phi}^H\mathbf{R}_{ris}^{1/2}) \\
		=&c_{k}c_{i}\delta f_{k,1,1}(\mathbf{\Phi})f_{i,3,1}(\mathbf{\Phi}).
	\end{aligned}
\end{align}

Thirdly, we consider the terms with $\omega=3$. When $\psi=1$, we have
\begin{align}\label{A.45}
	\begin{aligned}
		&\mathbb{E}\left\{\left(\mathbf{q}_k^3{}^H\mathbf{q}_i^1\right)\left(\mathbf{q}_k^3{}^H\mathbf{q}_i^1\right)^H\right\} \\
		=&Mc_{k}c_{i}\delta\varepsilon_{k}\varepsilon_{i}|f_{i}(\mathbf\Phi)|^{2}f_{kk,2}(\mathbf\Phi).
	\end{aligned}
\end{align}

When $\psi=2$, we have
\begin{align}\label{A.46}
	\begin{aligned}
		&\mathbb{E}\left\{\left(\mathbf{q}_k^3{}^H\mathbf{q}_i^2\right)\left(\mathbf{q}_k^3{}^H\mathbf{q}_i^2\right)^H\right\} \\
		=&c_kc_i\delta\varepsilon_kf_{i,1,1}(\mathbf\Phi)f_{kk,2}(\mathbf\Phi).
	\end{aligned}
\end{align}

When $\psi=4$, we have
\begin{align}\label{A.47}
	\begin{aligned}
		&\mathbb{E}\left\{\left(\mathbf{q}_k^3{}^H\mathbf{q}_i^4\right)\left(\mathbf{q}_k^3{}^H\mathbf{q}_i^4\right)^H\right\} \\
		=&c_kc_i\varepsilon_k\mathbb{E}\left\{|\bar{\mathbf{h}}_k^H\mathbf{D}_k^{1/2}\mathbf{\Phi}^H\mathbf{R}_{ris}^{1/2}\tilde{\mathbf{H}}_2^H\tilde{\mathbf{H}}_2\mathbf{R}_{ris}^{1/2}\boldsymbol{\Phi}\mathbf{R}_{VR,i}^{1/2}\tilde{\mathbf{h}}_i|^2\right\} \\
		=&c_{k}c_{i}\varepsilon_{k}\mathbb{E}\left\{\bar{\mathbf{h}}_{k}^{H}\mathbf{D}_{k}^{1/2}\boldsymbol{\Phi}^{H}\mathbf{R}_{ris}^{1/2}\tilde{\mathbf{H}}_2^H\tilde{\mathbf{H}}_2\mathbf{R}_{ris}^{1/2}\boldsymbol{\Phi}\mathbf{R}_{VR,i}\right.\\
		&\left.\times\boldsymbol{\Phi}^{H}\mathbf{R}_{ris}^{1/2}\tilde{\mathbf{H}}_2^H\tilde{\mathbf{H}}_2\mathbf{R}_{ris}^{1/2}\boldsymbol{\Phi}\mathbf{D}_{k}^{1/2}\bar{\mathbf{h}}_{k}\right\} \\
		=&c_{k}c_{i}\varepsilon_{k}\bar{\mathbf{h}}_{k}^{H}\mathbf{D}_{k}^{1/2}\mathbf{\Phi}^{H}\mathbf{R}_{ris}^{1/2}\begin{Bmatrix}M\mathrm{Tr}\big(\mathbf{R}_{ris}^{1/2}\mathbf{\Phi}\mathbf{R}_{VR,i}\mathbf{\Phi}^{H}\mathbf{R}_{ris}^{1/2}\big)\mathbf{I}_{N}\\+M^{2}\mathbf{R}_{ris}^{1/2}\mathbf{\Phi}\mathbf{R}_{VR,i}\mathbf{\Phi}^{H}\mathbf{R}_{ris}^{1/2}\end{Bmatrix}\\
		&\times\mathbf{R}_{ris}^{1/2}\mathbf{\Phi}\mathbf{D}_{k}^{1/2}\mathbf{\bar{h}}_{k} \\
		=&c_kc_i\varepsilon_k\begin{Bmatrix}M\mathrm{Tr}\big(\mathbf{R}_{ris}\boldsymbol{\Phi}\mathbf{R}_{VR,i}\boldsymbol{\Phi}^H\big)\mathbf{\bar{h}}_k^H\mathbf{D}_k^{1/2}\boldsymbol{\Phi}^H\mathbf{R}_{ris}\boldsymbol{\Phi}\mathbf{D}_k^{1/2}\mathbf{\bar{h}}_k\\+\boldsymbol{M}^2\mathbf{\bar{h}}_k^H\mathbf{D}_k^{1/2}\boldsymbol{\Phi}^H\mathbf{R}_{ris}\boldsymbol{\Phi}\mathbf{R}_{VR,i}\boldsymbol{\Phi}^H\mathbf{R}_{ris}\boldsymbol{\Phi}\mathbf{D}_k^{1/2}\mathbf{\bar{h}}_k\end{Bmatrix} \\
		=&c_{k}c_{i}\varepsilon_{k}M\left(f_{kk,2}(\boldsymbol{\Phi})f_{i,3,1}(\boldsymbol{\Phi})+Mf_{ki,4}(\boldsymbol{\Phi})\right).
	\end{aligned}
\end{align}

Fourthly, we consider the terms with $\omega=4$. When $\psi=1$, we have
\begin{align}\label{A.48}
	\begin{aligned}
		&\mathbb{E}\left\{\left(\mathbf{q}_k^4{}^H\mathbf{q}_i^1\right)\left(\mathbf{q}_k^4{}^H\mathbf{q}_i^1\right)^H\right\} \\
		=&Mc_{k}c_{i}\delta\varepsilon_{i}|f_{i}(\mathbf\Phi)|^{2}f_{k,3,1}(\mathbf\Phi).
	\end{aligned}
\end{align}

When $\psi=2$, we have
\begin{align}\label{A.49}
	\begin{aligned}
		&\mathbb{E}\left\{\left(\mathbf{q}_k^4{}^H\mathbf{q}_i^2\right)\left(\mathbf{q}_k^4{}^H\mathbf{q}_i^2\right)^H\right\} \\
		=&c_kc_i\delta f_{i,1,1}(\mathbf\Phi)f_{k,3,1}(\mathbf\Phi).
	\end{aligned}
\end{align}

When $\psi=3$, we have
\begin{align}\label{A.50}
	\begin{aligned}
		&\mathbb{E}\left\{\left(\mathbf{q}_k^4{}^H\mathbf{q}_i^3\right)\left(\mathbf{q}_k^4{}^H\mathbf{q}_i^3\right)^H\right\} \\
		=&c_{k}c_{i}\varepsilon_{i}M\left(f_{ii,2}(\mathbf\Phi)f_{k,3,1}(\mathbf\Phi)+Mf_{ik,4}(\mathbf\Phi)\right).
	\end{aligned}
\end{align}

Next, we calculate the result of $2\sum_{\omega_1=1}^4\sum_{\psi_1=\omega_1+1}^4\mathbb{E}\left\{\left(\mathbf{q}_k^{\omega_1H}\mathbf{q}_i^{\omega_1}\right)\left(\mathbf{q}_k^{\psi_1H}\mathbf{q}_i^{\psi_1}\right)^H\right\}$ in (\ref{A.33}). Due to $\tilde{\mathbf{h}}_i$, $\tilde{\mathbf{h}}_k$, $\tilde{\mathbf{H}}_2$ are independent of each other, the expression can be further simplified as follows after removing the terms with zero expectation:
\begin{align}\label{A.51}
	\begin{aligned}
		&2\sum_{\omega_1=1}^4\sum_{\psi_1=\omega_1+1}^4\mathbb{E}\left\{\left(\mathbf{q}_k^{\omega_1H}\mathbf{q}_i^{\omega_1}\right)\left(\mathbf{q}_k^{\psi_1H}\mathbf{q}_i^{\psi_1}\right)^H\right\}\\
		=&2\mathrm{Re}\left\{\mathbb{E}\left\{\left(\mathbf{q}_{k}^{1}{}^{H}\mathbf{q}_{i}^{1}\right)\left(\mathbf{q}_{k}^{3}{}^{H}\mathbf{q}_{i}^{3}\right)^{H}\right\}\right\}\\
		&+2\mathrm{Re}\left\{\mathbb{E}\left\{\left(\mathbf{q}_{k}^{2}{}^{H}\mathbf{q}_{i}^{2}\right)\left(\mathbf{q}_{k}^{4}{}^{H}\mathbf{q}_{i}^{4}\right)^{H}\right\}\right\}.
	\end{aligned}
\end{align}

Therefore, we calculate the two items separately.
\begin{align}\label{A.52}
	\begin{aligned}
		&2\mathrm{Re}\left\{\mathbb{E}\left\{\left(\mathbf{q}_{k}^{1}{}^{H}\mathbf{q}_{i}^{1}\right)\left(\mathbf{q}_{k}^{3}{}^{H}\mathbf{q}_{i}^{3}\right)^{H}\right\}\right\}\\
		=&2c_{k}c_{i}\delta\varepsilon_{k}\varepsilon_{i}\mathrm{Re}\left\{\bar{\mathbf{h}}_{k}^{H}\mathbf{D}_{k}^{1/2}\mathbf{\Phi}^{H}\bar{\mathbf{H}}_{2}^{H}\bar{\mathbf{H}}_{2}\mathbf{\Phi}\mathbf{D}_{i}^{1/2}\bar{\mathbf{h}}_{i}\right.\\
		&\left.\times\bar{\mathbf{h}}_{i}^{H}\mathbf{D}_{i}^{1/2}\mathbf{\Phi}^{H}\mathbf{R}_{ris}^{1/2}\mathbb{E}\{\tilde{\mathbf{H}}_{2}^{H}\tilde{\mathbf{H}}_{2}\}\mathbf{R}_{ris}^{1/2}\mathbf{\Phi}\mathbf{D}_{k}^{1/2}\bar{\mathbf{h}}_{k}\right\}\\
		=&2Mc_{k}c_{i}\delta\varepsilon_{k}\varepsilon_{i}\mathrm{Re}\left\{\bar{\mathbf{h}}_{k}^{H}\mathbf{D}_{k}^{1/2}\mathbf{\Phi}^{H}\bar{\mathbf{H}}_{2}^{H}\bar{\mathbf{H}}_{2}\mathbf{\Phi}\mathbf{D}_{i}^{1/2}\bar{\mathbf{h}}_{i}\right.\\
		&\left.\times\bar{\mathbf{h}}_{i}^{H}\mathbf{D}_{i}^{1/2}\mathbf{\Phi}^{H}\mathbf{R}_{ris}\mathbf{\Phi}\mathbf{D}_{k}^{1/2}\bar{\mathbf{h}}_{k}\right\}\\
		=&2M^2c_kc_i\delta\varepsilon_k\varepsilon_i\mathrm{Re}\left\{\bar{\mathbf{h}}_k^H\mathbf{D}_k^{1/2}\mathbf{\Phi}^H\mathbf{a}_N\mathbf{a}_N^H\mathbf{\Phi}\mathbf{D}_i^{1/2}\mathbf{\bar{h}}_i\right.\\
		&\left.\times\bar{\mathbf{h}}_i^H\mathbf{D}_i^{1/2}\mathbf{\Phi}^H\mathbf{R}_{ris}\mathbf{\Phi}\mathbf{D}_k^{1/2}\mathbf{\bar{h}}_k\right\}\\
		=&2M^2c_kc_i\delta\varepsilon_k\varepsilon_i\mathrm{Re}\left\{f_{ki,7}(\boldsymbol{\Phi})f_{ik,2}(\boldsymbol{\Phi})\right\}.
	\end{aligned}
\end{align}
\begin{align}\label{A.53}
	\begin{aligned}
		&2\mathrm{Re}\left\{\mathbb{E}\left\{\left(\mathbf{q}_{k}^{2}{}^{H}\mathbf{q}_{i}^{2}\right)\left(\mathbf{q}_{k}^{4}{}^{H}\mathbf{q}_{i}^{4}\right)^{H}\right\}\right\}\\
		=&2c_{k}c_{i}\delta\varepsilon_{k}\varepsilon_{i}\mathrm{Re}\left\{\mathbb{E}\left\{\tilde{\mathbf{h}}_{k}^{H}\mathbf{D}_{k}^{1/2}\mathbf{\Phi}^{H}\bar{\mathbf{H}}_{2}^{H}\bar{\mathbf{H}}_{2}\mathbf{\Phi}\mathbf{D}_{i}^{1/2}\tilde{\mathbf{h}}_{i}\right.\right.\\
		&\left.\left.\times\tilde{\mathbf{h}}_{i}^{H}\mathbf{D}_{i}^{1/2}\mathbf{\Phi}^{H}\mathbf{R}_{ris}^{1/2}\mathbb{E}\{\tilde{\mathbf{H}}_{2}^{H}\tilde{\mathbf{H}}_{2}\}\mathbf{R}_{ris}^{1/2}\mathbf{\Phi}\mathbf{D}_{k}^{1/2}\tilde{\mathbf{h}}_{k}\right\}\right\}\\
		=&2Mc_{k}c_{i}\delta\mathrm{Re}\left\{\mathbb{E}\left\{\tilde{\mathbf{h}}_{k}^{H}\mathbf{R}_{VR,k}^{1/2}\boldsymbol{\Phi}^{H}\bar{\mathbf{H}}_{2}^{H}\bar{\mathbf{H}}_{2}\boldsymbol{\Phi}\mathbf{R}_{VR,i}\right.\right.\\
		&\left.\left.\times\boldsymbol{\Phi}^{H}\mathbf{R}_{ris}\boldsymbol{\Phi}\mathbf{R}_{VR,k}^{1/2}\tilde{\mathbf{h}}_{k}\right\}\right\} \\
		=&2Mc_kc_i\delta\mathrm{Re}\left\{\mathrm{Tr}\left\{\mathbf{R}_{VR,k}^{1/2}\boldsymbol{\Phi}^H\bar{\mathbf{H}}_2^H\bar{\mathbf{H}}_2\boldsymbol{\Phi}\mathbf{R}_{VR,i}\right.\right.\\
		&\left.\left.\times\boldsymbol{\Phi}^H\mathbf{R}_{ris}\boldsymbol{\Phi}\mathbf{R}_{VR,k}^{1/2}\right\}\right\} \\
		=&2c_{k}c_{i}\delta\mathrm{Re}\{f_{ki,5}(\boldsymbol{\Phi})\}.
	\end{aligned}
\end{align}

Finally, we focus on the remaining terms in (\ref{A.33}). The derivation process of the first item is as follows:
\begin{align}\label{A.54}
	\begin{aligned}
		&2\mathrm{Re}\left\{\mathbb{E}\left\{\left(\mathbf{q}_{k}^{1}{}^{H}\mathbf{q}_{i}^{2}\right)\left(\mathbf{q}_{k}^{3}{}^{H}\mathbf{q}_{i}^{4}\right)^{H}\right\}\right\}\\
		=&2\mathrm{Re}\left\{c_{k}c_{i}\delta\varepsilon_{k}\bar{\mathbf{h}}_{k}^{H}\mathbf{D}_{k}^{1/2}\mathbf{\Phi}^{H}\bar{\mathbf{H}}_{2}^{H}\bar{\mathbf{H}}_{2}\mathbf{\Phi}\mathbf{R}_{VR,i}^{1/2}\mathbb{E}\{\tilde{\mathbf{h}}_{i}\tilde{\mathbf{h}}_{i}^{H}\}\right.\\
		&\left.\times\mathbf{R}_{VR,i}^{1/2}\mathbf{\Phi}^{H}\mathbf{R}_{ris}^{1/2}\mathbb{E}\{\tilde{\mathbf{H}}_{2}^{H}\tilde{\mathbf{H}}_{2}\}\mathbf{R}_{ris}^{1/2}\mathbf{\Phi}\mathbf{D}_{k}^{1/2}\mathbf{\bar{\mathbf{h}}_{k}}\right\}\\
		=&2Mc_{k}c_{i}\delta\varepsilon_{k}\mathrm{Re}\{\bar{\mathbf{h}}_{k}^{H}\mathbf{D}_{k}^{1/2}\mathbf{\Phi}^{H}\bar{\mathbf{H}}_{2}^{H}\bar{\mathbf{H}}_{2}\mathbf{\Phi}\mathbf{R}_{VR,i}\mathbf{\Phi}^{H}\mathbf{R}_{ris}\mathbf{\Phi}\mathbf{D}_{k}^{1/2}\bar{\mathbf{h}}_{k}\} \\
		=&2Mc_{k}c_{i}\delta\varepsilon_{k}\mathrm{Re}\{f_{ki,6}(\boldsymbol{\Phi})\}.
	\end{aligned}
\end{align}

The derivation process of the second item is as follows:
\begin{align}\label{A.55}
	\begin{aligned}
		&2\mathrm{Re}\left\{\mathbb{E}\left\{\left(\mathbf{q}_{k}^{2}{}^{H}\mathbf{q}_{i}^{1}\right)\left(\mathbf{q}_{k}^{4}{}^{H}\mathbf{q}_{i}^{3}\right)^{H}\right\}\right\}\\
		=&2\mathrm{Re}\left\{\mathbb{E}\left\{c_{k}c_{i}\delta\varepsilon_{i}\tilde{\mathbf{h}}_{k}^{H}\mathbf{R}_{VR,k}^{1/2}\boldsymbol{\Phi}^{H}\bar{\mathbf{H}}_{2}^{H}\bar{\mathbf{H}}_{2}\boldsymbol{\Phi}\mathbf{D}_{i}^{1/2}\bar{\mathbf{h}}_{i}\right.\right.\\
		&\left.\left.\times\bar{\mathbf{h}}_{i}^{H}\mathbf{D}_{i}^{1/2}\boldsymbol{\Phi}^{H}\mathbf{R}_{ris}^{1/2}\mathbb{E}\{\tilde{\mathbf{H}}_{2}^{H}\tilde{\mathbf{H}}_{2}\}\mathbf{R}_{ris}^{1/2}\boldsymbol{\Phi}\mathbf{R}_{VR,k}^{1/2}\tilde{\mathbf{h}}_{k}\right\}\right\} \\
		=&2c_kc_i\delta\varepsilon_i\boldsymbol{M}\mathrm{Re}\left\{\mathbb{E}\left\{\tilde{\mathbf{h}}_k^H\mathbf{R}_{VR,k}^{1/2}\boldsymbol{\Phi}^H\bar{\mathbf{H}}_2^H\bar{\mathbf{H}}_2\boldsymbol{\Phi}\mathbf{D}_i^{1/2}\bar{\mathbf{h}}_i\right.\right.\\
		&\left.\left.\times\bar{\mathbf{h}}_i^H\mathbf{D}_i^{1/2}\boldsymbol{\Phi}^H\mathbf{R}_{ris}\boldsymbol{\Phi}\mathbf{R}_{VR,k}^{1/2}\tilde{\mathbf{h}}_k\right\}\right\} \\
		=&2c_kc_i\delta\varepsilon_iM\mathrm{Re}\{\bar{\mathbf{h}}_i^H\mathbf{D}_i^{1/2}\boldsymbol{\Phi}^H\mathbf{R}_{ris}\boldsymbol{\Phi}\mathbf{R}_{VR,k}\boldsymbol{\Phi}^H\bar{\mathbf{H}}_2^H\bar{\mathbf{H}}_2\boldsymbol{\Phi}\mathbf{D}_k^{1/2}\bar{\mathbf{h}}_k\} \\
		=&2Mc_{k}c_{i}\delta\varepsilon_{i}\mathrm{Re}\{f_{ik,6}(\mathbf{\Phi})\}.
	\end{aligned}
\end{align}

Therefore, by combining the above parts, the result of $I_{VR,ki}(\boldsymbol{\Phi})$ can be obtained as shown in (\ref{I}). Based on the value of $i$ and the superposition of various terms, the final result is $\sum_{i=1,i\neq k}^KI_{VR,ki}(\mathbf\Phi)$. Then, the uplink approximate achievable rate of user $k$ can be obtained by substituting the results of $E_{VR,k}^{\mathrm{noise}}(\boldsymbol{\Phi})$, $E_{VR,k}^{\mathrm{signal}}(\boldsymbol{\Phi})$ and $\sum_{i=1,i\neq k}^KI_{VR,ki}(\mathbf\Phi)$ into the original formula (\ref{rate}).

\section{}\label{appendixB}

Due to the derivation approach of the objective function $\frac{\partial f_{VR}(\boldsymbol{\theta})}{\partial\boldsymbol{\theta}}$ following (\ref{f_VR_gradient}) and (\ref{SINR_VR_gradient}), the final result can be obtained by calculating the gradients of $E_{VR,k}^{\mathrm{noise}}(\boldsymbol{\Phi})$, $E_{VR,k}^{\mathrm{signal}}(\boldsymbol{\Phi})$ and $\sum_{i=1,i\neq k}^KI_{VR,ki}(\mathbf\Phi)$. Therefore, using the conclusion of \cite[Lemma 1]{27}, we process each auxiliary function and calculate its gradient in this section.

For $f_k^{\prime}(\boldsymbol{\theta})$, we have
\begin{align}\label{C.1}
	\begin{aligned}
		f_k^{\prime}(\boldsymbol{\theta}) =&\frac{\partial\left(\mathbf{a}_N^H\mathbf{\Phi}\mathbf{D}_k^{1/2}\mathbf{\bar{h}}_k\mathbf{\bar{h}}_k^H\mathbf{D}_k^{1/2}\mathbf{\Phi}^H\mathbf{a}_N\right)}{\partial\boldsymbol{\theta}}  \\
		=&\frac{\partial\left(\mathrm{Tr}(\mathbf{a}_N^H\mathbf{\Phi}\mathbf{D}_k^{1/2}\bar{\mathbf{h}}_k\bar{\mathbf{h}}_k^H\mathbf{D}_k^{1/2}\mathbf{\Phi}^H\mathbf{a}_N)\right)}{\partial\boldsymbol{\theta}} \\
		=&2\operatorname{Im}\left\{\boldsymbol{\Phi}^H\left(\mathbf{a}_N\mathbf{a}_N^H\odot\left(\mathbf{D}_k^{1/2}\bar{\mathbf{h}}_k\bar{\mathbf{h}}_k^H\mathbf{D}_k^{1/2}\right)^T\right)\boldsymbol{c}\right\} \\
		=&\boldsymbol{f}_{a}(\mathbf{a}_{N}\mathbf{a}_{N}^{H},\mathbf{D}_{k}^{1/2}\mathbf{\bar{h}}_{k}\mathbf{\bar{h}}_{k}^{H}\mathbf{D}_{k}^{1/2}).
	\end{aligned}
\end{align}

For $f_{k,1,1}^{\prime}(\boldsymbol{\theta})$, we have
\begin{align}\label{C.2}
	\begin{aligned}
		f_{k,1,1}^{\prime}(\boldsymbol{\theta}) =&\frac{\partial\left(\mathrm{Tr}(\bar{\mathbf{H}}_2\boldsymbol{\Phi}\mathbf{R}_{VR,k}\boldsymbol{\Phi}^H\bar{\mathbf{H}}_2^H)\right)}{\partial\boldsymbol{\theta}}  \\
		=&2\operatorname{Im}\left\{\boldsymbol{\Phi}^H\left(\bar{\mathbf{H}}_2^H\bar{\mathbf{H}}_2\odot\left(\mathbf{R}_{VR,k}\right)^T\right)\boldsymbol{c}\right\} \\
		=&\boldsymbol{f}_{a}(\bar{\mathbf{H}}_{2}^{H}\bar{\mathbf{H}}_{2},\mathbf{R}_{VR,k}).
	\end{aligned}
\end{align}

For $f_{ki,1,2}^{\prime}(\boldsymbol{\theta})$, we have
\begin{align}\label{C.3}
	\begin{aligned}
		&f_{ki,1,2}^{\prime}(\boldsymbol{\theta})\\
		=&\frac{\partial\left(\mathrm{Tr}(\bar{\mathbf{H}}_2\boldsymbol{\Phi}\mathbf{R}_{VR,k}\boldsymbol{\Phi}^H\bar{\mathbf{H}}_2^H\bar{\mathbf{H}}_2\boldsymbol{\Phi}\mathbf{R}_{VR,i}\boldsymbol{\Phi}^H\bar{\mathbf{H}}_2^H)\right)}{\partial\boldsymbol{\theta}}  \\
		=&\frac{\partial\left(\mathrm{Tr}(\bar{\mathbf{H}}_2(\boldsymbol{\Phi}\mathbf{R}_{VR,k}\boldsymbol{\Phi}^H)\bar{\mathbf{H}}_2^H\bar{\mathbf{H}}_2(\boldsymbol{\Phi}\mathbf{R}_{VR,i}\boldsymbol{\Phi}^H)\bar{\mathbf{H}}_2^H)\right)}{\partial\boldsymbol{\theta}} \\
		=&2\operatorname{Im}\left\{\boldsymbol{\Phi}^H\left(\bar{\mathbf{H}}_2^H\bar{\mathbf{H}}_2\boldsymbol{\Phi}\mathbf{R}_{VR,i}\boldsymbol{\Phi}^H\bar{\mathbf{H}}_2^H\bar{\mathbf{H}}_2\odot\left(\mathbf{R}_{VR,k}\right)^T\right)\boldsymbol{c}\right\} \\
		&+2\operatorname{Im}\left\{\boldsymbol{\Phi}^H\left(\bar{\mathbf{H}}_2^H\bar{\mathbf{H}}_2\boldsymbol{\Phi}\mathbf{R}_{VR,k}\boldsymbol{\Phi}^H\bar{\mathbf{H}}_2^H\bar{\mathbf{H}}_2\odot\left(\mathbf{R}_{VR,i}\right)^T\right)\boldsymbol{c}\right\} \\
		=&\boldsymbol{f}_a\big(\bar{\mathbf{H}}_2^H\bar{\mathbf{H}}_2\boldsymbol{\Phi}\mathbf{R}_{VR,i}\boldsymbol{\Phi}^H\bar{\mathbf{H}}_2^H\bar{\mathbf{H}}_2,\mathbf{R}_{VR,k}\big)\\
		&+\boldsymbol{f}_a\big(\bar{\mathbf{H}}_2^H\bar{\mathbf{H}}_2\boldsymbol{\Phi}\mathbf{R}_{VR,k}\boldsymbol{\Phi}^H\bar{\mathbf{H}}_2^H\bar{\mathbf{H}}_2,\mathbf{R}_{VR,i}\big).
	\end{aligned}
\end{align}

For $f_{ki,2}^{\prime}(\boldsymbol{\theta})$, we have
\begin{align}\label{C.4}
	\begin{aligned}
		f_{ki,2}^{\prime}(\boldsymbol{\theta}) =&\frac{\partial(\bar{\mathbf{h}}_k^H\mathbf{D}_k^{1/2}\boldsymbol{\Phi}^H\mathbf{R}_{ris}\boldsymbol{\Phi}\mathbf{D}_i^{1/2}\bar{\mathbf{h}}_i)}{\partial\boldsymbol{\theta}}  \\
		=&\frac{\partial\left(\mathrm{Tr}(\bar{\mathbf{h}}_k^H\mathbf{D}_k^{1/2}\mathbf{\Phi}^H\mathbf{R}_{ris}\mathbf{\Phi}\mathbf{D}_i^{1/2}\bar{\mathbf{h}}_i)\right)}{\partial\boldsymbol{\theta}} \\
		=&2\operatorname{Im}\left\{\boldsymbol{\Phi}^H\left(\mathbf{R}_{ris}\odot\left(\mathbf{D}_i^{1/2}\bar{\mathbf{h}}_i\bar{\mathbf{h}}_k^H\mathbf{D}_k^{1/2}\right)^T\right)\boldsymbol{c}\right\} \\
		=&\boldsymbol{f}_{a}(\mathbf{R}_{ris},\mathbf{D}_{i}^{1/2}\mathbf{\bar{h}}_{i}\mathbf{\bar{h}}_{k}^{H}\mathbf{D}_{k}^{1/2}).
	\end{aligned}
\end{align}

For $f_{k,3,1}^{\prime}(\boldsymbol{\theta})$, we have
\begin{align}\label{C.5}
	\begin{aligned}
		f_{k,3,1}^{\prime}(\boldsymbol{\theta}) =&\frac{\partial\left(\mathrm{Tr}(\mathbf{R}_{ris}\mathbf{\Phi}\mathbf{R}_{VR,k}\mathbf{\Phi}^H)\right)}{\partial\boldsymbol{\theta}}  \\
		=&2\operatorname{Im}\left\{\mathbf{\Phi}^H\left(\mathbf{R}_{ris}\odot\left(\mathbf{R}_{VR,k}\right)^T\right)\mathbf{c}\right\} \\
		=&\boldsymbol{f}_{a}(\mathbf{R}_{ris},\mathbf{R}_{VR,k}).
	\end{aligned}
\end{align}

For $f_{ki,3,2}^{\prime}(\boldsymbol{\theta})$, we have
\begin{align}\label{C.6}
	\begin{aligned}
		&f_{ki,3,2}^{\prime}(\boldsymbol{\theta})\\
		=&\frac{\partial\left(\operatorname{Tr}(\mathbf{R}_{ris}\boldsymbol{\Phi}\mathbf{R}_{VR,k}\boldsymbol{\Phi}^H\mathbf{R}_{ris}\boldsymbol{\Phi}\mathbf{R}_{VR,i}\boldsymbol{\Phi}^H)\right)}{\partial\boldsymbol{\theta}}  \\
		=&2\operatorname{Im}\left\{\mathbf{\Phi}^H\left(\mathbf{R}_{ris}(\mathbf{\Phi}\mathbf{R}_{VR,i}\mathbf{\Phi}^H)\mathbf{R}_{ris}\odot\left(\mathbf{R}_{VR,k}\right)^T\right)\mathbf{c}\right\} \\
		&+2\operatorname{Im}\left\{\boldsymbol{\Phi}^H\left(\boldsymbol{R}_{ris}(\boldsymbol{\Phi}\mathbf{R}_{VR,k}\boldsymbol{\Phi}^H)\mathbf{R}_{ris}\odot\left(\mathbf{R}_{VR,i}\right)^T\right)\boldsymbol{c}\right\} \\
		=&\boldsymbol{f}_a(\mathbf{R}_{ris}(\mathbf{\Phi}\mathbf{R}_{VR,i}\mathbf{\Phi}^H)\mathbf{R}_{ris},\mathbf{R}_{VR,k}) \\
		&+\boldsymbol{f}_a(\mathbf{R}_{ris}(\mathbf{\Phi}\mathbf{R}_{VR,k}\mathbf{\Phi}^H)\mathbf{R}_{ris},\mathbf{R}_{VR,i}).
	\end{aligned}
\end{align}

For $f_{ki,4}^{\prime}(\boldsymbol{\theta})$, we have
\begin{align}\label{C.7}
	\begin{aligned}
		&f_{ki,4}^{\prime}(\boldsymbol{\theta})\\
		=&\frac{\partial\left(\mathrm{Tr}(\bar{\mathbf{h}}_k^H\mathbf{D}_k^{1/2}\mathbf{\Phi}^H\mathbf{R}_{ris}\mathbf{\Phi}\mathbf{R}_{VR,i}\mathbf{\Phi}^H\mathbf{R}_{ris}\mathbf{\Phi}\mathbf{D}_k^{1/2}\mathbf{\bar{h}}_k)\right)}{\partial\boldsymbol{\theta}}  \\
		=&2\operatorname{Im}\left\{\mathbf{\Phi}^H\left(\mathbf{R}_{ris}(\mathbf{\Phi}\mathbf{D}_k^{1/2}\mathbf{\bar{h}}_k\mathbf{\bar{h}}_k^H\mathbf{D}_k^{1/2}\mathbf{\Phi}^H)\mathbf{R}_{ris}\right.\right.\\
		&\left.\left.\odot(\mathbf{R}_{VR,i})^T\right)\boldsymbol{c}\right\} \\
		&+2\operatorname{Im}\left\{\boldsymbol{\Phi}^H\left(\mathbf{R}_{ris}(\boldsymbol{\Phi}\mathbf{R}_{VR,i}\boldsymbol{\Phi}^H)\mathbf{R}_{ris}\right.\right.\\
		&\left.\left.\odot\left(\mathbf{D}_k^{1/2}\bar{\mathbf{h}}_k\bar{\mathbf{h}}_k^H\mathbf{D}_k^{1/2}\right)^T\right)\boldsymbol{c}\right\} \\
		=&\boldsymbol{f}_a(\mathbf{R}_{ris}\mathbf{\Phi}\mathbf{D}_k^{1/2}\mathbf{\bar{h}}_k\mathbf{\bar{h}}_k^H\mathbf{D}_k^{1/2}\mathbf{\Phi}^H\mathbf{R}_{ris},\mathbf{R}_{VR,i}) \\
		&+\boldsymbol{f}_a(\mathbf{R}_{ris}\boldsymbol{\Phi}\mathbf{R}_{VR,i}\boldsymbol{\Phi}^H\mathbf{R}_{ris},\mathbf{D}_k^{1/2}\mathbf{\bar{h}}_k\mathbf{\bar{h}}_k^H\mathbf{D}_k^{1/2}).
	\end{aligned}
\end{align}

For $f_{ik,4}^{\prime}(\boldsymbol{\theta})$, we have
\begin{align}\label{C.8}
	\begin{aligned}
		f_{ik,4}^{\prime}(\boldsymbol{\theta}) =&\boldsymbol{f}_a(\mathbf{R}_{ris}\boldsymbol{\Phi}\mathbf{D}_i^{1/2}\mathbf{\bar{h}}_i\mathbf{\bar{h}}_i^H\mathbf{D}_i^{1/2}\boldsymbol{\Phi}^H\mathbf{R}_{ris},\mathbf{R}_{VR,k})  \\
		&+\boldsymbol{f}_a(\mathbf{R}_{ris}\boldsymbol{\Phi}\mathbf{R}_{VR,k}\boldsymbol{\Phi}^H\mathbf{R}_{ris},\mathbf{D}_i^{1/2}\bar{\mathbf{h}}_i\bar{\mathbf{h}}_i^H\mathbf{D}_i^{1/2}).
	\end{aligned}
\end{align}

For $f_{ki,5}^{\prime}(\boldsymbol{\theta})$, we have
\begin{align}\label{C.9}
	\begin{aligned}
		&f_{ki,5}^{\prime}(\boldsymbol{\theta})\\
		=&\frac{\partial\left(\mathrm{Tr}(\bar{\mathbf{H}}_2\boldsymbol{\Phi}\mathbf{R}_{VR,k}\boldsymbol{\Phi}^H\mathbf{R}_{ris}\boldsymbol{\Phi}\mathbf{R}_{VR,i}\boldsymbol{\Phi}^H\mathbf{\bar{H}}_2^H)\right)}{\partial\boldsymbol{\theta}}  \\
		=&2\operatorname{Im}\left\{\boldsymbol{\Phi}^H\left(\bar{\mathbf{H}}_2^H\bar{\mathbf{H}}_2\boldsymbol{\Phi}\mathbf{R}_{VR,k}\boldsymbol{\Phi}^H\mathbf{R}_{ris}\odot\left(\mathbf{R}_{VR,i}\right)^T\right)\boldsymbol{c}\right\} \\
		&+2\operatorname{Im}\left\{\boldsymbol{\Phi}^H\left(\mathbf{R}_{ris}\boldsymbol{\Phi}\mathbf{R}_{VR,i}\boldsymbol{\Phi}^H\bar{\mathbf{H}}_2^H\bar{\mathbf{H}}_2\odot\left(\mathbf{R}_{VR,k}\right)^T\right)\boldsymbol{c}\right\} \\
		=&\boldsymbol{f}_a(\bar{\mathbf{H}}_2^H\bar{\mathbf{H}}_2\boldsymbol{\Phi}\mathbf{R}_{VR,k}\boldsymbol{\Phi}^H\mathbf{R}_{ris},\mathbf{R}_{VR,i})&   \\
		&+\boldsymbol{f}_{a}(\mathbf{R}_{ris}\boldsymbol{\Phi}\mathbf{R}_{VR,i}\boldsymbol{\Phi}^{H}\mathbf{\bar{H}}_{2}^{H}\mathbf{\bar{H}}_{2},\mathbf{R}_{VR,k}).
	\end{aligned}
\end{align}

For $f_{ik,5}^{\prime}(\boldsymbol{\theta})$, we have
\begin{align}\label{C.10}
	\begin{aligned}
		f_{ik,5}^{\prime}(\boldsymbol{\theta}) =&\frac{\partial\left(f_{ki,5}^*(\boldsymbol{\theta})\right)}{\partial\boldsymbol{\theta}}  \\
		=&\boldsymbol{f}_{a}(\bar{\mathbf{H}}_{2}^{H}\bar{\mathbf{H}}_{2},\mathbf{R}_{VR,i}\boldsymbol{\Phi}^{H}\mathbf{R}_{ris}\boldsymbol{\Phi}\mathbf{R}_{VR,k}) \\
		&+\boldsymbol{f}_{a}(\mathbf{R}_{ris},\mathbf{R}_{VR,k}\mathbf{\Phi}^{H}\mathbf{\bar{H}}_{2}^{H}\mathbf{\bar{H}}_{2}\mathbf{\Phi}\mathbf{R}_{VR,i}).
	\end{aligned}
\end{align}

For $f_{ki,6}^{\prime}(\boldsymbol{\theta})$, we have
\begin{align}\label{C.11}
	\begin{aligned}
		&f_{ki,6}^{\prime}(\boldsymbol{\theta})\\
		=&\frac{\partial(\bar{\mathbf{h}}_k^H\mathbf{D}_k^{1/2}\mathbf{\Phi}^H\bar{\mathbf{H}}_2^H\bar{\mathbf{H}}_2\mathbf{\Phi}\mathbf{R}_{VR,i}\mathbf{\Phi}^H\mathbf{R}_{ris}\mathbf{\Phi}\mathbf{D}_k^{1/2}\bar{\mathbf{h}}_k)}{\partial\boldsymbol{\theta}}  \\
		=&2\operatorname{Im}\left\{\mathbf{\Phi}^H\left(\mathbf{R}_{ris}(\mathbf{\Phi}\mathbf{D}_k^{1/2}\mathbf{\bar{h}}_k\mathbf{\bar{h}}_k^H\mathbf{D}_k^{1/2}\mathbf{\Phi}^H)\mathbf{\bar{H}}_2^H\mathbf{\bar{H}}_2\right.\right. \\
		&\left.\left.\odot\left(\mathbf{R}_{VR,i}\right)^T\right)\mathbf{c}\right\}+2\operatorname{lm}\left\{\boldsymbol{\Phi}^H\left(\bar{\mathbf{H}}_2^H\bar{\mathbf{H}}_2(\boldsymbol{\Phi}\mathbf{R}_{VR,i}\boldsymbol{\Phi}^H)\mathbf{R}_{ris}\right.\right.\\
		&\left.\left.\odot\left(\mathbf{D}_k^{1/2}\bar{\mathbf{h}}_k\bar{\mathbf{h}}_k^H\mathbf{D}_k^{1/2}\right)^T\right)\boldsymbol{c}\right\} \\
		=&\boldsymbol{f}_a(\mathbf{R}_{ris}\boldsymbol{\Phi}\mathbf{D}_k^{1/2}\bar{\mathbf{h}}_k\bar{\mathbf{h}}_k^H\mathbf{D}_k^{1/2}\boldsymbol{\Phi}^H\bar{\mathbf{H}}_2^H\bar{\mathbf{H}}_2,\mathbf{R}_{VR,i}) \\
		&+\boldsymbol{f}_{a}(\bar{\mathbf{H}}_{2}^{H}\bar{\mathbf{H}}_{2}\boldsymbol{\Phi}\mathbf{R}_{VR,i}\boldsymbol{\Phi}^{H}\mathbf{R}_{ris},\mathbf{D}_{k}^{1/2}\bar{\mathbf{h}}_{k}\bar{\mathbf{h}}_{k}^{H}\mathbf{D}_{k}^{1/2}).
	\end{aligned}
\end{align}

For $f_{ik,6}^{\prime}(\boldsymbol{\theta})$, we have
\begin{align}\label{C.12}
	\begin{aligned}
		f_{ik,6}^{\prime}(\boldsymbol{\theta})& =\boldsymbol{f}_{a}(\mathbf{R}_{ris}\boldsymbol{\Phi}\mathbf{D}_{i}^{1/2}\bar{\mathbf{h}}_{i}\bar{\mathbf{h}}_{i}^{H}\mathbf{D}_{i}^{1/2}\boldsymbol{\Phi}^{H}\bar{\mathbf{H}}_{2}^{H}\bar{\mathbf{H}}_{2},\mathbf{R}_{VR,k})  \\
		&+\boldsymbol{f}_a(\bar{\mathbf{H}}_2^H\bar{\mathbf{H}}_2\boldsymbol{\Phi}\mathbf{R}_{VR,k}\boldsymbol{\Phi}^H\mathbf{R}_{ris},\mathbf{D}_i^{1/2}\bar{\mathbf{h}}_i\bar{\mathbf{h}}_i^H\mathbf{D}_i^{1/2}).
	\end{aligned}
\end{align}

For $f_{ik,6}^*(\boldsymbol{\theta})$, we have
\begin{align}\label{C.13}
	\begin{aligned}
		&f_{ik,6}^*(\boldsymbol{\theta})\\
		=&\frac{\partial(\bar{\mathbf{h}}_k^H\mathbf{D}_k^{1/2}\mathbf{\Phi}^H\mathbf{R}_{ris}\mathbf{\Phi}\mathbf{R}_{VR,i}\mathbf{\Phi}^H\bar{\mathbf{H}}_2^H\bar{\mathbf{H}}_2\mathbf{\Phi}\mathbf{D}_k^{1/2}\bar{\mathbf{h}}_k)}{\partial\boldsymbol{\theta}}  \\
		=&\frac{\partial\left(\mathrm{Tr}\left(\mathbf{R}_{ris}(\boldsymbol{\Phi}\mathbf{R}_{VR,i}\boldsymbol{\Phi}^H)\bar{\mathbf{H}}_2^H\bar{\mathbf{H}}_2(\boldsymbol{\Phi}\mathbf{D}_k^{1/2}\bar{\mathbf{h}}_k\bar{\mathbf{h}}_k^H\mathbf{D}_k^{1/2}\boldsymbol{\Phi}^H)\right)\right)}{\partial\boldsymbol{\theta}} \\
		=&2\operatorname{Im}\left\{\boldsymbol{\Phi}^H\left(\bar{\mathbf{H}}_2^H\bar{\mathbf{H}}_2(\boldsymbol{\Phi}\mathbf{D}_k^{1/2}\bar{\mathbf{h}}_k\bar{\mathbf{h}}_k^H\mathbf{D}_k^{1/2}\boldsymbol{\Phi}^H)\mathbf{R}_{ris}\right.\right.\\
		&\left.\left.\odot(\mathbf{R}_{VR,i})^T\right)\boldsymbol{c}\right\} \\
		&+2\operatorname{lm}\left\{\boldsymbol{\Phi}^H\left(\mathbf{R}_{ris}(\boldsymbol{\Phi}\mathbf{R}_{VR,i}\boldsymbol{\Phi}^H)\bar{\mathbf{H}}_2^H\bar{\mathbf{H}}_2\right.\right.\\
		&\left.\left.\odot\left(\mathbf{D}_k^{1/2}\bar{\mathbf{h}}_k\bar{\mathbf{h}}_k^H\mathbf{D}_k^{1/2}\right)^T\right)\boldsymbol{c}\right\} \\
		=&\boldsymbol{f}_a(\mathbf{R}_{ris}\boldsymbol{\Phi}\mathbf{R}_{VR,i}\boldsymbol{\Phi}^H\bar{\mathbf{H}}_2^H\bar{\mathbf{H}}_2,\mathbf{D}_k^{1/2}\bar{\mathbf{h}}_k\bar{\mathbf{h}}_k^H\mathbf{D}_k^{1/2}) \\
		&+\boldsymbol{f}_a(\bar{\mathbf{H}}_2^H\bar{\mathbf{H}}_2\mathbf{\Phi}\mathbf{D}_k^{1/2}\bar{\mathbf{h}}_k\bar{\mathbf{h}}_k^H\mathbf{D}_k^{1/2}\boldsymbol{\Phi}^H\mathbf{R}_{ris},\mathbf{R}_{VR,i}).
	\end{aligned}
\end{align}

For $f_{ki,7}^{\prime}(\boldsymbol{\theta})$, we have
\begin{align}\label{C.14}
	\begin{aligned}
		f_{ki,7}^{\prime}(\boldsymbol{\theta})& =\frac{\partial(\bar{\mathbf{h}}_k^H\mathbf{D}_k^{1/2}\mathbf{\Phi}^H\mathbf{a}_N\mathbf{a}_N^H\mathbf{\Phi}\mathbf{D}_i^{1/2}\bar{\mathbf{h}}_i)}{\partial\boldsymbol{\theta}}  \\
		&=\frac{\partial\left(\mathrm{Tr}(\bar{\mathbf{h}}_k^H\mathbf{D}_k^{1/2}\mathbf{\Phi}^H\mathbf{a}_N\mathbf{a}_N^H\mathbf{\Phi}\mathbf{D}_i^{1/2}\bar{\mathbf{h}}_i)\right)}{\partial\boldsymbol{\theta}} \\
		&=2\operatorname{Im}\left\{\mathbf{\Phi}^H\left(\mathbf{a}_N\mathbf{a}_N^H\odot\left(\mathbf{D}_i^{1/2}\bar{\mathbf{h}}_i\bar{\mathbf{h}}_k^H\mathbf{D}_k^{1/2}\right)^T\right)\mathbf{c}\right\} \\
		&=\boldsymbol{f}_{a}(\mathbf{a}_{N}\mathbf{a}_{N}^{H},\mathbf{D}_{i}^{1/2}\mathbf{\bar{h}}_{i}\mathbf{\bar{h}}_{k}^{H}\mathbf{D}_{k}^{1/2}).
	\end{aligned}
\end{align}

Therefore, we can obtain the gradients of the auxiliary functions for the required achievable rate expression as shown in (\ref{fk_VR_gradient}).

\end{appendices}

\nocite{*}
\bibliographystyle{IEEEtran}
\bibliography{mylib}

\begin{thebibliography}{10}
\providecommand{\url}[1]{#1}
\csname url@samestyle\endcsname
\providecommand{\newblock}{\relax}
\providecommand{\bibinfo}[2]{#2}
\providecommand{\BIBentrySTDinterwordspacing}{\spaceskip=0pt\relax}
\providecommand{\BIBentryALTinterwordstretchfactor}{4}
\providecommand{\BIBentryALTinterwordspacing}{\spaceskip=\fontdimen2\font plus
\BIBentryALTinterwordstretchfactor\fontdimen3\font minus
  \fontdimen4\font\relax}
\providecommand{\BIBforeignlanguage}[2]{{%
\expandafter\ifx\csname l@#1\endcsname\relax
\typeout{** WARNING: IEEEtran.bst: No hyphenation pattern has been}%
\typeout{** loaded for the language `#1'. Using the pattern for}%
\typeout{** the default language instead.}%
\else
\language=\csname l@#1\endcsname
\fi
#2}}
\providecommand{\BIBdecl}{\relax}
\BIBdecl

\bibitem{5}
M.~Di~Renzo, A.~Zappone, M.~Debbah, M.-S. Alouini, C.~Yuen, J.~de~Rosny, and
  S.~Tretyakov, ``Smart radio environments empowered by reconfigurable
  intelligent surfaces: How it works, state of research, and the road ahead,''
  \emph{IEEE Journal on Selected Areas in Communications}, vol.~38, no.~11, pp.
  2450--2525, 2020.

\bibitem{6}
C.~Pan, H.~Ren, K.~Wang, J.~F. Kolb, M.~Elkashlan, M.~Chen, M.~Di~Renzo,
  Y.~Hao, J.~Wang, A.~L. Swindlehurst, X.~You, and L.~Hanzo, ``Reconfigurable
  intelligent surfaces for 6g systems: Principles, applications, and research
  directions,'' \emph{IEEE Communications Magazine}, vol.~59, no.~6, pp.
  14--20, 2021.

\bibitem{a1}
W.~Tang, M.~Z. Chen, X.~Chen, J.~Y. Dai, Y.~Han, M.~Di~Renzo, Y.~Zeng, S.~Jin,
  Q.~Cheng, and T.~J. Cui, ``Wireless communications with reconfigurable
  intelligent surface: Path loss modeling and experimental measurement,''
  \emph{IEEE Transactions on Wireless Communications}, vol.~20, no.~1, pp.
  421--439, 2021.

\bibitem{a2}
Q.~Wu, S.~Zhang, B.~Zheng, C.~You, and R.~Zhang, ``Intelligent reflecting
  surface-aided wireless communications: A tutorial,'' \emph{IEEE Transactions
  on Communications}, vol.~69, no.~5, pp. 3313--3351, 2021.

\bibitem{a3}
C.~Pan, G.~Zhou, K.~Zhi, S.~Hong, T.~Wu, Y.~Pan, H.~Ren, M.~D. Renzo,
  A.~Lee~Swindlehurst, R.~Zhang, and A.~Y. Zhang, ``An overview of signal
  processing techniques for ris/irs-aided wireless systems,'' \emph{IEEE
  Journal of Selected Topics in Signal Processing}, vol.~16, no.~5, pp.
  883--917, 2022.

\bibitem{a4}
G.~Zhou, C.~Pan, H.~Ren, P.~Popovski, and A.~L. Swindlehurst, ``Channel
  estimation for ris-aided multiuser millimeter-wave systems,'' \emph{IEEE
  Transactions on Signal Processing}, vol.~70, pp. 1478--1492, 2022.

\bibitem{a5}
Z.~Peng, C.~Pan, G.~Zhou, H.~Ren, S.~Jin, P.~Popovski, R.~Schober, and X.~You,
  ``Two-stage channel estimation for ris-aided multiuser mmwave systems with
  reduced error propagation and pilot overhead,'' \emph{IEEE Transactions on
  Signal Processing}, vol.~71, pp. 3607--3622, 2023.

\bibitem{a6}
H.~Jiang, B.~Xiong, H.~Zhang, and E.~Basar, ``Physics-based 3d end-to-end
  modeling for double-ris assisted non-stationary uav-to-ground communication
  channels,'' \emph{IEEE Transactions on Communications}, vol.~71, no.~7, pp.
  4247--4261, 2023.

\bibitem{a7}
H.~Ren, X.~Liu, C.~Pan, Z.~Peng, and J.~Wang, ``Performance analysis for
  ris-aided secure massive mimo systems with statistical csi,'' \emph{IEEE
  Wireless Communications Letters}, vol.~12, no.~1, pp. 124--128, 2023.

\bibitem{a8}
Y.~Guo, P.~Sun, Z.~Yuan, C.~Huang, Q.~Guo, Z.~Wang, and C.~Yuen, ``Efficient
  channel estimation for ris-aided mimo communications with unitary approximate
  message passing,'' \emph{IEEE Transactions on Wireless Communications},
  vol.~22, no.~2, pp. 1403--1416, 2023.

\bibitem{21}
S.~Zhou, W.~Xu, K.~Wang, M.~Di~Renzo, and M.-S. Alouini, ``Spectral and energy
  efficiency of irs-assisted miso communication with hardware impairments,''
  \emph{IEEE Wireless Communications Letters}, vol.~9, no.~9, pp. 1366--1369,
  2020.

\bibitem{a9}
T.~Chen, M.~You, Y.~Zhang, G.~Zheng, J.~B. Gros, G.~Lerosey, Y.~Nasser,
  F.~Burton, and G.~Gradoni, ``Model-free optimization and experimental
  validation of ris-assisted wireless communications under rich multipath
  fading,'' \emph{IEEE Wireless Communications Letters}, vol.~13, no.~3, pp.
  627--631, 2024.

\bibitem{a10}
C.~Huang, S.~Hu, G.~C. Alexandropoulos, A.~Zappone, C.~Yuen, R.~Zhang, M.~D.
  Renzo, and M.~Debbah, ``Holographic mimo surfaces for 6g wireless networks:
  Opportunities, challenges, and trends,'' \emph{IEEE Wireless Communications},
  vol.~27, no.~5, pp. 118--125, 2020.

\bibitem{a11}
B.~Zheng, C.~You, and R.~Zhang, ``Double-irs assisted multi-user mimo:
  Cooperative passive beamforming design,'' \emph{IEEE Transactions on Wireless
  Communications}, vol.~20, no.~7, pp. 4513--4526, 2021.

\bibitem{27}
K.~Zhi, C.~Pan, H.~Ren, K.~Wang, M.~Elkashlan, M.~D. Renzo, R.~Schober, H.~V.
  Poor, J.~Wang, and L.~Hanzo, ``Two-timescale design for reconfigurable
  intelligent surface-aided massive mimo systems with imperfect csi,''
  \emph{IEEE Transactions on Information Theory}, vol.~69, no.~5, pp.
  3001--3033, 2023.

\bibitem{28}
Z.~Wang, L.~Liu, and S.~Cui, ``Channel estimation for intelligent reflecting
  surface assisted multiuser communications: Framework, algorithms, and
  analysis,'' \emph{IEEE Transactions on Wireless Communications}, vol.~19,
  no.~10, pp. 6607--6620, 2020.

\bibitem{12}
Y.~Han, W.~Tang, S.~Jin, C.-K. Wen, and X.~Ma, ``Large intelligent
  surface-assisted wireless communication exploiting statistical csi,''
  \emph{IEEE Transactions on Vehicular Technology}, vol.~68, no.~8, pp.
  8238--8242, 2019.

\bibitem{30}
Y.~Jia, C.~Ye, and Y.~Cui, ``Analysis and optimization of an intelligent
  reflecting surface-assisted system with interference,'' in \emph{ICC 2020 -
  2020 IEEE International Conference on Communications (ICC)}, 2020, pp. 1--6.

\bibitem{36}
A.~Abrardo, D.~Dardari, and M.~Di~Renzo, ``Intelligent reflecting surfaces:
  Sum-rate optimization based on statistical position information,'' \emph{IEEE
  Transactions on Communications}, vol.~69, no.~10, pp. 7121--7136, 2021.

\bibitem{37}
Y.~Gao, J.~Xu, W.~Xu, D.~W.~K. Ng, and M.-S. Alouini, ``Distributed irs with
  statistical passive beamforming for miso communications,'' \emph{IEEE
  Wireless Communications Letters}, vol.~10, no.~2, pp. 221--225, 2021.

\bibitem{b1}
C.~Hu, L.~Dai, S.~Han, and X.~Wang, ``Two-timescale channel estimation for
  reconfigurable intelligent surface aided wireless communications,''
  \emph{IEEE Transactions on Communications}, vol.~69, no.~11, pp. 7736--7747,
  2021.

\bibitem{b2}
Y.~Cao, T.~Lv, and W.~Ni, ``Two-timescale optimization for intelligent
  reflecting surface-assisted mimo transmission in fast-changing channels,''
  \emph{IEEE Transactions on Wireless Communications}, vol.~21, no.~12, pp.
  10\,424--10\,437, 2022.

\bibitem{b3}
X.~Gan, C.~Zhong, C.~Huang, Z.~Yang, and Z.~Zhang, ``Multiple riss assisted
  cell-free networks with two-timescale csi: Performance analysis and system
  design,'' \emph{IEEE Transactions on Communications}, vol.~70, no.~11, pp.
  7696--7710, 2022.

\bibitem{b4}
S.~Yang, W.~Lyu, Y.~Xiu, Z.~Zhang, and C.~Yuen, ``Active 3d double-ris-aided
  multi-user communications: Two-timescale-based separate channel estimation
  via bayesian learning,'' \emph{IEEE Transactions on Communications}, vol.~71,
  no.~6, pp. 3605--3620, 2023.

\bibitem{40}
Y.~Han, S.~Jin, C.-K. Wen, and T.~Q.~S. Quek, ``Localization and channel
  reconstruction for extra large ris-assisted massive mimo systems,''
  \emph{IEEE Journal of Selected Topics in Signal Processing}, vol.~16, no.~5,
  pp. 1011--1025, 2022.

\bibitem{c1}
K.~Zhi, C.~Pan, H.~Ren, K.~K. Chai, C.-X. Wang, R.~Schober, and X.~You,
  ``Performance analysis and low-complexity design for xl-mimo with near-field
  spatial non-stationarities,'' \emph{IEEE Journal on Selected Areas in
  Communications}, pp. 1--1, 2024.

\bibitem{c2}
E.~D. Carvalho, A.~Ali, A.~Amiri, M.~Angjelichinoski, and R.~W. Heath,
  ``Non-stationarities in extra-large-scale massive mimo,'' \emph{IEEE Wireless
  Communications}, vol.~27, no.~4, pp. 74--80, 2020.

\bibitem{43}
X.~Li, S.~Zhou, E.~Björnson, and J.~Wang, ``Capacity analysis for spatially
  non-wide sense stationary uplink massive mimo systems,'' \emph{IEEE
  Transactions on Wireless Communications}, vol.~14, no.~12, pp. 7044--7056,
  2015.

\bibitem{46}
E.~D. Carvalho, A.~Ali, A.~Amiri, M.~Angjelichinoski, and R.~W. Heath,
  ``Non-stationarities in extra-large-scale massive mimo,'' \emph{IEEE Wireless
  Communications}, vol.~27, no.~4, pp. 74--80, 2020.

\bibitem{47}
M.~Cui and L.~Dai, ``Channel estimation for extremely large-scale mimo:
  Far-field or near-field?'' \emph{IEEE Transactions on Communications},
  vol.~70, no.~4, pp. 2663--2677, 2022.

\bibitem{51}
A.~Ali, E.~D. Carvalho, and R.~W. Heath, ``Linear receivers in non-stationary
  massive mimo channels with visibility regions,'' \emph{IEEE Wireless
  Communications Letters}, vol.~8, no.~3, pp. 885--888, 2019.

\bibitem{48}
X.~Wei, L.~Dai, Y.~Zhao, G.~Yu, and X.~Duan, ``Codebook design and beam
  training for extremely large-scale ris: Far-field or near-field?''
  \emph{China Communications}, vol.~19, no.~6, pp. 193--204, 2022.

\bibitem{50}
X.~Yu, W.~Shen, R.~Zhang, C.~Xing, and T.~Q.~S. Quek, ``Channel estimation for
  xl-ris-aided millimeter-wave systems,'' \emph{IEEE Transactions on
  Communications}, vol.~71, no.~9, pp. 5519--5533, 2023.

\bibitem{52}
D.~Gunasinghe and G.~Amarasuriya, ``Achievable rate analysis for extra-large
  ris-aided massive mimo with visibility regions,'' in \emph{ICC 2023 - IEEE
  International Conference on Communications}, 2023, pp. 1542--1547.

\bibitem{58}
B.~Zheng, C.~You, and R.~Zhang, ``Double-irs assisted multi-user mimo:
  Cooperative passive beamforming design,'' \emph{IEEE Transactions on Wireless
  Communications}, vol.~20, no.~7, pp. 4513--4526, 2021.

\bibitem{d1}
K.~Zhi, C.~Pan, H.~Ren, and K.~Wang, ``Power scaling law analysis and phase
  shift optimization of ris-aided massive mimo systems with statistical csi,''
  \emph{IEEE Transactions on Communications}, vol.~70, no.~5, pp. 3558--3574,
  2022.

\bibitem{d2}
Y.~Jia, C.~Ye, and Y.~Cui, ``Analysis and optimization of an intelligent
  reflecting surface-assisted system with interference,'' in \emph{ICC 2020 -
  2020 IEEE International Conference on Communications (ICC)}, 2020, pp. 1--6.

\bibitem{67}
E.~Björnson and L.~Sanguinetti, ``Rayleigh fading modeling and channel
  hardening for reconfigurable intelligent surfaces,'' \emph{IEEE Wireless
  Communications Letters}, vol.~10, no.~4, pp. 830--834, 2021.

\bibitem{d3}
T.~L. Marzetta, E.~G. Larsson, and H.~Yang, \emph{Fundamentals of massive
  MIMO}.\hskip 1em plus 0.5em minus 0.4em\relax Cambridge University Press,
  2016.

\bibitem{a}
Q.~Zhang, Z.~Lu, S.~Jin, K.-K. Wong, H.~Zhu, and M.~Matthaiou, ``Power scaling
  of massive mimo systems with arbitrary-rank channel means and imperfect
  csi,'' in \emph{2013 IEEE Global Communications Conference (GLOBECOM)}, 2013,
  pp. 4157--4162.

\bibitem{d4}
L.~Xingsi, ``An entropy-based aggregate method for minimax optimization,''
  \emph{Engineering Optimization}, vol.~18, no.~4, pp. 277--285, 1992.

\end{thebibliography}

\end{document}